\documentclass[a4paper,11pt]{article}
\pdfoutput=1
\usepackage[utf8x]{inputenc}
\usepackage{jheppub}
\usepackage{epsf}
\usepackage{graphicx}	
\usepackage{dcolumn}	
\usepackage{bm}			
\usepackage{amssymb, amsmath}
\usepackage{wasysym}
\usepackage{lipsum, color}
\usepackage{physics}
\usepackage{bbm}		
\usepackage{multirow}
\usepackage{float}

\usepackage{slashed}
\usepackage{hyperref}

\let\oldquote\quote
\renewcommand\quote{\scriptsize\oldquote}
\let\oldquotation\quotation
\renewcommand\quotation{\scriptsize\oldquotation}
\renewcommand{\vec}{\mathbf}

\newcommand{\gsim}{\lower.7ex\hbox{$\;\stackrel{\textstyle>}{\sim}\;$}}
\newcommand{\lsim}{\lower.7ex\hbox{$\;\stackrel{\textstyle<}{\sim}\;$}}

\def\beq{\begin{equation}}
\def\eeq{\end{equation}}
\def\be{\begin{equation}}
\def\ee{\end{equation}}
\def\bea{\begin{eqnarray}}
\def\eea{\end{eqnarray}}
\def\bmat{\begin{pmatrix}}
	\def\emat{\end{pmatrix}}
\def\bei{\begin{itemize}}
	\def\eei{\end{itemize}}

\newcommand{\vk}{\mathbf{k}}
\newcommand{\vq}{\mathbf{q}}
\newcommand{\te}{\tilde \eta}

\usepackage[normalem]{ulem}

\makeatletter
\def\section{\@startsection {section}{1}{\z@}{-3.5ex plus -1ex minus
		-.2ex}{2.3ex plus .2ex}{\large\bf}}
\def\subsection{\@startsection{subsection}{2}{\z@}{-3.25ex plus -1ex
		minus -.2ex}{1.5ex plus .2ex}{\normalsize\bf}}
\makeatother
\makeatletter

\@addtoreset{equation}{section}

\makeatother

\usepackage{graphicx}  
\usepackage{dcolumn}   
\usepackage{bm,relsize}        
\usepackage{amssymb, amsmath}
\usepackage{wasysym}
\usepackage{slashed}
\usepackage{bbold} 
\usepackage{mathtools}
\usepackage{comment}
\usepackage{cancel}
\usepackage{ulem}

\usepackage{feyn}

\def\beq{\begin{equation}}
\def\eeq{\end{equation}}

\begin{document}

\global\long\def\com#1#2{\underset{{\scriptstyle #2}}{\underbrace{#1}}}

\global\long\def\comtop#1#2{\overset{{\scriptstyle #2}}{\overbrace{#1}}}

\global\long\def\ket#1{\left|#1\right\rangle }

\global\long\def\bra#1{\left\langle #1\right|}

\global\long\def\braket#1#2{\left\langle #1|#2\right\rangle }

\global\long\def\op#1#2{\left|#1\right\rangle \left\langle #2\right|}

\global\long\def\opk#1#2#3{\left\langle #1|#2|#3\right\rangle }

\global\long\def\L{\mathcal{L}}

\title{The Effective Field Theory of Large Scale Structure\\ for Mixed Dark Matter Scenarios}

\author[a,b,c,d]{Francesco Verdiani,}
\author[e,f]{Emanuele Castorina,}
\author[g,h,i]{Ennio Salvioni,$\,$}
\author[b,c,d]{\mbox{Emiliano Sefusatti$\,$}}
\affiliation[a]{Scuola Internazionale Superiore di Studi Avanzati (SISSA),\\ Via Bonomea 265, 34136 Trieste, Italy}
\affiliation[b]{INFN Sezione di Trieste, Via Valerio 2, 34127 Trieste, Italy}
\affiliation[c]{Istituto Nazionale di Astrofisica, Osservatorio Astronomico di Trieste,\\ Via Tiepolo 11, 34143 Trieste, Italy}
\affiliation[d]{Institute for Fundamental Physics of the Universe, Via Beirut 2, 34151 Trieste, Italy}
\affiliation[e]{Dipartimento di Fisica ``Aldo Pontremoli'', Universit\`a degli Studi di Milano,\\ Via Celoria 16, 20133 Milan, Italy}
\affiliation[f]{INFN, Sezione di Milano, Via Celoria 16, 20133 Milan, Italy}
\affiliation[g]{Department of Physics and Astronomy, University of Sussex,\\ Sussex House, BN1 9RH Brighton, United Kingdom}
\affiliation[h]{Departament de F\'isica, Universitat Aut\`onoma de Barcelona, 08193 Bellaterra, Barcelona, Spain}
\affiliation[i]{IFAE and BIST, Campus UAB, 08193 Bellaterra, Barcelona, Spain}

\date{\today}


\emailAdd{fverdian@sissa.it, emanuele.castorina@unimi.it, \mbox{esalvioni@ifae.es, emiliano.sefusatti@inaf.it}}

\abstract{
We initiate a systematic study of the perturbative nonlinear dynamics of cosmological fluctuations in dark sectors comprising a fraction of non-cold dark matter, for example ultra-light axions or light thermal relics. These mixed dark matter scenarios exhibit suppressed growth of perturbations below a characteristic, cosmologically relevant, scale associated with the microscopic nature of the non-cold species. As a consequence, the scale-free nonlinear solutions developed for pure cold dark matter and for massive neutrinos do not, in general, apply. We thus extend the Effective Field Theory of Large Scale Structure to model the coupled fluctuations of the cold and non-cold dark matter components, describing the latter as a perfect fluid with finite sound speed at linear level. We provide new analytical solutions wherever possible and devise an accurate and computationally tractable prescription for the evaluation of the one-loop galaxy power spectrum, which can be applied to probe mixed dark matter scenarios with current and upcoming galaxy survey data. As a first application of this framework, we derive updated constraints on the energy density in ultra-light axions using a combination of Planck and BOSS data. Our refined theoretical modeling leads to somewhat weaker bounds compared to previous analyses.}

\maketitle

\section{Introduction and Main Results}\label{sec:intro}
Dark matter (DM) is one of the pillars of the cosmological model. In the standard picture, the so-called $\Lambda$CDM scenario, DM is described on large scales as a collisionless, cold fluid whose energy density redshifts like the volume while the Universe expands. These properties imply that the formation and evolution of the Large Scale Structure (LSS) proceeds hierarchically, from the smaller scales first to the larger ones later~\cite{Peebles84,PeeblesLSS}. In other words, the clustering of DM particles is a process with no physical scale, fully controlled by gravity, and structure formation can proceed indefinitely down to the smallest scales.\footnote{Baryonic physics alters this picture on very small scales, but it can be taken into account using hydrodynamical simulations.} A mathematical formulation of these statements is that the root mean square amplitude of DM fluctuations is a monotonically increasing function of $k$, where smaller physical scales correspond to larger values of $k$.

This sharp prediction of the $\Lambda$CDM scenario can be tested against observations. If DM growth is strongly suppressed below some characteristic scale, we expect $\mathcal{O}(1)$ changes in the abundance and clustering properties of astrophysical objects whose size is right at or smaller than that scale. Counts of the Milky Way satellites \cite{Nadler:2021dft,Newton:2020cog,Dekker:2021scf,Tan:2024cek}, measurements of the Lyman-$\alpha$ transmitted flux through the Intergalactic Medium~\cite{Irsic:2023equ,Rogers:2020ltq,Kobayashi:2017jcf,Garcia-Gallego:2025kiw, Armengaud:2017nkf}, or strong lensing of galaxies \cite{Zelko:2022tgf,Vegetti:2023mgp,Keeley:2023ive} already place important bounds on scenarios where all of DM has vanishing fluctuations at wavenumbers larger than some characteristic value. Prototypical models realizing such suppressed growth are thermal warm DM, where the DM particles have a sufficiently high temperature that their random velocities prevent structure from forming below the free-streaming scale, and ultra-light scalar DM, where power is suppressed below a scale determined by quantum pressure support. The above-mentioned astrophysical probes are sensitive to characteristic wavenumbers $k_s$ in the ballpark of $10\,h \,\mathrm{Mpc}^{-1}$. 

If the scale of the DM power suppression is cosmologically relevant, i.e.~much larger than individual astrophysical objects, then it can be probed using the anisotropy of the Cosmic Microwave Background (CMB) and measurements of the LSS obtained from the clustering of galaxies. In this regime, the possibility that all of DM has suppressed fluctuations above $k_s$ is already ruled out by data. This is reminiscent of the constraints on massive neutrinos~\cite{Lesgourgues:2006nd}, whose characteristic wavenumber is roughly $0.01\,h \,\mathrm{Mpc}^{-1}$: given current uncertainties, neutrinos can form at most a $\sim 0.5\%$ fraction of the matter density~\cite{Planck:2018vyg,DESI:2025zgx,ACT:2025tim,DESI:2025gwf,SPT-3G:2025bzu}. 

Nevertheless, scenarios where a fraction $f_\chi < 1$ of the DM energy density has suppressed growth of fluctuations above a certain cosmologically relevant $k_s$,\footnote{To be precise, in this paper we define $f_\chi$ as the fraction of the {\it total} matter with characteristic wavenumber $k_s$. However, that distinction is neglected in the present section.} while the remainder is cold, are still allowed and highly plausible from a particle physics perspective. Using a common terminology, in this paper we refer to such models as mixed DM scenarios. The non-cold component can be composed, to name but a few possibilities, of ultra-light axions produced via vacuum misalignment in the early Universe~\cite{Preskill:1982cy,Abbott:1982af,Dine:1982ah}, perhaps originating from string theory~\cite{,Arvanitaki:2009fg}, or by light but massive thermal relics, including fermions such as gravitinos~\cite{Weinberg:1982zq,Feng:2010ij} or bosons such as hot axions~\cite{Baumann:2016wac}. Furthermore, a suppression of clustering can be realized even for a cold DM component, in the presence of interactions with additional dark sector species~\cite{Cyr-Racine:2015ihg,Becker:2020hzj}.

As the size of $f_\chi$ probed by cosmological observations decreases with increasing sensitivity, it becomes imperative to refine the theoretical predictions in order to avoid a biased inference. This is especially true in the current era of galaxy surveys, with DESI~\cite{DESI:2025fxa}, Euclid~\cite{Euclid:2024yrr}, and the Vera Rubin Observatory~\cite{Rubin} poised to revolutionize our understanding of the dark sector over the next few years~\cite{Boddy:2022knd,Mao:2022fyx,Euclid:2024pwi,Dvorkin:2022jyg}. Of particular interest are scenarios where the new characteristic scale $k_s$ is within the perturbative range, and the theory is therefore amenable to ab-initio theoretical predictions for the LSS observables. 

In this paper, focusing on such perturbative scales, we extend the Effective Field Theory of Large Scale Structure (EFTofLSS)~\cite{Baumann:2010tm,Carrasco:2013sva,Lewandowski:2015ziq,Vlah:2015sea,Porto:2013qua} to mixed DM scenarios where a fraction $f_\chi$ of the DM energy density is not cold. Our working assumption is that the non-cold species can be modeled, within linear perturbation theory, as a perfect fluid with non-negligible pressure perturbation, parametrized by a finite sound speed $c_s^2$. As a consequence, its fluctuations are suppressed below a characteristic scale $\sim c_s / \mathcal{H}$, where $\mathcal{H}$ is the Hubble scale. We consider general forms for the time and scale dependences of $c_s^2$. These encompass, for example, both ultra-light axions and light thermal relics, which we adopt as benchmarks to illustrate our results, as well as many other microscopic dark sector models whose detailed study we leave for future work.

The novel two-fluid extension of the EFTofLSS we develop here is most impactful when the characteristic wavenumber is within the observational window of galaxy surveys, very roughly $0.01\,h \,\mathrm{Mpc}^{-1} \lesssim k_s \lesssim 1\,h \,\mathrm{Mpc}^{-1}$. It is constructed via a perturbative expansion for small $f_\chi$, which is justified a posteriori by the constraints we derive. While we are, perhaps, the first to approach this problem with physics beyond the Standard Model (BSM) in mind, similar questions have been studied in previous literature. In particular, our work heavily draws inspiration from, and is indebted to, the nearly two-decades-long effort of the LSS community to model the clustering of massive neutrinos in perturbation theory beyond the linear regime~\cite{Wong:2008ws,Lesgourgues:2009am,Upadhye:2013ndm,Blas:2014hya,Fuhrer:2014zka,Archidiacono:2015ota,Levi:2016tlf,Senatore:2017hyk,Villaescusa-Navarro:2017mfx,Ruggeri:2017dda,deBelsunce:2018xtd,Chen:2019cfu,Zhu:2019kzb,Aviles:2020cax,Garny:2020ilv,Aviles:2021que,Garny:2022fsh,Nascimento:2023ezc,Verdiani:2025znc}. For related efforts in the context of the CDM-baryon system see Refs.~\cite{Lewandowski:2014rca,Braganca:2020nhv}.

As already mentioned, at linear level we model the non-cold species as a perfect fluid, i.e., the presence of a finite sound speed is the only difference with respect to the CDM description. It should, however, be kept in mind that the phenomenology of some BSM scenarios goes beyond the perfect fluid regime. For instance, light relics that decoupled from thermal equilibrium while relativistic have a non-zero anisotropic shear. Although it decays with time, the shear can still contribute to the evolution of the density perturbations of the warm species, see Ref.~\cite{Lesgourgues:2006nd} for a review.\footnote{For ultra-light scalar fields, the linear theory dynamics can be exactly mapped onto that of a perfect fluid with non-zero sound speed~\cite{Hu:1998kj}.} In this work we assume that such additional effects, unless they can be approximately reabsorbed into an effective value for $c_s^2$ (as it is indeed the case for the shear of thermal relics), are negligible.


Beyond linear level, in general each BSM model can introduce intrinsic nonlinearities associated with its microphysical description. These need to be added to the nonlinearities of pure gravitational origin, which are universal and are the only ones included in our framework. Nevertheless, in Appendix~\ref{app:app_4} we study the intrinsic nonlinearities in the two scenarios we adopt as benchmarks in this paper, namely ultra-light scalar fields and thermal relics. In the former case the intrinsic nonlinearities are associated with the wave-like nature of the field~\cite{Li:2018kyk}, whereas in the latter they originate from the Vlasov equation. We show that the effects of intrinsic nonlinearities are minor in both scenarios, thus supporting our approach.


Our main results are briefly summarized as follows. First, even though our focus is on the nonlinear regime of clustering, in Sec.~\ref{sec:linear} we present new analytical linear solutions that accurately describe the evolution of the perturbations of the two coupled fluids, for any power-law dependence of the sound speed on both time and scale. These linear solutions, which are obtained by means of an expansion for small $f_\chi$, are useful to elucidate the dependence of the observables on the underlying model parameters, but also represent the input for the subsequent nonlinear calculations. 

Moving beyond linear order, it is well known that analytical solutions in the presence on a finite sound speed cannot be attained for all kinematic configurations. Nevertheless, in Sec.~\ref{sec:matter_nonl} we present nonlinear analytical expressions valid in the kinematic regimes where the involved momenta are well separated from $k_s$, which could be relevant for future measurements of the matter and galaxy bispectra. We do so for a specific choice of the time dependence of $c_s^2$, which applies to both ultra-light axions and light thermal relics. Furthermore, in Sec.~\ref{sec:IR} we identify for the first time the exact solutions in the infrared limit. In Sec.~\ref{sec:UV} we clarify the ultraviolet structure of the two-fluid model and the associated new EFTofLSS counterterms required to absorb any possible divergences.

An important goal of this paper is to apply the new two-fluid EFTofLSS to perform cosmological tests of mixed DM scenarios, in particular using measurements of the galaxy power spectrum by current and upcoming redshift surveys. For this purpose, in Sec.~\ref{sec:prescription} we devise a prescription, motivated by the exact infrared results presented earlier, which allows us to evaluate the perturbative solutions for both CDM and the non-cold species with sufficient speed to be implemented in a Markov Chain Monte Carlo parameter exploration. When compared to the exact, but computationally very expensive, numerical solutions, our prescription makes an error significantly smaller than the uncertainties in the measurements by current and upcoming surveys, as we show in Sec.~\ref{sec:p_accuracy_P}.

As a first, proof-of-principle application of our framework, in Sec.~\ref{sec:constraints} we present updated constraints on the energy density in ultra-light axions from a combination of Planck CMB and BOSS LSS data. In the range of axion masses where the characteristic wavenumber $k_s$ falls within the BOSS observational window, our bounds are somewhat weaker than found in previous single-fluid analyses~\cite{Lague:2021frh,Rogers:2023ezo}, due to the new EFTofLSS parameters present in our refined theoretical modeling of the galaxy power spectrum. These results highlight that a consistent description of BSM effects is essential to ensure the robustness of potential discoveries by galaxy surveys. An outlook to future directions is given in Sec.~\ref{sec:outlook}, while a complete description of our constraints on ultra-light axions can be found in Appendix~\ref{app:app_3}.

\section{Framework and Linear Cosmology}\label{sec:linear}
We assume the presence of a non-cold DM species $\chi$ and a cold component $c$. For simplicity we will always refer to $\chi$ as the ``warm'' component, even though this may technically be a misnomer for some microscopic realizations. Assuming the starting point of the evolution $a_{\rm ini}$ to be in matter domination, the coupled equations governing the dynamics of the perturbations for the two fluids~\cite{Ma:1995ey} read, in the sub-horizon limit $k \gg \mathcal{H}$, 
\begin{subequations}\label{eq:initial}
\begin{align}
\delta_\chi^\prime + \theta_\chi =&\; - \nabla_i (\delta_\chi v_\chi^i)\,, \label{eq:continuity_chi}\\
\theta_\chi^\prime + \mathcal{H} \theta_\chi +  \frac{1}{\bar{\rho}_\chi}\nabla^2 \delta \mathcal{P}_\chi + \frac{3}{2}\mathcal{H}^2 \Omega_m [f_\chi \delta_\chi + (1 - f_\chi) \delta_c] =&\; - \nabla_i (v_\chi^j \nabla_j v_\chi^i )\,, \label{eq:Euler_chi} \\
\delta_c^\prime + \theta_c =&\; - \nabla_i (\delta_c v_c^i)\,, \\
\theta_c^\prime + \mathcal{H} \theta_c + \frac{3}{2}\mathcal{H}^2 \Omega_m [f_\chi \delta_\chi + (1 - f_\chi) \delta_c] =&\; - \nabla_i (v_c^j \nabla_j v_c^i )\,,
\end{align}
\end{subequations}
where primes denote conformal time derivatives and $\mathcal{H} = a'/a$. We split the density and pressure for the two species as $\rho_x = \bar{\rho}_x + \delta \rho_x$ and $\mathcal{P}_x = \overline{\mathcal{P}}_x + \delta \mathcal{P}_x$ for $x = \chi,c$ and define the density contrast $\delta_x \equiv \delta \rho_x / \bar{\rho}_x$ and velocity divergence $\theta_x \equiv \nabla_i v_x^i$. In addition, we denote with
\begin{equation}
f_\chi \equiv \frac{ \Omega_\chi} {\Omega_m}
\end{equation}
the fraction of the matter energy density in the warm species, where $\Omega_i \equiv \bar{\rho}_i/\bar{\rho}_{\rm tot}$ and $\Omega_m = \Omega_\chi + \Omega_c$. Throughout this paper the subset $c$ refers to the combined CDM+baryon fluid: $\delta_c = (f_{\rm cdm} \delta_{\rm cdm} + f_b \delta_b)/(f_{\rm cdm} + f_b)$ and similarly for $v_c^i$. The background pressure is assumed to be negligible for both species, $w_{x} = \overline{\mathcal{P}}_x/\bar{\rho}_x = 0$, hence the background energy densities redshift as $\bar{\rho}_{x} \propto a^{-3}$ at all times $a > a_{\rm ini}$ where the description in Eqs.~\eqref{eq:initial} applies. It is also useful to define the total and relative density perturbations as 
\begin{equation}\label{eq:tot_rel}
\delta_m \equiv f_\chi \delta_\chi + (1 - f_\chi) 
\delta_c\,, \qquad \delta_r \equiv \delta_\chi - \delta_c\,,
\end{equation}
with analogous expressions for the velocity fields.

The perturbed Euler equation for the warm component~\eqref{eq:Euler_chi} contains a pressure term $\nabla^2 \delta \mathcal{P}_\chi / \bar{\rho}_\chi$. We do not include a possible anisotropic shear stress term, which would manifest as $-\nabla^2 \sigma_\chi$ on the left-hand side of Eq.~\eqref{eq:Euler_chi}. This is fully accurate for some microscopic scenarios described by our framework (such as ultra-light axions~\cite{Marsh:2015xka}), whereas in other scenarios it is possible to absorb at least partially the effect of shear stress into a suitable modification of the pressure term. The latter possibility is discussed in Sec.~\ref{eq:relics} for light thermal relics, including massive neutrinos~\cite{Shoji:2010hm}.

To solve Eqs.~\eqref{eq:initial} we change the time variable to $\eta \equiv \log D_{\Lambda \rm CDM}$, where $D_{\Lambda \rm CDM}$ is the linear growth factor of a $\Lambda$CDM cosmology with the same total matter energy density as the mixed warm/cold scenario we are considering~\cite{Garny:2020ilv}. We obtain
\begin{subequations}\label{eq:eta}
\begin{align}
\partial_\eta \delta_\chi  - \Theta_\chi& = \nabla_i (\delta_\chi V_\chi^i),  \\ \label{eq:eta_2}
\partial_\eta \Theta_\chi +  \Big(\frac{3\Omega_m}{2f_{\Lambda \rm CDM}^2} - 1\Big)\Theta_\chi -  \frac{\nabla^2 \delta \mathcal{P}_\chi}{f_{\Lambda \rm CDM}^2\mathcal{H}^2 \bar{\rho}_\chi}  - \frac{3\Omega_m}{2f_{\Lambda \rm CDM}^2}& [f_\chi \delta_\chi + (1 - f_\chi) \delta_c] \nonumber \\ &=\nabla_i (V_\chi^j \nabla_j V_\chi^i ),
\end{align}
\begin{align}
\partial_\eta \delta_c  - \Theta_c& = \nabla_i (\delta_c V_c^i), \label{eq:eta_3}  \\
\partial_\eta \Theta_c +  \Big(\frac{3\Omega_m}{2f_{\Lambda \rm CDM}^2} - 1\Big)\Theta_c - \frac{3\Omega_m}{2f_{\Lambda \rm CDM}^2} [f_\chi \delta_\chi + (1 - f_\chi) \delta_c] &= \nabla_i (V_c^j \nabla_j V_c^i ), \label{eq:eta_4}
\end{align}
\end{subequations}
where we introduced the definitions
\begin{equation}
f_{\Lambda \rm CDM} = \frac{d \log D_{\Lambda \rm CDM}}{d \log a}\,, \qquad V_x^i = - \frac{v_x^i}{f_{\Lambda \rm CDM} \mathcal{H}}\,, \qquad \Theta_x = \nabla_i V_x^i\,.
\end{equation}
Henceforth, we work in Fourier space. Setting $\Omega_m/ f_{\Lambda \rm CDM}^2 \to 1$, which is known to be a highly accurate approximation in $\Lambda$CDM cosmologies even during vacuum energy domination (see e.g.~Ref.~\cite{Garny:2020ilv} for an appraisal), we rewrite the equations as
\begin{subequations}\label{eq:eta_kspace_nonlinear_initial}
\begin{align}
&\partial_{\eta} \delta_\chi (\vec{k},\eta)   - \Theta_\chi(\vec{k},\eta)  =\int_{\bf{k}} \mathrm{d} k_{12} \alpha(\vec{k}_2, \vec{k}_1) \delta_\chi(\vec{k}_1, \eta)\Theta_\chi(\vec{k}_2, \eta)\,,  \\
&\partial_{\eta} \Theta_\chi(\vec{k},\eta) +  \frac{1}{2}\Theta_\chi(\vec{k},\eta) + \frac{3}{2} \, c_{s,\mathrm{eff}}^2 k^2 \delta_\chi(\vec{k},\eta) -\frac{3}{2} [f_\chi \delta_\chi(\vec{k},\eta) + (1 - f_\chi) \delta_c(\vec{k},\eta)] \nonumber \\
&\hspace{3.5cm} =\int_{\bf{k}} \mathrm{d} k_{12} \beta(\vec{k}_1, \vec{k}_2)\Theta_\chi(\vec{k}_1, \eta)\Theta_\chi(\vec{k}_2, \eta)\,, \label{eq:eta_kspace_nonlinear_initial_chi_euler}\\
&\partial_{\eta} \delta_c(\vec{k},\eta)  - \Theta_c(\vec{k},\eta)\, =\int_{\bf{k}} \mathrm{d} k_{12} \alpha(\vec{k}_2, \vec{k}_1)\delta_c(\vec{k}_1, \eta)\Theta_c(\vec{k}_2, \eta)\,,  \\
&\partial_{\eta} \Theta_c (\vec{k},\eta) + \frac{1}{2}\Theta_c (\vec{k},\eta) - \frac{3}{2} [f_\chi \delta_\chi (\vec{k},\eta) + (1 - f_\chi) \delta_c (\vec{k},\eta)]  \nonumber\\ 
&\hspace{3.5cm} = \int_{\bf{k}} \mathrm{d} k_{12} \beta(\vec{k}_1, \vec{k}_2)\Theta_c(\vec{k}_1, \eta)\Theta_c(\vec{k}_2, \eta)\,,
\end{align}
\end{subequations}
where $c_{s, \mathrm{eff}}^2 = 2 c_s^2 / ( 3 \mathcal{H}^2 \Omega_m)$ and
\begin{equation}\label{eq:cs2}
c_s^2 \equiv \frac{ \delta P_\chi (\vec{k}, \eta) } { \delta \rho_\chi (\vec{k}, \eta)}
\end{equation}
is the sound speed of the warm fluid. In Eqs.~\eqref{eq:eta_kspace_nonlinear_initial} we also introduced the standard mode-coupling functions
\begin{equation}
\alpha(\vec{k}_1, \vec{k}_2) = 1 + \frac{\vec{k}_1 \cdot \vec{k}_2}{k_1^2}\,,\qquad \beta(\vec{k}_1, \vec{k}_2) = \frac{(\vec{k}_1 + \vec{k}_2)^2 \vec{k}_1 \cdot \vec{k}_2}{2 k_1^2 k_2^2}\,,
\end{equation}
and the shorthand notation
\begin{equation}
\int_{\mathbf{k}} \mathrm{d}k_{1\ldots j} \equiv \int \frac{\mathrm{d}^3 k_1 \dots \mathrm{d}^3 k_j}{(2\pi)^{3(j - 1)}} \, \delta^{D}\Big(\mathbf{k} - \sum_{i\,=\,1}^{j} \mathbf{k}_i\Big), 
\end{equation}
where $\delta^D$ denotes the Dirac delta.

The temporal and spatial variations of the sound speed in Eq.~\eqref{eq:cs2} encode the model dependence. We assume power-law scalings with wavenumber and scale factor,
\begin{equation}
c_s^2 \propto k^p a^{-\gamma}\,,
\end{equation}
where the coefficients $p$ and $\gamma$ are a priori arbitrary. The matching to several concrete microphysical scenarios will be discussed later in Sec.~\ref{sec:micro_models}. Since $\mathcal{H}^2\Omega_m \propto a^{-1}$, we parametrize
\begin{equation}\label{eq:cs_eff2}
c_{s, \mathrm{eff}}^2 \equiv \frac{k^p}{k_{s\ast}^{2+p}}\, \Big( \frac{a_\ast}{a} \Big)^{\gamma - 1} \,\simeq\; \frac{k^p}{k_{s\ast}^{2+p}}\, e^{(\eta_\ast - \eta)(\gamma - 1)}
\end{equation}
where we found it convenient to define a reference time $a_\ast$. In the last equality we have made the Einstein-de Sitter (EdS) approximation $D_{\Lambda \rm CDM} = a$, which implies $\eta = \log a$. This is the only instance where the EdS approximation (which also implies $\Omega_m/f^2_{\Lambda \rm CDM}\to 1$, but is more restrictive) is used in the present paper and it will be justified a posteriori, by showing in Sec.~\ref{sec:accuracy} that our analytical linear solutions accurately reproduce exact numerical results down to the redshift $z\sim 0.5$ of observational relevance for galaxy surveys.

After substituting Eq.~\eqref{eq:cs_eff2} into Eqs.~\eqref{eq:eta_kspace_nonlinear_initial}, we observe that the explicit dependence on $k$ introduced by the $\chi$ sound speed term can be removed (for $\gamma \neq 1$) by switching to the new momentum-dependent time variable
\begin{equation}
\tilde{\eta} \equiv \eta - \eta_\ast - \frac{2+p}{\gamma - 1} \log \frac{k}{k_{s\ast}}\;.
\end{equation}
Since $d\tilde{\eta} = d\eta$, the equations are then cast in their final form
\begin{subequations}\label{eq:eta_kspace_nonlinear}
\begin{align}
&\partial_{\tilde{\eta}} \delta_\chi (\vec{k},\tilde{\eta})   - \Theta_\chi(\vec{k},\tilde{\eta})  =\int_{\bf{k}} \mathrm{d} k_{12} \alpha(\vec{k}_2, \vec{k}_1) \delta_\chi(\vec{k}_1, \tilde{\eta})\Theta_\chi(\vec{k}_2, \tilde{\eta})\,, \label{eq:eta_kspace_nonlinear_chi_cont}  \\
&\partial_{\tilde{\eta}} \Theta_\chi(\vec{k},\tilde{\eta}) +  \frac{1}{2}\Theta_\chi(\vec{k},\tilde{\eta}) + \frac{3}{2} \, e^{-(\gamma - 1)\tilde{\eta}} \delta_\chi(\vec{k},\tilde{\eta}) -\frac{3}{2} [f_\chi \delta_\chi(\vec{k},\tilde{\eta}) + (1 - f_\chi) \delta_c(\vec{k},\tilde{\eta})] \nonumber \\
&\hspace{3.5cm} =\int_{\bf{k}} \mathrm{d} k_{12} \beta(\vec{k}_1, \vec{k}_2)\Theta_\chi(\vec{k}_1, \tilde{\eta})\Theta_\chi(\vec{k}_2, \tilde{\eta})\,, \label{eq:eta_kspace_nonlinear_chi_euler}\\
&\partial_{\tilde{\eta}} \delta_c(\vec{k},\tilde{\eta})  - \Theta_c(\vec{k},\tilde{\eta})\, =\int_{\bf{k}} \mathrm{d} k_{12} \alpha(\vec{k}_2, \vec{k}_1)\delta_c(\vec{k}_1, \tilde{\eta})\Theta_c(\vec{k}_2, \tilde{\eta})\,, \label{eq:eta_kspace_nonlinear_c_cont}  \\
&\partial_{\tilde{\eta}} \Theta_c (\vec{k},\tilde{\eta}) + \frac{1}{2}\Theta_c (\vec{k},\tilde{\eta}) - \frac{3}{2} [f_\chi \delta_\chi (\vec{k},\tilde{\eta}) + (1 - f_\chi) \delta_c (\vec{k},\tilde{\eta})]  \nonumber\\ 
&\hspace{3.5cm} = \int_{\bf{k}} \mathrm{d} k_{12} \beta(\vec{k}_1, \vec{k}_2)\Theta_c(\vec{k}_1, \tilde{\eta})\Theta_c(\vec{k}_2, \tilde{\eta})\,, \label{eq:eta_kspace_nonlinear_c_euler}
\end{align}
\end{subequations}
In what follows, we solve Eqs.~\eqref{eq:eta_kspace_nonlinear} perturbatively in the density and velocity fluctuations. At each perturbative order, we further perform an expansion for $f_\chi \ll 1$. Hence, we write for the density fields 
\begin{equation}\label{eq:pt-expansion-notation}
\delta_x (\vec{k}, \tilde{\eta}) = \sum_{n\, =\, 1}^{\infty} \delta_x^{[n]}(\vec{k},\tilde{\eta}) = \sum_{n\, =\, 1}^{\infty} \Big[ \delta_x^{(0)[n]}(\vec{k},\tilde{\eta}) + f_\chi \delta_x^{(1)[n]}(\vec{k},\tilde{\eta}) + \mathcal{O}(f_\chi^2) \Big] \,,
\end{equation}
and analogously for the velocity fields $\Theta_x$, with $x = \chi, c$. The $[n]$ index denotes the perturbative order, while the $(0)$ and $(1)$ indices denote the order in the small $f_\chi$ expansion. In practice, we will find that the key physical effects are captured by truncating the small $f_\chi$ expansion to the first nontrivial order for each species: $\mathcal{O}(f_\chi^0)$ for $\chi$ and $\mathcal{O}(f_\chi)$ for $c$, which behaves as CDM at zeroth order. 

As common practice, we also define a time-dependent characteristic wavenumber $k_s(a)$ via the relation $ c_{s, \mathrm{eff}}^2 k^2 \equiv [k / k_s (a)]^{2+p}$, so that the sound speed term in the Euler equation for $\chi$ takes the form $\tfrac{3}{2}[ k/k_s(a)]^{2+p}\delta_\chi$. From Eq.~\eqref{eq:cs_eff2} we thus obtain
\begin{equation}\label{eq:Jeans}
k_s(a) = k_{s\ast} \Big(\frac{a}{a_\ast}\Big)^{\frac{\gamma - 1}{2 + p}}
\end{equation}
and $\tilde{\eta}$ can be written as
\begin{equation}\label{eq:etatilde_approx}
\tilde{\eta}\simeq - \frac{2+p}{\gamma - 1} \log \frac{k}{k_s(a)}
\end{equation}
by using the EdS approximation. From a microscopic perspective, $k_s(a)$ can have a variety of origins. Some examples will be provided in Sec.~\ref{sec:micro_models}.

\subsection{General linear solutions}\label{sec:lin_sol}
First, we focus on solving Eqs.~\eqref{eq:eta_kspace_nonlinear} at linear level, keeping the parameters $\gamma$ and $p$ arbitrary. At zeroth order in $f_\chi$, the equations for $c$ are combined into
\begin{equation}
\Big( \frac{\partial^2 }{\partial \tilde{\eta}^2} + \frac{1}{2}\frac{\partial }{\partial \tilde{\eta}} - \frac{3}{2} \Big) \delta_c^{(0)[1]} = 0\,,
\end{equation}
where we recall that $[1]$ indicates the first (linear) perturbative order, while $(0)$ indicates the zeroth order in $f_\chi$. The above equation is solved by the standard growing mode for CDM, $\delta_c^{(0)[1]}({\bf k}, \tilde{\eta}) = \Theta_c^{(0)[1]}({\bf k},\tilde{\eta}) = e^{\tilde{\eta}-\tilde{\eta}_{\rm ini}} \delta_0(\bf{k})$.
Plugging this result into the equations for $\chi$ gives, still at zeroth order in $f_\chi$,
\begin{equation}\label{eq:lin_g0}
\Big(\frac{\partial^2}{\partial \tilde{\eta}^2} + \frac{1}{2}\frac{\partial}{\partial \tilde{\eta}} + \frac{3}{2} e^{-(\gamma - 1) \tilde{\eta}} \Big) \delta_\chi^{(0)[1]} = \frac{3}{2} e^{\tilde{\eta}-\tilde{\eta}_{\rm ini}} \delta_0(\bf{k})\,.
\end{equation}
This is solved by
\begin{align}\label{eq:delta_chi_0}
\delta_\chi^{(0)[1]} ( \mathbf{k}, \tilde{\eta}) =&\; \int_{- \infty}^{+ \infty} \mathrm{d} \tilde{\eta}^\prime G_\chi (\tilde{\eta}, \tilde{\eta}^\prime) \frac{3}{2} e^{\tilde{\eta}^\prime - \tilde{\eta}_{\rm ini}} \delta_0({\bf{k}})\,, \qquad \Theta_\chi^{(0)[1]} (\mathbf{k},\tilde{\eta}) = \partial_{\tilde{\eta}}\, \delta_\chi^{(0)[1]} (\mathbf{k},\tilde{\eta})\,,
\end{align}
where the Green's function is expressed in terms of the homogeneous solutions as
\begin{equation}
G_\chi (\tilde{\eta}, \tilde{\eta}^\prime) = \frac{ A_2 (\tilde{\eta}) - A_1 (\tilde{\eta})\frac{A_2(\tilde{\eta}^\prime)}{A_1(\tilde{\eta}^\prime)}  }{\Big( \frac{\partial A_2}{\partial \tilde{\eta}} - \frac{\partial A_1}{\partial \tilde{\eta}} \frac{A_2 }{A_1} \Big) (\tilde{\eta}^\prime)}\,\theta_{\rm H} (\tilde{\eta} - \tilde{\eta}^\prime)\,, \quad\;\; A_{1,\,2} (\tilde{\eta}) = e^{- \tilde{\eta}/4}\, J_{\, \mp \frac{1}{2 (\gamma - 1)}} \bigg( \frac{\sqrt{6}\, e^{ - ( \gamma - 1 )\tilde{\eta} / 2 } }{ \gamma - 1} \bigg),
\end{equation}
with $J_\alpha$ denoting the Bessel function of order $\alpha$ and $\theta_{\rm H}$ the Heaviside step function. Introducing a notation that will be used throughout the paper, we define the ratios of linear perturbations
\begin{equation}\label{eq:ratios_tf}
g(\tilde{\eta}) \equiv \frac{\delta_\chi^{[1]}({\bf k},\tilde{\eta})}{\delta_c^{[1]}({\bf k},\tilde{\eta})}\,, \qquad  h(\tilde{\eta}) \equiv \frac{\Theta_\chi^{[1]}({\bf k},\tilde{\eta})}{\delta_c^{[1]}({\bf k},\tilde{\eta})}\,, \qquad s(\tilde{\eta}) \equiv \frac{\Theta_c^{[1]}({\bf k},\tilde{\eta})}{\delta_c^{[1]}({\bf k},\tilde{\eta})}\,.
\end{equation}
Assuming adiabatic initial conditions, as we do in this work, the quantities above are simply ratios of linear transfer functions. Notice that 
\begin{equation}\label{eq:s_def}
s(\tilde{\eta}) = \partial_{\tilde{\eta}} \log \delta_c^{[1]} \equiv \frac{f(k)}{f_{\Lambda \rm CDM}}\,,
\end{equation}
where the last equality defines the scale-dependent linear growth rate $f(k)$.

\begin{figure}
    \centering
    \includegraphics[width = \textwidth]{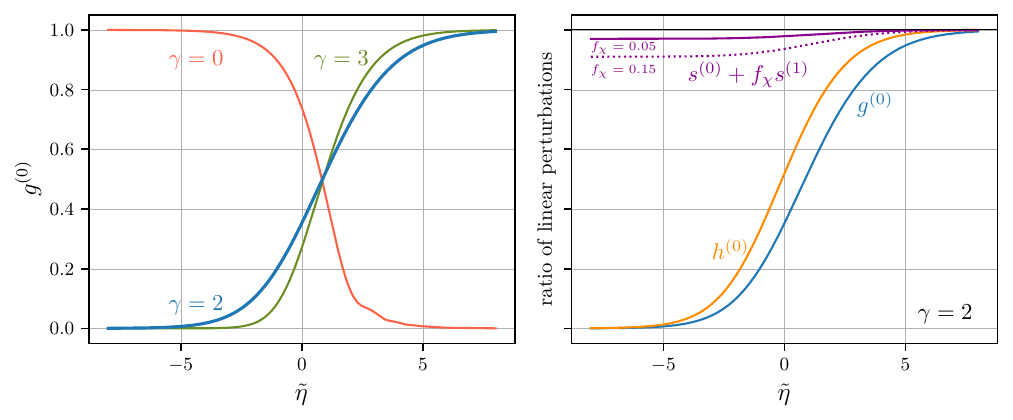}
    \vspace{-0.5cm}\caption{{\it Left:} The ratio of the linear density perturbations of $\chi$ and $c$, $g = \delta_\chi^{[1]}/\delta_c^{[1]}$, evaluated at zeroth order in the small $f_\chi$ expansion as a function of the $k$-dependent time variable $\tilde{\eta}$, for some illustrative choices of the parameter $\gamma$ that sets the time dependence of the characteristic scale $k_s(a)$. {\it Right:} For $\gamma = 2$, the ratios of linear transfer functions $g$ and $h = \Theta_\chi^{[1]}/\delta_c^{[1]}$ at zeroth order in $f_\chi$, while $s = \Theta_c^{[1]}/\delta_c^{[1]}$ is evaluated up to first order with $f_\chi = 0.05$ or $0.15$. One has $s^{(0)} = 1$ identically, whereas $s^{(1)} = -3/5$ for $k \gg k_s(a)$ (namely,  $\tilde{\eta}\ll 0$).} 
    \label{fig:g0}
\end{figure}

In the left panel of Fig.~\ref{fig:g0} we show $g^{(0)}$, evaluated at the leading order in the small $f_\chi$ expansion, for some (integer) choices of $\gamma$. We see that our framework encompasses very different possible time evolutions of the characteristic scale $k_s(a)$ in Eq.~\eqref{eq:Jeans}, which can either increase with time (as observed for $\gamma = 2,3$ in the figure) or decrease $(\gamma = 0)$.\footnote{For $\gamma = 1$ the characteristic scale is time-independent. In this case we solve the equations directly in $\eta$, obtaining at zeroth order in $f_\chi$ the solutions $\delta_c^{(0)[1]}({\bf k},\eta) = \Theta_c^{(0)[1]}({\bf k},\eta) = e^{\eta - \eta_{\rm ini}} \delta_0(\bf{k})$ and $\delta_\chi^{(0)[1]}({\bf k},\eta) = \Theta_\chi^{(0)[1]}({\bf k},\eta) = \delta_c^{(0)[1]}({\bf k},\eta) / \big[ 1 + (k/k_{s\ast})^{2+p}\big]$.}
We also have $h^{(0)} = g^{(0)} + \partial_{\tilde{\eta}}g^{(0)}$, whereas for the cold species we simply have $s^{(0)} = 1$ at this order.

Next, we solve the equations for the cold fluid at $\mathcal{O}(f_\chi)$. The correction to the growth rate, $s(\tilde{\eta}) = 1 + f_\chi s^{(1)}(\tilde{\eta})$, satisfies the first order differential equation
\begin{equation}\label{eq:s1}
\Big(\partial_{\tilde{\eta}} + \frac{5}{2}\Big) s^{(1)} = \frac{3}{2} \big( g^{(0)} - 1 \big)\,,
\end{equation}
whose solution reads
\begin{equation}
s^{(1)} (\tilde{\eta} ) = \int_{- \infty}^{+\infty} \mathrm{d}\tilde{\eta}^\prime G_s (\tilde{\eta}, \tilde{\eta}^\prime) \frac{3}{2} \big( g^{(0)}(\tilde{\eta}^\prime) - 1 \big)\,, \qquad G_s (\tilde{\eta}, \tilde{\eta}^\prime) = e^{- \frac{5}{2} (\tilde{\eta} - \tilde{\eta}^\prime)} \theta_{\rm H} (\tilde{\eta} - \tilde{\eta}^\prime) \,.
\end{equation}
We then invert the relation $s(\tilde{\eta}) = \partial_{\tilde{\eta}} \log \delta_c^{[1]}$ to arrive at
\begin{subequations}
\begin{align}
\delta_c^{(1) [1]} ({\bf k},\tilde{\eta}) =&\; \int_{\tilde{\eta}_{\rm ini}}^{\tilde{\eta}} \mathrm{d}\tilde{\eta}^\prime s^{(1)}(\tilde{\eta}^\prime) \;\delta_c^{(0)[1]} ({\bf k},\tilde{\eta})\,, \\
\Theta_c^{(1)[1]} ({\bf k},\tilde{\eta}) =&\; \Big( \int_{\tilde{\eta}_{\rm ini}}^{\tilde{\eta}} \mathrm{d}\tilde{\eta}^\prime s^{(1)}(\tilde{\eta}^\prime) + s^{(1)}(\tilde{\eta}) \Big) \delta_c^{(0)[1]} ({\bf k},\tilde{\eta}) \,.
\end{align}
\end{subequations}
The linear solutions for the $\chi$ fluid at $\mathcal{O}(f_\chi)$ can be derived similarly, however their expressions are not especially illuminating and we do not report them here.

\subsection{Linear solutions for $\gamma = 2$ and matching to microscopic models}\label{sec:micro_models}
In the remainder of this section we focus the discussion on $\gamma = 2$, corresponding to a time dependence of the sound speed $c_s^2 \propto k^p a^{-2}$. This case encompasses two widely-studied scenarios, which in this paper we adopt as benchmarks to illustrate our results: (i) ultra-light axions (ULAs, $p = 2$), where $k_s$ is identified with the Jeans scale induced by quantum pressure support; and (ii) thermal relics ($p = 0$), including the standard neutrinos, where $k_s$ is interpreted as the free streaming scale remnant of a thermal distribution. Before proceeding, we emphasize that we expect the framework developed in this paper to be applicable to a much broader range of dark sector models, whose exploration is left for future work.

For $\gamma = 2$, Eq.~\eqref{eq:delta_chi_0} gives the closed-form expressions
\begin{equation}\label{eq:sols_gamma2}
\delta_\chi^{(0)[1]} ( \mathbf{k}, \tilde{\eta}) = g^{(0)}(\tilde{\eta}) \delta_c^{(0)[1]} ( \mathbf{k}, \tilde{\eta})\,,\qquad \Theta_\chi^{(0)[1]} ( \mathbf{k}, \tilde{\eta}) = h^{(0)}(\tilde{\eta}) \delta_c^{(0)[1]} ( \mathbf{k}, \tilde{\eta})\,,
\end{equation}
where $h^{(0)} = g^{(0)} + \partial_{\tilde{\eta}}g^{(0)}$ and
\begin{equation}\label{eq:g0_gamma2}
g^{(0)} (\tilde{\eta}) = G^{(0)}(\sqrt{6}e^{-\tilde{\eta}/2})\,,\qquad G^{(0)}(x) \equiv  1 + x^2 \Big[ \Big(\mathrm{Si} (x) - \frac{\pi}{2}\Big)\sin  x + \mathrm{Ci}(x)\cos x\Big].
\end{equation}
$\mathrm{Si}$ and $\mathrm{Ci}$ denote the sine integral and cosine integral functions.\footnote{Solutions similar to Eqs.~\eqref{eq:sols_gamma2} and~\eqref{eq:g0_gamma2} were given in Ref.~\cite{Kamalinejad:2022yyl} where, however, Heaviside function-like initial conditions were assumed. Our results are consistent with Eq.~(4.9) in Ref.~\cite{Kamalinejad:2022yyl}, but we automatically implement the natural initial conditions, which yield the far simpler expression for $g^{(0)}$ in Eq.~\eqref{eq:g0_gamma2} above.} For $k \gg k_s(a)$ (namely $\tilde{\eta} \ll 0$) the $\chi$ perturbations are extremely suppressed and $g^{(0)}, h^{(0)} \to 0$ with $h^{(0)}/g^{(0)} \to 2$. Conversely, for $k \ll k_s(a)$ (corresponding to $\tilde{\eta} \gg 0$) the $\chi$ species becomes indistinguishable from the cold component and $g^{(0)}, h^{(0)} \to 1$.

Including $\mathcal{O}(f_\chi)$ corrections, the growth rate of the cold species takes the constant value $s = 1 - 3 f_\chi /5$ for $k\gg k_s(a)$. This result is familiar from the study of massive neutrinos (see Ref.~\cite{Lesgourgues:2006nd} for a review): on small enough scales a fraction $f_\chi$ of the total matter does not cluster, hence the gravitational potential is reduced by a factor $1 - f_\chi$, which in turn suppresses the growth of the cold perturbations as $\delta_c^{[1]}({\bf{k}},\tilde{\eta}) \simeq e^{(1-3f_\chi/5)(\tilde{\eta} - \tilde{\eta}_{\rm ini})} \delta_0 ({\bf{k}})$. By contrast, on large scales $k \ll k_s(a)$ the growth rate takes the $\Lambda\mathrm{CDM}$ value, $s = 1$. In the right panel of Fig.~\ref{fig:g0} we show the behavior of the ratios of transfer functions $g, h$ and $s$ for $\gamma = 2$. Each of them is evaluated to the first non-trivial order in the small $f_\chi$ expansion, namely $\mathcal{O}(f_\chi^0)$ for $g$ and $h$ and $\mathcal{O}(f_\chi)$ for $s$. 

We now turn to a brief description of our two benchmarks of choice, ULAs and light thermal relics, as well as their matching to our framework.

\subsubsection{Ultra-light axions}
ULAs develop a Jeans scale due to quantum pressure support~\cite{Hu:2000ke} and correspond to $\gamma = p = 2$. While the characteristic scale $k_s(a)$ is fixed by the ULA mass $m_a$, see Eq.~\eqref{eq:Jeans_ULAs} below, the fraction $f_\chi = \Omega_a / \Omega_m$ of energy density made of ULAs is determined by the earlier cosmological history and is model-dependent. For instance, the $\chi$ energy density can be generated by an interaction of the ULA with CDM~\cite{Bottaro:2024pcb}, or by an initial condition if the ULA is assumed not to couple to other species~\cite{Marsh:2010wq,Hlozek:2014lca,Lague:2021frh,Rogers:2023ezo}. In this paper we focus on the latter scenario, where $f_\chi$ can be traded for an initial condition on the ULA field value.

We model the ULA as a real scalar field $\phi$ with potential $V(\phi) = m_a^2 f_a^2 [1 - \cos(\phi/f_a)]$, where $f_a$ is chosen to be large enough that $\phi \ll f_a$ is always satisfied during the cosmological evolution, hence the potential is well approximated by a pure mass term $V(\phi) = m_a^2 \phi^2 / 2\,$. At the background level, the dynamics of the scalar field is governed by the Klein-Gordon equation $\overline{\phi}^{\,\prime\prime} + 2 \mathcal{H} \overline{\phi}^{\,\prime} + a^2 \partial V/\partial \phi = 0$, with initial conditions $\overline{\phi}^{\,\prime}(a_i) = 0$ and $\overline{\phi} (a_i)$. The latter can be traded for the present-day fraction of energy density in ULAs via a shooting procedure. The cosmological evolution of both background and perturbations is computed using the Einstein-Boltzmann solver~\texttt{AxiCLASS}~\cite{Poulin:2018dzj,Smith:2019ihp}. 

At early times, the dynamical equations for the scalar field are solved exactly. Eventually $\phi$ begins to oscillate when $a = a_{m_a}$, with $3 H(a_{m_a}) = m_a$, with a timescale much shorter than the expansion of the Universe. This allows us to accurately describe the evolution of the energy density and pressure of $\phi$ using an effective fluid approximation~\cite{Hu:2000ke,Hlozek:2014lca,Cookmeyer:2019rna,Passaglia:2022bcr}. The description in our Eqs.~\eqref{eq:eta_kspace_nonlinear_initial}, with $\gamma = p = 2$, starts applying at $a_{\rm ini} = a_{m_a}$ if the ULA begins to oscillate in matter domination (MD), whereas it starts at $a_{\rm ini} = a_{\rm eq}$ if the oscillations begin during radiation domination (RD). Taking $c_s^2 = k^2 / (4m_a^2 a^2)$~\cite{Hu:2000ke,Hlozek:2014lca,Urena-Lopez:2023ngt}, we obtain for the characteristic (Jeans) scale
\begin{equation}\label{eq:Jeans_ULAs}
k_s(a) \approx a^{1/4}\; 3.9\times 10^{-4}\, \frac{h}{\mathrm{Mpc}} \Big( \frac{m_a}{H_0}\Big)^{1/2} \Big( \frac{\Omega_m^0}{0.3} \Big)^{1/4},
\end{equation}
where $H_0 = 2.1\times 10^{-33}\,h\,\mathrm{eV}$. Notice that the assumption $\sigma(\vec{k},\tau) = 0$ made in our framework is fully justified for ULAs, because a scalar field minimally coupled to gravity has zero anisotropic stress at linear order in perturbation theory \cite{Hu:1998kj}. However, a complete fluid-dynamical description for ULAs would also include intrinsic wave nonlinearities, which stem from the higher-order terms in a perturbative expansion of the quantum pressure (see for instance Ref.~\cite{Li:2018kyk}). In Appendix~\ref{app:app_4_ULAs} we show that the impact on our results of these intrinsic nonlinearities is mild, which justifies our choice to omit them.

The matter power spectrum for ULA cosmologies has a drop-like feature, as a consequence of the Jeans scale. The location of the feature is determined by $m_a$, whereas its depth is fixed by $f_\chi$. To obtain an analytical estimate of the location, we need to distinguish between ULAs that became nonrelativistic during MD or RD. For ULAs that became nonrelativistic during MD, the condition $3 H(a_{m_a}) = m_a$ is satisfied at $a_{m_a} = [3 (\Omega_m^0)^{1/2} H_0/m_a]^{2/3}$. Imposing that $a_{m_a} > a_{\rm eq}$ then sets the upper edge of this region as $m_a/H_0 < 3 \times 10^5$ or equivalently $m_a < 4 \times 10^{-28}\;\mathrm{eV}$. The drop in the power spectrum is located at the smallest $k$ that is affected by the suppressed growth~\cite{Amendola:2005ad}, namely $k_s(a_{m_a})$. From Eq.~\eqref{eq:Jeans_ULAs} we find\
\begin{equation}\label{eq:kdrop-MD}
k_{\rm drop} = k_s (a_{m_a}) \approx 4.2\times 10^{-4}\, \frac{h}{\mathrm{Mpc}} \left( \frac{m_a}{H_0}\right)^{1/3} \left( \frac{\Omega_m^0}{0.3} \right)^{1/3}  \,, \qquad \left(m_a/H_0 < 3 \times 10^{5}\right)
\end{equation}
with a $m_a^{1/3}$ scaling.\footnote{An alternative definition of the location of the drop is the horizon crossing scale $k_m = a_{m_a} H(a_{m_a})$, which was used for instance in Ref.~\cite{Lague:2021frh}. As it was pointed out in Ref.~\cite{Marsh:2011bf}, the two definitions agree aside from $\mathcal{O}(1)$ numbers: we find $k_{m} = (2/9)^{2/3} \,k_{\rm drop} \approx 0.37\,k_{\rm drop}$.} For ULAs that became nonrelativistic during RD, $k_{\rm drop}$ saturates at the value of $k_s$ at matter-radiation equality, which yields
\begin{equation}\label{eq:kdrop-RD}
k_{\rm drop} = k_s (a_{\rm eq}) \approx 5.2\times 10^{-2}\, \frac{h}{\mathrm{Mpc}} \left( \frac{m_a}{10^6 H_0}\right)^{1/2} \,, \qquad \big(m_a/H_0 > 3 \times 10^{5}\big)
\end{equation}
scaling as $m_a^{1/2}$.

\subsubsection{Light but massive thermal relics}\label{eq:relics}
Light but massive thermal relics~\cite{Munoz:2018ajr,DePorzio:2020wcz,Xu:2021rwg} correspond to $\gamma = 2$ and $p = 0$. These species decouple from the thermal bath while relativistic, preserving their unperturbed phase space distribution, which takes a Fermi-Dirac or Bose-Einstein form. Eventually, they become nonrelativistic at $a = a_{\rm nr}$, but they retain a cosmologically relevant free-streaming scale. In general, light relics are characterized by their mass, number of degrees of freedom and temperature. The temperature, in turn, depends on when the species decoupled from the thermal bath. The application of the two-fluid EFTofLSS proposed in this paper to general light relics, going beyond Refs.~\cite{Munoz:2018ajr,Xu:2021rwg} where the clustering of the warm species was not included, will be presented elsewhere~\cite{modelspaper}.

In this work, to illustrate the main features of the application to light relics we focus exclusively on the most familiar among them, namely the standard massive neutrinos \cite{Lesgourgues:2006nd}. These decoupled from the thermal bath at temperature around $T\sim \mathrm{MeV}$ and therefore have $T_{\nu,0}/T_{\gamma,0} \approx 0.716$ today. One neutrino of mass $m_\nu$ is described in our framework by starting the fluid description at $z_{\rm ini} = z_{\rm nr}$, where $z_{\rm nr} \approx 1890\, (m_\nu / \mathrm{eV})$ is the point of the matter-dominated era when $\langle p \rangle = \frac{7 \pi^4}{180\, \zeta(3)}\, T_\nu = m_\nu$. For massive neutrinos the shear stress $\sigma(\vec{k},\tau)$ does not vanish, and in principle accounting for its evolution would require one to truncate the Boltzmann hierarchy at a multipole $l_{\rm max} \geq 2$, going beyond the fluid approximation based on $\delta$ and $\theta$. However, it has been shown that at late times the effect of the shear can be approximated in the fluid description as an additional contribution to the pressure term, effectively rescaling it by a factor $9/5$~\cite{Shoji:2010hm}. We follow this approach and furthermore we approximate $c_s^2$ with the ($k$-independent) adiabatic sound speed $c_g^2 = \overline{\mathcal{P}}^\prime / \overline{\rho}^\prime \simeq 5\hspace{0.2mm} \sigma_{\nu}^2/9$, where $\sigma_{\nu}$ is the velocity dispersion~\cite{Shoji:2010hm,Garny:2020ilv}. Thus we arrive at the choice $c_s^2 = \frac{9}{5} c_g^2 \simeq \sigma_\nu^2 = \tfrac{15\, \zeta(5)}{\zeta(3)} T_{\nu, 0}^2 / (m_\nu^2 a^2)$.\footnote{Alternative definitions of the velocity dispersion, differing by $\mathcal{O}(1)$ factors, are common in the literature. In particular, $\sigma_\nu^2 = \tfrac{3\zeta(3)}{2\log 2} T_{\nu, 0}^2 / (m_\nu^2 a^2)$ in Refs.~\cite{Munoz:2018ajr,Xu:2021rwg,Nascimento:2023psl}, which would lead to $k_s(a)$ being larger by a factor $\approx 2.2$ compared to our expression in Eq.~\eqref{eq:k_s_thermal}.} Then the characteristic wavenumber in Eq.~\eqref{eq:Jeans}, which is interpreted here as the neutrino free streaming scale, reads
\begin{equation}\label{eq:k_s_thermal}
k_s(a) \approx a^{1/2}\; 3.7\times 10^{-2}\, \frac{h}{\mathrm{Mpc}}   \Big(  \frac{m_\nu}{0.1\;\mathrm{eV}} \Big)  \Big( \frac{\Omega_m^0}{0.3} \Big)^{1/2}.
\end {equation}
A general fluid description of thermal relics would also include intrinsic nonlinearities, which arise upon taking moments of the Vlasov equation to derive the continuity and Euler equations. In Appendix~\ref{app:app_4_Vlasov} we show that these intrinsic nonlinearities have a mild impact on our results, supporting our approach where only nonlinearities of pure gravitational origin are accounted for.

Thermal freeze-out determines the warm matter fraction as
\begin{equation}\label{eq:f_nu}
f_\chi \approx \frac{1}{h^2 \Omega_m^0} \frac{\sum m_\nu}{93.14\;\mathrm{eV}} \approx 1.1 \times 10^{-2}\, h^{-2} \left(  \frac{\sum m_\nu}{0.3\;\mathrm{eV}} \right)  \left( \frac{0.3}{\Omega_m^0} \right)\,.
\end{equation}
Similarly to ULAs, in cosmologies involving light thermal relics the matter power spectrum displays a drop-like feature. For massive neutrinos, its location $k_{\rm drop}$ is approximately given by $k_s(a_{\rm nr})$ if the transition to the nonrelativistic regime happens during MD and by $k_s(a_{\rm eq})$ if it happens during RD, with $k_s(a)$ defined in Eq.~\eqref{eq:k_s_thermal}.

Since in this work we view the standard neutrinos as a convenient proxy for more general light thermal relics, for the sake of illustration we show results for a variety of neutrino masses, including values up to $m_\nu \sim 1\; \mathrm{eV}$ that have already been excluded by data.

\subsubsection{Accuracy of the analytical linear solutions and discussion}\label{sec:accuracy}

\begin{figure}
    \centering
    \includegraphics[width = \textwidth]{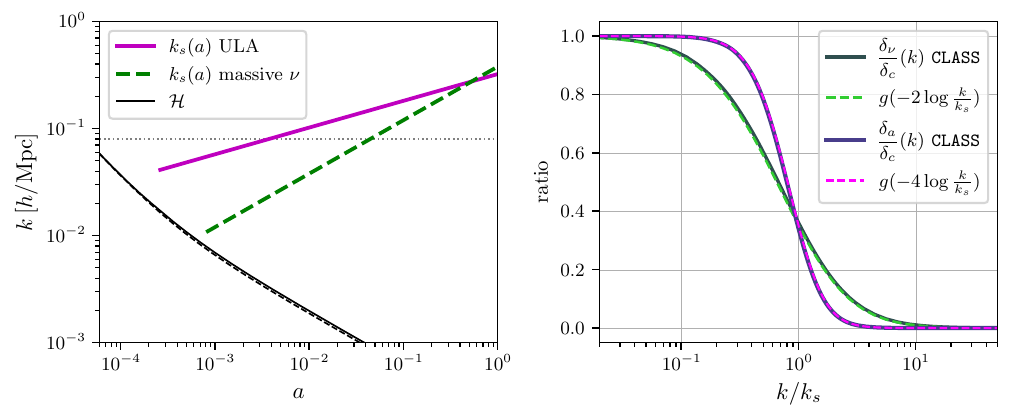}
    \vspace{-0.5cm}\caption{{\it Left:} Time evolution of the characteristic wavenumber $k_s(a)$ for two example models described by our framework: an ULA field with $m_a = 10^{-27}\;\mathrm{eV}$ and $f_\chi = 0.1$ (purple line) and a thermal relic massive neutrino with $m_\nu = 1\;\mathrm{eV}$ and $f_\chi \approx 0.08$ (dashed green line). The black curves show the evolution of the Hubble scale $\mathcal{H}$ for the ULA (solid) and massive neutrino (dashed) cosmologies. The dotted horizontal line indicates a representative $k$ mode that evolves differently in the two models. {\it Right:} The ratio of linear perturbations $\delta_\chi^{[1]}/\delta_c^{[1]}$ as a function of $k/k_s$, evaluated at $z = 0.5$, for the same cosmologies shown in the left panel. Dashed curves correspond to our universal closed-form approximation $g^{(0)}\big(\hspace{-1mm} - \frac{2+p}{\gamma - 1} \log \frac{k} {k_s}\big)$, with $\gamma = 2$ and $p = 2$~($p = 0$) for ULAs~(massive neutrinos). Solid curves show the same ratios as obtained from~\texttt{CLASS}.}
    \label{fig:ks_a_evolution}
\end{figure}

The time evolution of $k_s$ is illustrated in the left panel of Fig.~\ref{fig:ks_a_evolution} for two example cosmologies, one containing ULAs and the other containing thermal relics in the form of massive neutrinos. For the former we assume $m_a = 10^{-27}\;\mathrm{eV}$ with $f_\chi = 0.1$, whereas for the latter we take one neutrino of mass $m_\nu = 1\;\mathrm{eV}$, corresponding to a warm fraction $f_\chi \approx 0.08$ according to Eq.~\eqref{eq:f_nu}.

For the same two cosmologies, in the right panel of Fig.~\ref{fig:ks_a_evolution} we compare our closed-form approximation for the ratio of transfer functions $\delta_\chi^{[1]}/\delta_c^{[1]}$, which recalling Eq.~\eqref{eq:etatilde_approx} we write as $g^{(0)}\big( - \frac{2+p}{\gamma - 1} \log \frac{k} {k_s}\big)$ with the $g^{(0)}$ function defined in Eq.~\eqref{eq:g0_gamma2}, to the numerical value obtained from the numerical Einstein-Boltzmann solver \texttt{CLASS}~\cite{Lesgourgues:2011re,Blas:2011rf}. Specifically, for massive neutrinos we use the standard \texttt{CLASS} release~\cite{Lesgourgues:2011rh} with~\texttt{ncdm\_fluid\_approximation}~$=1$~\cite{Hu:1998kj}, while for ULAs we employ \texttt{AxiCLASS}~\cite{Poulin:2018dzj,Smith:2019ihp}. Excellent agreement is observed in both cases: our universal description of linear fluctuations accurately predicts the ratios of linear transfer functions. The ULA model has larger $p$, therefore a sharper transition around $k\sim k_s$ is observed.

\begin{figure*}
    \centering
     \includegraphics[width = \textwidth]{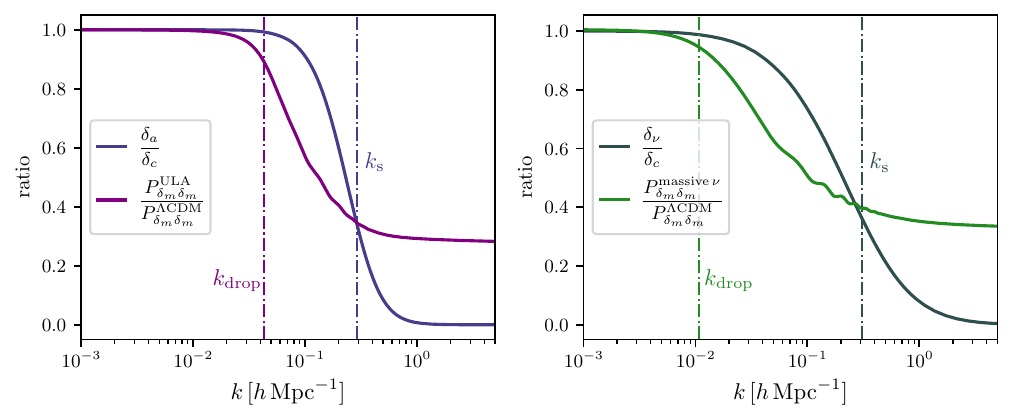}
    \vspace{-0.5cm}\caption{{\it Left:} Ratio of the linear matter power spectrum to $\Lambda$CDM (solid purple) compared to the ratio of linear perturbations $\delta_\chi^{[1]}/\delta_c^{[1]}$ (solid blue) for the same ULA cosmology as in Fig.~\ref{fig:ks_a_evolution}: $m_a = 10^{-27}\;\mathrm{eV}$ and $f_\chi = 0.1$. Both ratios are evaluated at $z = 0.5$ and obtained from~\texttt{CLASS}. The dot-dashed vertical lines show analytical approximations for the two relevant dynamical scales: $k_{\rm drop} = k_s(z_{\rm eq})$, which determines the beginning of the drop in the linear power spectrum, and $k_s(z)$, which marks the end of the step. {\it Right:} Same as in the left panel, but for the massive neutrino cosmology considered in Fig.~\ref{fig:ks_a_evolution}: $m_\nu = 1\;\mathrm{eV}$ and $f_\chi \approx 0.08$. Here the dot-dashed vertical lines show $k_{\rm drop} \simeq k_s(z_{\rm nr})$ and $k_s(z)$.}
    \label{fig:comparison_CLASS}
\end{figure*}

As can be seen in the left panel of Fig.~\ref{fig:ks_a_evolution}, the parameters of the two benchmark cosmologies were chosen to yield similar values of $k_s (z_{\rm obs})$ at the low redshift $z_{\rm obs} \sim 0.5$ observed by galaxy surveys. The warm fraction $f_\chi$ also takes similar values in the two scenarios. Nevertheless, we expect different imprints on the matter power spectrum due to the different time dependence of $k_s$: for instance, the $k = 0.08\,h\;\mathrm{Mpc}^{-1}$ mode (dotted horizontal line in the left panel of Fig.~\ref{fig:ks_a_evolution}) starts growing later in the massive neutrino model, hence the power spectrum will be more suppressed at this wavenumber than in the ULA scenario. These expectations are confirmed by Fig.~\ref{fig:comparison_CLASS}, where the \texttt{CLASS} output for the linear power spectrum of total matter, $P_{\delta_m \delta_m}^{L}$, is shown for the ULA (left panel) and massive neutrino (right panel) cosmologies, both normalized to the $\Lambda$CDM prediction. For comparison, we also show the ratio of transfer functions $\delta_\chi^{[1]}/\delta_c^{[1]}$ that appeared in the right panel of Fig.~\ref{fig:ks_a_evolution}.

Figure~\ref{fig:comparison_CLASS} highlights the clear separation between the two scales at play: $k_{\rm drop}$, which corresponds to the smallest $k$ where $P_{\delta_m \delta_m}^{L}$ is suppressed relative to $\Lambda$CDM, and $k_s(z_{\rm obs})$, which is relevant for dynamics at the low redshift $z_{\rm obs}$ and marks the end of the step-like feature. The ratio between the two scales, which determines the width of the step, is bounded from above by
\begin{equation} \label{eq:ratio_drop_ks}
\frac{k_s(z_{\rm obs})}{k_{\rm drop}} \leq \bigg( \frac{1 + z_{\rm eq}}{1 + z_{\rm obs}} \bigg)^{\frac{1}{2+p}}\,,
\end{equation}
where the inequality is saturated when $k_{\rm drop} = k_s(z_{\rm eq})$. Taking a representative $z_{\rm obs} = 0.5$, we find an upper bound of $\sim 7$ for ULAs ($p = 2$), which is indeed saturated in the left panel of Fig.~\ref{fig:comparison_CLASS} because for the considered mass the axion field starts oscillating before matter-radiation equality. The hierarchy between the two scales can be much larger for light thermal relics ($p = 0$), with an upper bound of $\sim 50$. In the cosmology assumed in the right panel of Fig.~\ref{fig:comparison_CLASS} the neutrino becomes nonrelativistic during MD, hence $k_{\rm drop} \simeq k_s (z_{\mathrm{nr}})$ and the ratio is about $30$.

On the other hand, the depth of the step in the matter power spectrum is determined by the fraction $f_\chi$. At the linear level, using our analytical solutions we readily estimate the depth as
\begin{equation}\label{eq:P_depth}
\frac{P_{\delta_m \delta_m}^L}{P_{\delta_m \delta_m}^{L,\, \Lambda\mathrm{CDM}}} \Bigg|_{k\, \gg\, k_s(z_{\rm obs})} =\, (1 - f_\chi)^2\, \exp \Big( \hspace{-0.5mm} - \frac{6}{5}f_\chi \log \frac{1 + z_{\rm ini}}{1 + z_{\rm obs}} \Big)\,.
\end{equation}
For the ULA~(neutrino) cosmology in Fig.~\ref{fig:comparison_CLASS} this gives $\approx 0.32~(0.42)$, in $\sim 20\%$ agreement with the numerical result $\approx 0.28~(0.34)$. This is reasonable, if one considers that in the estimate for the ULA we assumed instantaneous transition from radiation to matter domination ($z_{\rm ini} = z_{\rm eq}$), whereas for the neutrino we assumed instantaneous transition from the relativistic to the nonrelativistic regime ($z_{\rm ini} = z_{\rm nr}$).

Importantly, we note that expanding the linear power spectrum for small $f_\chi$ would not be appropriate, due to the large logarithm appearing in the exponent of Eq.~\eqref{eq:P_depth} (and the relatively large size of $f_\chi$ considered). It is well known that such large logarithm, which corresponds to a time-integrated effect, can invalidate the expansion of the linear perturbations of the cold species~\cite{Senatore:2017hyk}. For this reason, in our approach the expansion in $f_\chi \ll 1$ is never performed on $\delta_c^{[1]}$, but only on quantities that are by construction free from large time logarithms, namely {\it ratios} of linear perturbations ($\delta_\chi^{[1]}/\delta_c^{[1]}$ and $\Theta_x^{[1]}/\delta_c^{[1]}$ for $x = \chi, c$) and nonlinear kernels. In our framework and results the linear power spectrum of the cold species, $P_{\delta_c \delta_c}^L$, is always obtained from a full numerical solution computed with~\texttt{CLASS}.

\section{Galaxy Power Spectrum in Redshift Space}\label{sec:nonlinear_real}

In this section we present all the theoretical ingredients needed to predict the $1$-loop power spectrum of galaxies in redshift space, which can be compared to data.

\subsection{Matter nonlinearities}\label{sec:matter_nonl}
We begin by discussing solutions beyond the linear order to Eqs.~\eqref{eq:eta_kspace_nonlinear}, which describe the dynamics of the two coupled fluids $\chi$ and $c$. Using the notation introduced in Eq.~\eqref{eq:pt-expansion-notation}, we define the perturbative expansion of the solutions as
\begin{subequations}\label{eq:kernel_expansion}
 \begin{align}
     \delta^{[n]}_\chi(\vk,\te) &= \int_\vk \mathrm{d}k_{1\ldots n} \,F^{[n]}_\chi(\vk_1,\ldots,\vk_n;\te) \,\delta_\chi^{[1]}(\vk_1,\te)\ldots\delta_\chi^{[1]}(\vk_n,\te)\,,\\
     \Theta^{[n]}_\chi(\vk,\te) &= \int_\vk \mathrm{d}k_{1\ldots n} \,G^{[n]}_\chi(\vk_1,\ldots,\vk_n;\te) \,\delta_\chi^{[1]}(\vk_1,\te)\ldots\delta_\chi^{[1]}(\vk_n,\te)\,, \\
     \delta^{[n]}_c(\vk,\te) &= \int_\vk \mathrm{d}k_{1\ldots n}\, F^{[n]}_c(\vk_1,\ldots,\vk_n;\te)\, \delta_c^{[1]}(\vk_1,\te)\ldots\delta_c^{[1]}(\vk_n,\te)\,,\\
     \Theta^{[n]}_c(\vk,\te) &= \int_\vk \mathrm{d}k_{1\ldots n}\, G^{[n]}_c(\vk_1,\ldots,\vk_n;\te)\, \delta_c^{[1]}(\vk_1,\te)\ldots\delta_c^{[1]}(\vk_n,\te)\,.
 \end{align}
 \end{subequations}
Note that we chose to expand the nonlinear corrections for the $\chi$ species in terms of its own linear field $\delta_\chi^{[1]}$, rather than $\delta_c^{[1]}$. Importantly, our definition of the momentum-dependent time variable $\tilde{\eta}$ always refers to the external momentum $\vec{k} = \vec{k}_1 + \ldots + \vec{k}_n\,$. The linear solutions found in Sec.~\ref{sec:lin_sol} set 
\begin{equation}
F_\chi^{[1]}(\vec{k}_1;\tilde{\eta}_{k_1}) = F_c^{[1]}(\vec{k}_1;\tilde{\eta}_{k_1}) = 1\,,\quad G_\chi^{[1]}(\vec{k}_1;\tilde{\eta}_{k_1}) = \frac{h(\tilde{\eta}_{k_1})}{g(\tilde{\eta}_{k_1})}\,,\quad G_c^{[1]}(\vec{k}_1;\tilde{\eta}_{k_1}) = s(\tilde{\eta}_{k_1})\,,
\end{equation}
where we defined
\begin{equation}
\tilde{\eta}_{q} \equiv \tilde{\eta} - \frac{2+p}{\gamma - 1} \log \frac{q}{ k }
\end{equation}
for an arbitrary vector $\vec{q}$.

At nonlinear order, differential equations for the kernels are found by substituting the expansions~\eqref{eq:kernel_expansion} into Eqs.~\eqref{eq:eta_kspace_nonlinear}. For example, at second order ($n= 2$) we obtain
\begin{subequations}\label{eq:2nd}
\begin{align}\label{eq:chi_2nd_A}
 &\Big(\partial_{\tilde{\eta}} + \frac{h}{g}(\tilde{\eta}_{k_1})+\frac{h}{g}(\tilde{\eta}_{k_2}) \Big) F_\chi^{[2]} - G_\chi^{[2]} = \frac{1}{2}\Big( \alpha(\vec{k}_1, \vec{k}_2)\frac{h}{g}(\tilde{\eta}_{k_1}) + \alpha(\vec{k}_2, \vec{k}_1)\frac{h}{g}(\tilde{\eta}_{k_2}) \Big), \\
&\Big(\partial_{\tilde{\eta}} + \frac{1}{2} + \frac{h}{g}(\tilde{\eta}_{k_1})+\frac{h}{g}(\tilde{\eta}_{k_2}) \Big) G_\chi^{[2]} + \frac{3}{2}\Big(e^{-(\gamma - 1)\tilde{\eta}} - f_\chi\Big) F_\chi^{[2]} -\frac{3}{2}(1 - f_\chi) \frac{F_c^{[2]}}{g(\tilde{\eta}_{k_1}) g(\tilde{\eta}_{k_2})} \nonumber \\ \label{eq:chi_2nd_B}
&\hspace{8.0cm} = \beta(\vec{k}_1, \vec{k}_2) \frac{h}{g}(\tilde{\eta}_{k_1})\frac{h}{g} (\tilde{\eta}_{k_2})\,, \\ \label{eq:c_2nd_A}
&\Big(\partial_{\tilde{\eta}} + s(\tilde{\eta}_{k_1})+s(\tilde{\eta}_{k_2}) \Big) F_c^{[2]} - G_c^{[2]}=\frac{1}{2}\Big( \alpha(\vec{k}_1, \vec{k}_2) s(\tilde{\eta}_{k_1}) + \alpha(\vec{k}_2, \vec{k}_1) s(\tilde{\eta}_{k_2}) \Big),\\ \label{eq:c_2nd_B}
&\Big(\partial_{\tilde{\eta}} + \frac{1}{2} + s(\tilde{\eta}_{k_1})+s(\tilde{\eta}_{k_2}) \Big) G_c^{[2]} - \frac{3}{2} f_\chi F_\chi^{[2]} g(\tilde{\eta}_{k_1}) g(\tilde{\eta}_{k_2}) -\frac{3}{2}(1 - f_\chi) F_c^{[2]} \nonumber \\
&\hspace{8cm} = \beta(\vec{k}_1, \vec{k}_2) s(\tilde{\eta}_{k_1})s(\tilde{\eta}_{k_2})\,,
\end{align}
\end{subequations}
where the argument of the kernels is understood to be $(\vec{k}_1, \vec{k}_2; \tilde{\eta})$.

At $\mathcal{O}(f_\chi^0)$ we have $s^{(0)}(\tilde{\eta}) = 1$ and the equations for $c$ are solved by the standard, $\tilde{\eta}$-independent EdS kernels for all $n$,
\begin{equation}
F_c^{(0)[n]}(\vec{k}_1, \ldots, \vec{k}_n ; \tilde{\eta}) = F_n (\vec{k}_1, \ldots, \vec{k}_n)\,, \quad G_c^{(0)[n]}(\vec{k}_1, \ldots, \vec{k}_n ; \tilde{\eta}) = G_n (\vec{k}_1, \ldots, \vec{k}_n)\,,
\end{equation}
where the $F_n$ and $G_n$ functions can be found e.g.~in Ref.~\cite{Bernardeau:2001qr}. All the remaining pieces of the nonlinear solutions depend on the value of $\gamma$. In the rest of this section we focus on $\gamma = 2$, building upon the linear results presented in Sec.~\ref{sec:micro_models}.

\subsubsection{Analytical solutions for $\gamma = 2$}\label{sec:analytical}
Although a general solution for the nonlinear kernels defined in Eq.~\eqref{eq:kernel_expansion} requires numerical integration, analytical expressions can be derived for kinematic configurations where all the momenta are well separated from the characteristic scale $k_s$. In that respect, three distinct physical regimes can be identified depending on the size of $k_s$ relative to the observational window of galaxy surveys, which is roughly given by $k_{\rm eq} \lesssim k \lesssim k_{\rm NL}$ (with $k_{\rm eq} \equiv \mathcal{H}_{\rm eq}$): 
\begin{itemize}
\item When $k_{\rm drop} \gtrsim k_{\rm NL}$ (and therefore $k_s (z_{\rm obs}) \gg k_{\rm NL}$), the $\chi$ and $c$ species are indistinguishable on all scales probed by galaxy surveys, hence a standard description in terms of a single CDM fluid is adequate.
\item When $k_s(z_{\rm obs}) \lesssim k_{\rm eq}$, the $\chi$ perturbations are strongly suppressed in the whole region of wavenumbers relevant to LSS data. Massive neutrinos with $m_\nu \lesssim 0.1\;\mathrm{eV}$ are a well-known realization of this case, as Eq.~\eqref{eq:k_s_thermal} shows. A description in terms of a single fluid $c$ is appropriate, however the presence of the non-clustering species $\chi$ affects the evolution of the $c$ perturbations in a nontrivial way.
\item When $k_{\rm eq} < k_s(z_{\rm obs})$ and $k_{\rm drop} < k_{\rm NL}$ are both satisfied,\footnote{For instance, in the case of ULAs this regime corresponds, recalling Eq.~\eqref{eq:ratio_drop_ks}, to $k_{\rm eq} < k_s(z_{\rm obs}) < 7\, k_{\rm NL}$.} the characteristic scale leaves direct imprints in the fluctuations of both $\chi$ and $c$. A computationally efficient yet accurate modeling of the $1$-loop galaxy power spectrum in this regime is the main goal of this paper.
\end{itemize}
We now present analytical results for the nonlinear matter kernels corresponding to each of the three regimes.

If all momenta are much smaller than $k_s(a)$, then the $\chi$ and $c$ species are indistinguishable. This is a trivial limit where we just have a single fluid behaving as CDM, with kernels given by the standard $F_n$ and $G_n$. 

If all momenta are much larger than $k_s(a)$, the $\chi$ perturbations are entirely negligible and we only need to solve for the dynamics of the cold species $c$. In Appendix~\ref{sec:app_1} we show that, perhaps surprisingly, in this limit the equations obeyed by the $c$ perturbations are tightly related to the equations found in Refs.~\cite{Archidiacono:2022iuu,Bottaro:2023wkd} for a different type of new dynamics affecting the CDM+baryon fluid: DM self-interactions mediated by a massless (scalar) field. As a consequence, and as detailed in Appendix~\ref{sec:app_1}, solutions for this regime where all momenta are large are immediately obtained from the results of Ref.~\cite{Bottaro:2023wkd}. For instance, up to second perturbative order for $k_1, k_2, k \gg k_s(a)$, with $\vec{k} = \vec{k}_1 + \vec{k}_2$, we find
\begin{subequations}\label{eq:c_2nd_hot}
\begin{align}
F_c^{[2]}(\vec{k}_1, \vec{k}_2; \tilde{\eta}) =&\; F_2(\vec{k}_1, \vec{k}_2) + \frac{6f_\chi}{245} \big[ \alpha_s (\vec{k}_1, \vec{k}_2) - \beta(\vec{k}_1, \vec{k}_2) \big] + \mathcal{O}(f_\chi^2)\,, \\ \label{eq:c_2nd_hot_2}
G_c^{[2]}(\vec{k}_1, \vec{k}_2 ; \tilde{\eta}) =&\; \Big(1 - \frac{3f_\chi}{5}\Big) G_2(\vec{k}_1, \vec{k}_2) + \frac{12 f_\chi}{245} \big[ \alpha_s (\vec{k}_1, \vec{k}_2) - \beta(\vec{k}_1, \vec{k}_2) \big] + \mathcal{O}(f_\chi^2)\,.
\end{align}
\end{subequations}
These expressions can, of course, also be obtained by solving directly Eqs.~\eqref{eq:2nd} in the appropriate limit. The dominant beyond-$\Lambda$CDM imprint is the suppression of the linear growth rate $f(k)/f_{\Lambda \mathrm{CDM}} \simeq 1 - 3f_\chi/5$, which results in the factor multiplying $G_2$ in Eq.~\eqref{eq:c_2nd_hot_2}. Additional effects are strongly suppressed by accidentally small coefficients, such as $6 /245 \approx 0.024$, which make them practically unobservable. This picture is consistent with the results obtained in Refs.~\cite{Aviles:2021que,Noriega:2022nhf} for massive neutrino cosmologies.\footnote{Equations~\eqref{eq:c_2nd_hot} also agree with Eq.~(15) in Ref.~\cite{Kamalinejad:2020izi}, after correcting obvious typos in that reference.}

Finally, we turn to the regime where some momenta are large relative to $k_s(a)$, while others are small. This is the most relevant situation for our purposes, as well as the most complex one. The analytical results we present below are, to our knowledge, entirely new. 

At second perturbative order and $\mathcal{O}(f_\chi^0)$, Eqs.~\eqref{eq:chi_2nd_A} and~\eqref{eq:chi_2nd_B} yield in the $k_1 \ll k_s(a) \ll k_2$ configuration
\begin{subequations} \label{eq:limit_chi}
\begin{align}\label{eq:limit_chi_A}
F_\chi^{(0)[2]}(\vec{k}_1, \vec{k}_2; \tilde{\eta}) =\;& F_2(\vec{k}_1, \vec{k}_2)\,, \\ G_\chi^{(0)[2]}(\vec{k}_1, \vec{k}_2; \tilde{\eta}) =\;& 2\hspace{0.3mm} G_2(\vec{k}_1, \vec{k}_2) - \frac{3}{14}\alpha_s(\vec{k}_1, \vec{k}_2) + \frac{1}{2}\alpha_a (\vec{k}_1, \vec{k}_2) - \frac{2}{7}\beta(\vec{k}_1, \vec{k}_2) \,, \label{eq:limit_chi_B}
\end{align}
\end{subequations}
where we have defined the symmetric and antisymmetric components of $\alpha$ as
\begin{equation}
\alpha_{s,a}(\vec{k}_1, \vec{k}_2) \equiv \frac{1}{2} \big[\alpha(\vec{k}_1, \vec{k}_2) \pm \alpha(\vec{k}_2, \vec{k}_1) \big]\,.
\end{equation}
For the cold species at $\mathcal{O}(f_\chi)$ we obtain
\begin{subequations} \label{eq:limit_c}
\begin{align} \label{eq:limit_c_A}
F_c^{(1)[2]}(\vec{k}_1, \vec{k}_2; \tilde{\eta}) =\;&\hspace{-1mm} - \frac{69}{490} \alpha_s(\vec{k}_1, \vec{k}_2) + \frac{3}{14}\alpha_a(\vec{k}_1, \vec{k}_2) - \frac{18}{245}\beta(\vec{k}_1, \vec{k}_2) \,, \\ G_c^{(1)[2]}(\vec{k}_1, \vec{k}_2; \tilde{\eta}) =\;&\hspace{-1mm} -\frac{3}{5} G_2(\vec{k}_1, \vec{k}_2) - \frac{15}{98}\alpha_s(\vec{k}_1, \vec{k}_2) + \frac{9}{70}\alpha_a (\vec{k}_1, \vec{k}_2) + \frac{6}{245}\beta(\vec{k}_1, \vec{k}_2) \,. \label{eq:limit_c_B}
\end{align}
\end{subequations}
We have chosen to write the right-hand sides of Eqs.~\eqref{eq:limit_chi} and~\eqref{eq:limit_c} so that the infrared (IR) divergence for $\vec{k}_1 \to 0$, where present, is fully captured by the term proportional to $F_2$ or $G_2$. The appearance of the antisymmetric mode coupling function $\alpha_a$, which is absent in $\Lambda$CDM, is enabled by the presence of the new scale $k_s(a)$. The source terms of the continuity equations~\eqref{eq:chi_2nd_A} and~\eqref{eq:c_2nd_A}, which must be symmetric under $\vec{k}_1 \leftrightarrow \vec{k}_2$ exchange, contain products of $\alpha_a$ times antisymmetric combinations of the new functions of momentum $h/g$ and $s$. For example, for the $\chi$ species one finds
\begin{equation}
\alpha_a (\vec{k}_1, \vec{k}_2) \Big[ \frac{h}{g}(\tilde{\eta}_{k_1}) - \frac{h}{g}(\tilde{\eta}_{k_2}) \Big] \;\; \subset \;\; \alpha(\vec{k}_1, \vec{k}_2)\frac{h}{g}(\tilde{\eta}_{k_1}) + \alpha(\vec{k}_2, \vec{k}_1)\frac{h}{g}(\tilde{\eta}_{k_2}) \,,
\end{equation}
which leads to the appearance of $\alpha_a$ in Eq.~\eqref{eq:limit_chi} after using the limits $h^{(0)}/g^{(0)} \to 1$ for $k_1 \ll k_s(a)$ and $h^{(0)}/g^{(0)} \to 2$ for $k_2 \gg k_s(a)$. It should be noted that this momentum coupling is not present in bootstrap approaches to LSS~\cite{DAmico:2021rdb,Marinucci:2024add}, as the latter are constructed for scale-free theories where such a structure cannot appear.

The simplest observables where the new dynamics we derived above for the $k_1 \ll k_s(a) \ll k_2$ configuration will appear, are tree-level bispectra. However, since the LSS data analysis of Sec.~\ref{sec:constraints} considers only the galaxy power spectrum, we defer a detailed study of the signatures of mixed DM scenarios in higher-point correlators to future work.

\subsubsection{Infrared structure}\label{sec:IR}
Here we discuss the IR behavior of the matter solutions. We begin with $\chi$ at $\mathcal{O}(f_\chi^0)$. Taking $\vec{k}_1 = \vec{q}$ and $\vec{k}_2 = \vec{k} - \vec{q}$ with $\vec{q} \to 0$, the second-order equations~\eqref{eq:chi_2nd_A} and~\eqref{eq:chi_2nd_B} are solved by
\begin{equation}\label{eq:chi_2_IR}
F_\chi^{(0)[2]}(\vec{q}, \vec{k} - \vec{q}; \tilde{\eta}) \stackrel{\vec{q} \,\to\, 0}{=} F_2 (\vec{q}, \vec{k} - \vec{q})\,,\quad\; G_\chi^{(0)[2]}(\vec{q}, \vec{k} - \vec{q}; \tilde{\eta}) \stackrel{\vec{q} \,\to\, 0}{=} \frac{h^{(0)}}{g^{(0)}}(\tilde{\eta})\, G_2 (\vec{q}, \vec{k} - \vec{q})\,,
\end{equation}
which should be understood as equalities of the $1/q$ poles.\footnote{Equation~\eqref{eq:chi_2_IR} is found to be consistent with Eq.~\eqref{eq:limit_chi} after recalling that in the former $h^{(0)}/g^{(0)} \to 2$ for $k \gg k_s(a)$, while in the latter only the terms proportional to $F_2$ or $G_2$ are IR divergent.} They were derived by setting $g^{(0)}(\tilde{\eta}_{q}), h^{(0)}(\tilde{\eta}_{q}) \stackrel{\vec{q}\,\to\, 0}{\to} 1$ and employing the linear equation~\eqref{eq:lin_g0} for $g^{(0)}(\tilde{\eta})$. Hence, at $\mathcal{O}(f_\chi^0)$ the IR divergence in $F_\chi^{[2]}$ is the same as in a standard EdS cosmology. By contrast, the residual of the IR pole in $G_\chi^{[2]}$ is time dependent and proportional to the $\chi$ linear growth rate $h/g$. A similar calculation at the third perturbative order results in the equalities
\begin{equation}\label{eq:chi_3_IR}
F_\chi^{(0)[3]}(\vec{k},\vec{q}, -\vec{q}; \tilde{\eta}) \hspace{-0.8mm} \stackrel{\vec{q} \,\to\, 0}{=} \hspace{-1mm}F_3 (\vec{k},\vec{q}, -\vec{q})\,,\quad G_\chi^{(0)[3]}(\vec{k},\vec{q}, -\vec{q}; \tilde{\eta}) \hspace{-0.1mm}\stackrel{\vec{q} \,\to\, 0}{=}\hspace{-0.1mm} \frac{h^{(0)}}{g^{(0)}}(\tilde{\eta})\, G_3 (\vec{k},\vec{q}, -\vec{q})\,,
\end{equation}
which hold for the $1/q^2$ poles. 

An analogous IR structure is observed in the solutions for the cold species. At $\mathcal{O}(f_\chi)$, the second-order equations~\eqref{eq:c_2nd_A} and~\eqref{eq:c_2nd_B} are solved by
\begin{equation}\label{eq:c_2_IR}
F_c^{(1)[2]}(\vec{q}, \vec{k} - \vec{q}; \tilde{\eta}) \stackrel{\vec{q} \,\to\, 0}{=} 0\,, \qquad
G_c^{(1)[2]}(\vec{q}, \vec{k} - \vec{q}; \tilde{\eta}) \stackrel{\vec{q} \,\to\, 0}{=}  s^{(1)}(\tilde{\eta}) \,G_2 (\vec{q}, \vec{k} - \vec{q})\,,
\end{equation}
obtained by setting $s^{(1)}(\tilde{\eta}_q)\stackrel{\vec{q}\,\to\, 0}{\to} 0$ and $g^{(0)}(\tilde{\eta}_q)\stackrel{\vec{q}\,\to\, 0}{\to} 1$ and making use of the linear equation~\eqref{eq:s1} for $s^{(1)}(\tilde{\eta})$. Therefore, up to $\mathcal{O}(f_\chi)$ the IR divergence of $F_c^{[2]}$ is not modified relative to standard cosmology, whereas the residual of the IR pole in $G_c^{[2]}$ is proportional to the linear growth rate $s$. Similarly, at third order we find
\begin{equation}\label{eq:c_3_IR}
F_c^{(1)[3]} (\vec{k}, \vec{q}, - \vec{q}; \tilde{\eta}) \stackrel{\vec{q} \,\to\, 0}{=} 0\,, \qquad
G_c^{(1)[3]} (\vec{k}, \vec{q}, - \vec{q}; \tilde{\eta}) \stackrel{\vec{q} \,\to\, 0}{=} s^{(1)}(\tilde{\eta}) \,G_3 (\vec{k}, \vec{q}, - \vec{q})\,.
\end{equation}
It is easy to guess, then, that Eqs.~\eqref{eq:chi_2_IR}~-~\eqref{eq:c_3_IR} generalize to all orders in $f_\chi$, namely
\begin{equation}\label{eq:IR_limits_general}
 F_{\chi}^{[n]}\stackrel{\vec{q} \,\to\, 0}{=} F_n\,, \quad  G_{\chi}^{[n]}\stackrel{\vec{q}\, \to\, 0}{=} \frac{h}{g}(\tilde{\eta})\, G_n \,,\qquad  F_{c}^{[n]}\stackrel{\vec{q} \,\to\, 0}{=} F_n\,,\quad  G_{c}^{[n]}\stackrel{\vec{q}\, \to\, 0}{=} s(\tilde{\eta})\, G_n\,,
\end{equation}
where the arguments of the kernels are understood to be $(\vec{q}, \vec{k} - \vec{q};\tilde{\eta})$ for $n=2$ and $(\vec{k}, \vec{q}, -\vec{q};\tilde{\eta})$ for $n=3$. 

Equation~\eqref{eq:IR_limits_general} guarantees, at all orders in $f_\chi$, the absence of IR divergences in all 1-loop power spectra, as a direct consequence of their well-known cancellation in $\Lambda$CDM. For instance, the $1$-loop correction to the $\delta_c\,$-$\,\Theta_\chi$ cross power spectrum reads
\begin{align}
&\,P_{\delta_c \Theta_\chi}^{1\text{-}\mathrm{loop}}(k) = 4\hspace{-1mm} \int\hspace{-1mm} \frac{d^3 q}{(2\pi)^3}  F_{c}^{[2]} (\vec{q}, \vec{k} - \vec{q}) G_{\chi}^{[2]} (\vec{q}, \vec{k} - \vec{q}) P_{\delta_c \delta_\chi}^L (q) P_{\delta_c \delta_\chi}^L (| \vec{k} - \vec{q} \,|) \Theta_{\rm H} (| \vec{k} - \vec{q}| - q ) \nonumber \\
+& 3 P_{\delta_c \Theta_\chi}^L (k) \hspace{-1.5mm} \int \hspace{-1.5mm}\frac{d^3 q}{(2\pi)^3} F_{c}^{[3]} (\vec{k}, \vec{q}, - \vec{q} ) P_{\delta_c \delta_c}^L (q) + 3 P_{\delta_c \delta_\chi}^L (k) \hspace{-1.5mm}\int \hspace{-1.5mm} \frac{d^3 q}{(2\pi)^3} G_{\chi}^{[3]} (\vec{k}, \vec{q}, - \vec{q} ) P_{\delta_\chi \delta_\chi}^L (q)\,, \label{eq:P_c_Thchi_1l}
\end{align}
where the dependence on the time variable $\tilde{\eta}$ is understood. Exploiting Eq.~\eqref{eq:IR_limits_general}, it is immediate to verify that in the $\vec{q}\to 0$ limit the integrand in $P_{\delta_c \Theta_\chi}^{1\text{-}\mathrm{loop}}$ is proportional to
\begin{equation}
4 F_2 (\vec{q}, \vec{k} - \vec{q}) G_2 (\vec{q}, \vec{k} - \vec{q}) + 3 F_3(\vec{k}, \vec{q}, - \vec{q} ) + 3 G_3(\vec{k}, \vec{q}, - \vec{q} )\,, 
\end{equation}
where divergences cancel exactly. 

Furthermore, Eq.~\eqref{eq:IR_limits_general} ensures that all tree-level bispectra are free from IR poles. For example, consider
\begin{align}
B_{\delta_c \delta_c \Theta_\chi}(\vec{q}, \vec{p}_1, \vec{p}_2) = 2 \Big[ F_c^{[2]}(\vec{p}_1, \vec{p}_2; \tilde{\eta}_q) P_{\delta_c \delta_c}^L &(p_1) P_{\delta_c \Theta_\chi}^L (p_2) + F_c^{[2]}(\vec{q}, \vec{p}_2; \tilde{\eta}_{p_1}) P_{\delta_c \delta_c}^L (q) P_{\delta_c \Theta_\chi}^L (p_2) \nonumber \\ 
&+ G_\chi^{[2]}(\vec{q}, \vec{p}_1; \tilde{\eta}_{p_2}) P_{\delta_c \delta_\chi}^L (q) P_{\delta_c \delta_\chi}^L (p_1) \Big]\,.  
\end{align}
Making use of Eq.~\eqref{eq:IR_limits_general} one easily verifies that $B_{\delta_c \delta_c \Theta_\chi}$ is proportional to $F_2(\vec{q}, -\vec{p}_1) + G_2(\vec{q},\vec{p}_1)$ in the $\vec{q}\to 0$ limit, hence the IR divergence cancels.

\subsubsection{Ultraviolet structure}\label{sec:UV}
The sensitivity of the $1$-loop matter power spectra to UV modes is dominated by the $P^{13}$ contributions, which involve the third-order kernels. For the $\chi$ species we find the following behavior at $\mathcal{O}(f_\chi^0)$,
\begin{subequations}
\begin{align}
   F_\chi^{(0)[3]}(\vec{k},\vec{q},-\vec{q}; \tilde{\eta}) & \stackrel{k_s(a) \, \ll\, q}{=} \frac{u^{(0)} (\tilde{\eta})F_3(\vk,\vq,-\vq)}{g^{(0)}(\tilde{\eta})\, g^{(0)}(\tilde{\eta}_q)^2} \,,\\
   G_\chi^{(0)[3]}(\vec{k},\vec{q},-\vec{q};\tilde{\eta}) & \stackrel{k_s(a) \, \ll\, q}{=} \frac{(\partial_{\tilde{\eta}}u^{(0)}+3 u^{(0)})(\tilde{\eta}) F_3(\vk,\vq,-\vq)}{g^{(0)}(\tilde{\eta})\,g^{(0)}(\tilde{\eta}_q)^2}\,,
\end{align}
\end{subequations}
where 
\begin{equation}
u^{(0)}(\tilde{\eta}) = U^{(0)}(\sqrt{6} e^{- \tilde{\eta}/2}),\quad \mathrm{with}\quad U^{(0)}(x) = \frac{1}{7} -\frac{x^2}{140} +\frac{x^4}{840}G^{(0)}(x)\,.
\end{equation}
Accordingly, the 1-loop density power spectra involving the $\chi$ species scale as
\begin{align} \label{P13_c_chi}
P^{(0)13}_{\delta_c \delta_\chi}(k)& \;\xrightarrow{\mathrm{UV}}\; - \,\frac{61}{630\pi^2} \frac{1}{2} \bigg(1 +  \frac{u^{(0)}}{g^{(0)}}(\tilde{\eta})\bigg) k^2 P_{\delta_c \delta_\chi}^{(0)L}(k) \int^{\Lambda_{\rm UV}}  \mathrm{d}q P_{\delta_c \delta_c}^{(0)L}(q)\,, \\ \label{P13_chi_chi}
P^{(0)13}_{\delta_\chi \delta_\chi}(k)& \;\xrightarrow{\mathrm{UV}}\; - \,\frac{61}{630\pi^2} \frac{u^{(0)}}{g^{(0)}}(\tilde{\eta})\, k^2 P_{\delta_\chi \delta_\chi}^{(0)L}(k) \int^{\Lambda_{\rm UV}}  \mathrm{d}q P_{\delta_c \delta_c}^{(0)L}(q)\,.
\end{align}
Thus, the standard UV dependence $\sim k^2 P^L (k)$ is further modulated, already at $\mathcal{O}(f_\chi^0)$, by the ratio $u^{(0)}/g^{(0)}$, which smoothly transitions from $1/7$ at $k \ll k_s(a)$ to $1$ at $k \gg k_s(a)$.

\begin{figure}
    \centering
    \includegraphics[]{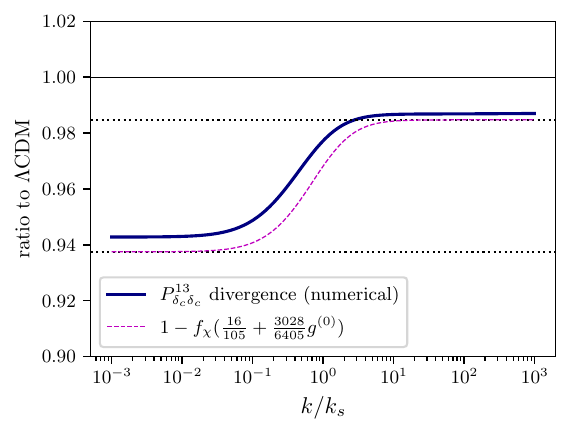}
    \caption{Coefficient of the UV divergence for the $P^{13}_{\delta_c \delta_c}(k)$ loop integral, as a function of $k/k_s(a)$, normalized to the standard $\Lambda$CDM value. The solid blue line shows the numerical result for $p = 0$, obtained by setting $f_\chi=0.1$. The two dotted lines show the analytical limits for $k\ll k_s(a)$ and $k\gg k_s(a)$, calculated using Eqs.~\eqref{eq:Fc13_analytical}. The dashed purple line corresponds to the analytical interpolation given within curly brackets in Eq.~\eqref{eq:Pcc-UV}. The residual discrepancies observed between numerical and analytical results are due to terms of $\mathcal{O}(f_\chi^2) \sim 1\%$, which were neglected in the analytical expansion for small $f_\chi$.}
    \label{fig:Pcc-UV-div}
\end{figure}

For the cold species, we can evaluate analytically the UV behavior at $\mathcal{O}(f_\chi)$ when the external wavenumber is well separated from $k_s(a)$, 
\begin{subequations}\label{eq:Fc13_analytical}
\begin{align}
F_c^{(1)[3]}(\vk,\vq,-\vq)&\;\xrightarrow{k\,\ll\, k_s(a)\, \ll\, q}\, \frac{1}{1890}(-116 \bar{\mu}^4+397 \bar{\mu}^2-71)\left(\frac{k}{q}\right)^2+\mathcal{O}\big(\left(k/q\right)^4\big), \\
F_c^{(1)[3]}(\vk,\vq,-\vq)&\;\xrightarrow{k_s(a)\, \ll\, k\, \ll\, q}\, \frac{4}{6615}\left(49 \bar{\mu}^4 - 71 \bar{\mu}^2 + 22 \right)\left(\frac{k}{q}\right)^2+\mathcal{O}\big(\left(k/q\right)^4\big),
\end{align}
\end{subequations}
where $\bar{\mu} \equiv \vec{k}\cdot \vec{q}/(kq)$. As a consequence, at $\mathcal{O}(f_\chi)$ the UV limit of the $1$-loop power spectrum of $\delta_c$ deviates from the standard $k^2 P^L_{\delta_c \delta_c}(k)$ dependence. Interpolating between the two limits in Eq.~\eqref{eq:Fc13_analytical} gives the approximate expression
\begin{align}
    P^{13}_{\delta_c \delta_c}(k) \xrightarrow{\mathrm{UV}} -\, \frac{61}{630 \pi^2}\Big\{1 - f_\chi\Big[\frac{16}{105} + \frac{3028}{6405}\,g^{(0)}&\,\Big( - (2 + p)\log \frac{k}{k_s(a)}\Big)\Big]\Big\} \nonumber \\ 
    &\,\times k^2  P_{\delta_c \delta_c}^{L}(k)\hspace{-1.25mm}\int^{\Lambda_{\rm UV}}  \hspace{-1mm}\mathrm{d} q P_{\delta_c \delta_c}^{L}(q)\,, \label{eq:Pcc-UV}
\end{align}
which indeed agrees with the numerical solution up to tiny $\mathcal{O}(f_\chi^2)$ corrections, as shown in Fig.~\ref{fig:Pcc-UV-div}. In principle, then, the standard $k^2 P_{\delta_c \delta_c}^L(k)$ counterterm should be modified to account for this new UV sensitivity at first order in $f_\chi$.

Finally, analogous UV structures are found for the power spectra involving the velocity divergences: for instance,
\begin{equation}\label{P13_deltac_Thchi}
P^{(0)13}_{\delta_c \Theta_\chi}(k) \xrightarrow{\mathrm{UV}} - \,\frac{61}{630\pi^2} \frac{1}{2} \bigg(1 +  \frac{\partial_{\tilde{\eta}}u^{(0)} + 3 u^{(0)}}{h^{(0)}}(\tilde{\eta})\bigg) k^2 P_{\delta_c \Theta_\chi}^{(0)L}(k) \int^{\Lambda_{\rm UV}}  \mathrm{d}q P_{\delta_c \delta_c}^{(0)L}(q)\,,
\end{equation}
where $(\partial_{\tilde{\eta}}u^{(0)} + 3 u^{(0)})/h^{(0)}$ smoothly transitions from $3/7$ at $k \ll k_s(a)$ to $2$ at $k \gg k_s(a)$.

\subsection{Bias expansion with two fluids}\label{sec:nonlinear_zspace}
The response of galaxy (and halo) formation to the presence of multiple fluid species has been studied extensively~\cite{Castorina:2013wga,LoVerde:2014pxa,Angulo:2015eqa,Schmidt:2016coo,Munoz:2018ajr,Chiang:2018laa,Chen:2019cfu,Bottaro:2023wkd}. The bias expansion for two fluids is naturally expressed in terms of the total matter ($m$) and relative $(r)$ perturbations, defined in Eq.~\eqref{eq:tot_rel}. However, in this work we limit a full description of bias to linear operators, for which the map between the $(m,r)$ and $(\chi,c)$ field bases is straightforward. Therefore we prefer to work with the latter basis, which we have used thus far to express our solutions. 

The galaxy number density perturbation $\delta_g \equiv n_g / \bar{n}_g - 1$ admits the bias expansion~\cite{Schmidt:2016coo,Chen:2019cfu,Bottaro:2023wkd}
\begin{equation}\label{eq:delta_g}
\delta_{g} = b_c \delta_c + b_\chi \delta_\chi + b_{\Theta} \left(\Theta_\chi - \Theta_c\right) + \ldots\,,
\end{equation}
where the dots stand for nonlinear operators. What is the expected size of $b_\chi$? A simple argument indicates that $b_\chi$ must vanish as \mbox{$f_\chi \to 0\,$.} If this were not the case, the galaxy perturbation would experience a finite jump from $\delta_g = (b_c + b_\chi)\delta_c$ to $\delta_g = b_c \delta_c$ around $k_s(a)$ despite the $\chi$ energy density being negligible, which is unphysical. Furthermore, studies of the scale dependence of bias in the presence of massive neutrinos~\cite{Castorina:2013wga,LoVerde:2014pxa,Munoz:2018ajr} indicate that $b_\nu$ is of order $f_\nu$ and has negative sign, $b_\nu = -\,  f_\nu \widehat{b}_\nu$ with $0 < \widehat{b}_\nu < b_c$. 

However, in general it is not guaranteed that an expansion of $b_\chi$ for $f_\chi \ll 1$ is meaningful when $f_\chi$ is as large as $\mathcal{O}(0.1)$, much bigger than the realistic values for the neutrino fraction $f_\nu$, as it will be considered in this work. For this reason, as our baseline choice we conservatively treat $b_\chi$ as an $\mathcal{O}(1)$ parameter and similarly for $b_\Theta$. Nonetheless, in Sec.~\ref{sec:constraints} we will also discuss the impact of tightening the prior on $b_\chi$ to $\mathcal{O}(f_\chi)$ size. Under our baseline assumptions, a formally consistent analytical calculation of galaxy correlators would require one to solve for the $\chi$ perturbations up to $\mathcal{O}(f_\chi)$, which is beyond the scope of this paper. In practice, however, we expect the key physical effects to be already captured by the leading non-trivial order, which for $\chi$ is $\mathcal{O}(f_\chi^0)$ and which we have retained throughout. 

\begin{figure}
    \centering
    \includegraphics[width=0.6\linewidth]{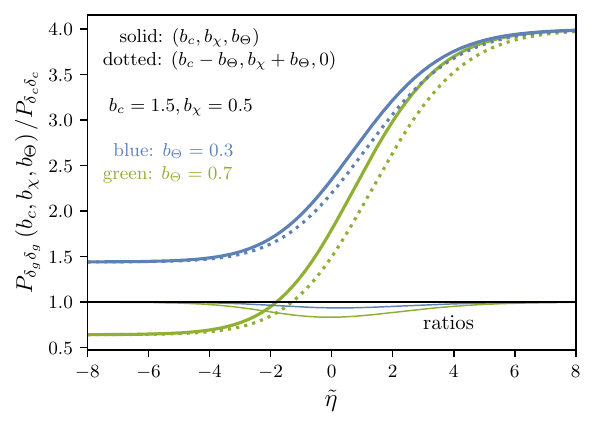}
    \caption{The ratio $P_{\delta_g \delta_g}^L/P_{\delta_c \delta_c}^L$ calculated at linear order from the bias expansion in Eq.~\eqref{eq:delta_g}, as a function of $\tilde{\eta}$. Solid lines correspond to a set of bias coefficients $(b_c, b_\chi, b_\Theta)$. Dotted lines correspond to a different set, where $b_\Theta = 0$ but $b_c$ and $b_\chi$ have been appropriately shifted, which reproduces well the shape of the galaxy power spectrum. The small residual difference depends on the size of $b_\Theta$.}
    \label{fig:bTheta_impact}
\end{figure}

Expanding the galaxy density perturbation as
\begin{equation}
\delta^{[n]}_g(\vk,\te) = \int_\vk \mathrm{d}k_{1\ldots n} \,F^{[n]}_g(\vk_1,\ldots,\vk_n;\te) \,\delta_c^{[1]}(\vk_1,\te)\ldots\delta_c^{[1]}(\vk_n,\te)\,, \label{eq:F_g^n}
\end{equation}
we find at linear level (recall that $s(\tilde{\eta}) = f(k)/f_{\Lambda \mathrm{CDM}}$)
\begin{equation}
F_g^{[1]}(\vec{k}_1;\tilde{\eta}_{k_1}) = b_c + b_\chi g(\tilde{\eta}_{k_1}) + b_\Theta \Big[ h(\tilde{\eta}_{k_1}) - \frac{f(k_1)}{f_{\Lambda \rm CDM}}\Big]\,. \label{eq:Fg1}
\end{equation}
The $\mathcal{O}(f_\chi^0)$ results $s = 1$ and $h \sim g$, the latter meaning that the two functions have similar qualitative behaviors (see the right panel of Fig.~\ref{fig:g0}), suggest that the effect of $b_\Theta$ can be largely absorbed into redefinitions of $b_c$ and $b_\chi$. This is confirmed by Fig.~\ref{fig:bTheta_impact}, which shows that the impact of a non-zero $b_\Theta$ on the linear galaxy power spectrum can be approximately captured by shifting $b_c \to b_c - b_\Theta$ and $b_\chi \to b_\chi + b_\Theta$.

Nonlinear bias operators can be systematically included in Eq.~\eqref{eq:delta_g} by applying the general results of Ref.~\cite{Bottaro:2023wkd} (which built upon Refs.~\cite{Schmidt:2016coo,Chen:2019cfu}), and are constructed using the perturbations of both species. However, given that the impact of the nonlinear bias coefficients on the extraction of cosmological parameters is, in general, numerically smaller than for linear bias, in this paper we adopt a simplified approach where only nonlinear bias operators containing the cold species alone are included. The prescription for kernels presented in Sec.~\ref{sec:prescription} below also implies that nonlinear bias operators involving $\delta_\chi$ and $v_\chi^i$ can be viewed as scale-dependent corrections to the already largely uncertain nonlinear bias parameters with respect to the CDM plus baryon fluid.

\subsection{Redshift space distortions}
Our next step is to perform the mapping between real and redshift space, see e.g. Ref.~\cite{Bernardeau:2001qr}. To derive the redshift-space expression of the galaxy number density perturbation, which we denote $\delta_{g,s}$, we also require the bias expansion of the galaxy velocity $v_g^i\,$,
\begin{equation}\label{eq:v_g}
    v_{g}^i = (1-b_v) v_c^i + b_v v_\chi^i + \ldots \,,
\end{equation}
where the dots indicate again nonlinear operators. The presence of a single linear velocity bias coefficient $b_v$ is fixed by invariance under generalized Galilean transformations, which boost all velocities by the same (spatially constant) amount. 

Expanding the divergence of the galaxy velocity in analogy with Eq.~\eqref{eq:F_g^n}, we write 
\begin{equation}
\Theta_g^{[n]}(\vec{k},\tilde{\eta}) = \int_\vk \mathrm{d}k_{1\ldots n} \,G^{[n]}_g(\vk_1,\ldots,\vk_n;\te) \,\delta_c^{[1]}(\vk_1,\te)\ldots\delta_c^{[1]}(\vk_n,\te)\,. \label{eq:G_g^n}
\end{equation}
At linear level we obtain
\begin{equation}
G_g^{[1]}(\vec{k}_1;\tilde{\eta}_{k_1}) = (1 - b_v) \frac{f(k_1)}{f_{\Lambda \rm CDM}} + b_v h (\tilde{\eta}_{k_1})\,, \label{eq:Gg1}
\end{equation}
and by comparison with Eq.~\eqref{eq:Fg1} we expect the effect of $b_v$ to be largely degenerate with the one of $b_\Theta$. We do not consider any nonlinear velocity bias operators in this work, since their effects are expected to be very suppressed. However, if desired they could be included in Eq.~\eqref{eq:v_g} by exploiting previous results~\cite{Schmidt:2016coo,Bottaro:2023wkd}.

Defining the expansion of $\delta_{g,s}$ as in Eq.~\eqref{eq:F_g^n}, but with redshift space galaxy kernels labeled by $F_{g,s}^{[n]}$, we obtain up to third perturbative order
\begin{subequations}\label{eq:RSD-kernels}
\begin{align}
F_{g,s}^{[1]}(\vec{k}_1 ; \tilde{\eta}_{k_1}) =&\; F_g^{[1]}(\vec{k}_1; \tilde{\eta}_{k_1}) + f_{\Lambda \rm CDM}\mu_1^2 G_g^{[1]}(\vec{k}_1 ; \tilde{\eta}_{k_1})\,,
\end{align}
\begin{align}
F_{g,s}^{[2]}(\vec{k}_1, \vec{k}_2 ; \tilde{\eta}) =&\; F_{g}^{[2]}(\vec{k}_1, \vec{k}_2 ; \tilde{\eta}) + f_{\Lambda \rm CDM}\mu^2 G_{g}^{[2]}(\vec{k}_1, \vec{k}_2 ; \tilde{\eta}) \nonumber \\
+&\; f_{\Lambda \rm CDM} k \mu \Big[\frac{\mu_{1}}{2 k_1} F_g^{[1]}(\vec{k}_2;\tilde{\eta}_{k_2}) G_g^{[1]}(\vec{k}_1;\tilde{\eta}_{k_1}) + \vec{k}_1 \leftrightarrow \vec{k}_2 \Big] \nonumber \\
+&\; f_{\Lambda \rm CDM}^2 k^2 \mu^2 \frac{\mu_{1}\mu_{2}}{2k_1 k_2} G_g^{[1]}(\vec{k}_1;\tilde{\eta}_{k_1}) G_g^{[1]}(\vec{k}_2;\tilde{\eta}_{k_2})\,, \\
 F_{g,s}^{[3]}(\vec{k}_1, \vec{k}_2, \vec{k}_3; \tilde{\eta}) =&\;F_{g}^{[3]}(\vec{k}_1, \vec{k}_2 , \vec{k}_3; \tilde{\eta}) + f_{\Lambda \rm CDM}\mu^2 G_{g}^{[3]}(\vec{k}_1, \vec{k}_2 , \vec{k}_3; \tilde{\eta})\nonumber\\
+\; \frac{1}{3}\Big\{ f_{\Lambda \rm CDM}k\mu  \Big[&\frac{\mu_{1}}{k_1}G_{g}^{[1]}(\vec{k}_1; \tilde{\eta}_{k_1})F_{g}^{[2]}(\vec{k}_2, \vec{k}_3; \tilde{\eta}_{k_{23}}) +\frac{\mu_{23}}{k_{23}}F_{g}^{[1]}(\vec{k}_1; \tilde{\eta}_{k_1})G_{g}^{[2]}(\vec{k}_2, \vec{k}_3; \tilde{\eta}_{k_{23}})\Big]\nonumber\\
+\, f_{\Lambda \rm CDM}^2 k^2 \mu^2  \Big[&\frac{\mu_{1}}{k_{1}}\frac{\mu_{23}}{k_{23}} G_{g}^{[1]}(\vec{k}_1;\tilde{\eta}_{k_1})G_{g}^{[2]}(\vec{k}_2, \vec{k}_3; \tilde{\eta}_{k_{23}}) \nonumber \\
+\frac{\mu_2 \mu_3}{2 k_2 k_3}&\, F_{g}^{[1]}(\vec{k}_1; \tilde{\eta}_{k_1}) G_{g}^{[1]}(\vec{k}_2;\tilde{\eta}_{k_2})G_{g}^{[1]}(\vec{k}_3;\tilde{\eta}_{k_3})\Big]\Big\} + \vec{k}_1 \leftrightarrow \vec{k}_2 + \vec{k}_1 \leftrightarrow \vec{k}_3\nonumber\\
 + f_{\Lambda \rm CDM}^3 & k^3 \mu^3\frac{\mu_1 \mu_2 \mu_3}{6 k_1 k_2 k_3}G_{g}^{[1]}(\vec{k}_1; \tilde{\eta}_{k_1})G_{g}^{[1]}(\vec{k}_2 ; \tilde{\eta}_{k_2})G_{g}^{[1]}(\vec{k}_3; \tilde{\eta}_{k_3})\,, \label{eq:Fgs3}
\end{align}
\end{subequations}
where $\mu \equiv \vec{k} \cdot \widehat{\vec{z}}/k$ with $\vec{k} = \vec{k}_1 + \ldots + \vec{k}_n$, $\mu_{i_1 \ldots i_m} \equiv \vec{k}_{i_1\ldots i_m} \cdot \widehat{\vec{z}}/k_{i_1 \ldots i_m}$ with $\vec{k}_{i_1\ldots i_m} = \vec{k}_{i_1} + \ldots + \vec{k}_{i_m}$ and $\widehat{\vec{z}}$ is the unit vector along the line of sight, which we take as our $z$ axis. In the $\Lambda$CDM limit, Eqs.~\eqref{eq:RSD-kernels} reproduce standard results~\cite{DAmico:2019fhj,Ivanov:2019pdj}. The $1$-loop contribution to the redshift-space galaxy power spectrum is therefore
\begin{align}
P_{\delta_g \delta_g, s}^{1\text{-}\mathrm{loop}}(k,\mu) = 2\hspace{-1mm} \int\hspace{-1mm} \frac{d^3 q}{(2\pi)^3}&\, \big[ F_{g,s}^{[2]} (\vec{q}, \vec{k} - \vec{q}; \tilde{\eta}) \big]^2 P_{\delta_c \delta_c}^L (q) P_{\delta_c \delta_c}^L (| \vec{k} - \vec{q} \,|) \nonumber \\ &\,+ 6\, F_{g,s}^{[1]}(\vec{k}; \tilde{\eta}) P_{\delta_c \delta_c}^L (k) \int\hspace{-1mm} \frac{d^3 q}{(2\pi)^3} F_{g,s}^{[3]} (\vec{k}, \vec{q}, - \vec{q}; \tilde{\eta}) P_{\delta_c \delta_c}^L (q)\,. \label{eq:Pggs_1loop}
\end{align}
The redshift-space galaxy power spectrum is then expanded in its multipoles
\begin{equation}
P_{\ell}(k) \equiv \frac{2\ell + 1}{2} \int_{-1}^{+1}\mathrm{d} \mu\, P_{\delta_g \delta_g, s}(k,\mu) \mathcal{L}_\ell (\mu)\,,
\end{equation}
where $\mathcal{L}_\ell$ are the Legendre polynomials. As customary, in this work we consider the $\ell = 0, 2, 4$ multipoles.

\subsection{Prescription for nonlinear kernels}\label{sec:prescription}

The analytical understanding of the matter kernels we obtained in Sec.~\ref{sec:matter_nonl} by means of the small $f_\chi$ expansion, allows us to design a prescription that strongly simplifies the evaluation of $P_{\delta_g \delta_g,s}(k, \mu)$ up to $1$-loop order, while retaining sufficient accuracy. The prescription is obtained by imposing that all IR limits in Eq.~\eqref{eq:IR_limits_general} are reproduced exactly, leading us to define
\begin{subequations}\label{eq:prescription}
\begin{align} 
F_{\chi}^{[n]} (\vec{k}_1, \ldots, \vec{k}_n; \tilde{\eta})_{\mathrm p} \,=&\;   \frac{g(\tilde{\eta})\,F_n(\vec{k}_1, \ldots, \vec{k}_n) }{g(\tilde{\eta}_{k_1})\ldots g(\tilde{\eta}_{k_n})}\,, \label{eq:Fchin_p} \\
G_{\chi}^{[n]}(\vec{k}_1, \ldots, \vec{k}_n;  \tilde{\eta})_{\mathrm p} \,=&\; \frac{h(\tilde{\eta})\, G_n(\vec{k}_1, \ldots, \vec{k}_n)}{g(\tilde{\eta}_{k_1})\ldots g(\tilde{\eta}_{k_n})} \,, \label{eq:Gchin_p} \\
F_{c}^{[n]} (\vec{k}_1, \ldots, \vec{k}_n; \tilde{\eta})_{\mathrm p} \,=&\; F_n(\vec{k}_1, \ldots, \vec{k}_n)\,, \label{eq:Fcn_p} \\ 
G_{c}^{[n]} (\vec{k}_1, \ldots, \vec{k}_n; \tilde{\eta})_{\mathrm p} \,=&\; s(\tilde{\eta}) G_n(\vec{k}_1, \ldots, \vec{k}_n)\,.  \label{eq:Gcn_p}
\end{align}
\end{subequations}
We will denote any quantities evaluated according to this prescription with a ``$\mathrm{p}$'' subscript.

We see that the standard $\Lambda$CDM kernels $F_n$ and $G_n$ are the only nonlinear quantities appearing on the right-hand sides of Eqs.~\eqref{eq:prescription}. They multiply the ratios of linear transfer functions $g$, $h$ and $s$ evaluated on the {\it external} momentum $\vec{k} = \vec{k}_1 + \ldots + \vec{k}_n$, whereas the products of $g$ functions in the denominators of Eqs.~\eqref{eq:Fchin_p} and~\eqref{eq:Gchin_p} simply rephrase the perturbative expansions of Eqs.~\eqref{eq:kernel_expansion} in terms of only one linear field, namely $\delta_c^{[1]}$. As a consequence of these properties, the application of the prescription casts the 1-loop corrections to the matter power spectra into a simple factorized form. For instance, Eq.~\eqref{eq:P_c_Thchi_1l} becomes
\begin{align}
&\,P_{\delta_c \Theta_\chi}^{1\text{-}\mathrm{loop}}(k)_{\rm p} = h(\tilde{\eta})\,\Big\{ 4\hspace{-1mm} \int\hspace{-1mm} \frac{d^3 q}{(2\pi)^3}  F_{2} (\vec{q}, \vec{k} - \vec{q}) G_{2} (\vec{q}, \vec{k} - \vec{q}) P_{\delta_c \delta_c}^L (q) P_{\delta_c \delta_c}^L (| \vec{k} - \vec{q} \,|) \nonumber \\
&\,\times\Theta_{\rm H} (| \vec{k} - \vec{q}| - q ) + 3 P_{\delta_c \delta_c}^L (k) \hspace{-1.5mm} \int \hspace{-1.5mm}\frac{d^3 q}{(2\pi)^3} \big[ F_{3} (\vec{k}, \vec{q}, - \vec{q} ) + G_{3} (\vec{k}, \vec{q}, - \vec{q} ) \big] P_{\delta_c \delta_c}^L (q) \Big\}\,, \label{eq:P_c_Thchi_1l_prescr}
\end{align}
where the quantity in curly brackets is simply the standard form of the 1-loop correction calculated with $\Lambda$CDM kernels, taking as input the linear power spectrum for the cold species, $P^L_{\delta_c\delta_c}(k)$, as obtained from~\texttt{CLASS}. Standard methods developed for $\Lambda$CDM then enable a fast evaluation of Eq.~\eqref{eq:P_c_Thchi_1l_prescr} suitable for MCMC analysis.

One immediate consequence of the prescription is that, at all perturbative orders, the $\chi$ fields are obtained from the $c$ fields through multiplication by ratios of linear transfer functions, namely
\begin{equation}\label{eq:prescr_implications}
\delta_\chi (\vec{k}, \tilde{\eta})_{\rm p} = g(\tilde{\eta}) \delta_c (\vec{k}, \tilde{\eta})_{\rm p}\,,\qquad \Theta_\chi (\vec{k}, \tilde{\eta})_{\rm p} = \frac{h}{s}(\tilde{\eta}) \Theta_c (\vec{k}, \tilde{\eta})_{\rm p}\,.
\end{equation}
The first relation in Eq.~\eqref{eq:prescr_implications} was indeed assumed in some previous studies of massive neutrino cosmologies~\cite{Lesgourgues:2009am,Levi:2016tlf,Aviles:2021que}. 

In the rest of this section we thoroughly analyze the accuracy and implications of our prescription in Eqs.~\eqref{eq:prescription}. Before doing so, however, we stress again that this prescription is motivated by the need to efficiently compute the galaxy power spectrum at $1$-loop order. Indeed, since analytical expressions for the kernels covering the entire kinematic regime where the loops have support are not available, we are naturally prompted to adopt the above ansatz in order to manageably perform the required hundreds of thousands of evaluations. Nevertheless, the analytical expressions we presented in Sec.~\ref{sec:matter_nonl} (which were based on the small $f_\chi$ expansion, but did not rely on any ans\"atze) may be used to compute other observables exactly, for instance the tree-level galaxy bispectrum in some kinematic configurations, as argued in Sec.~\ref{sec:analytical}. 

\subsubsection{Accuracy for kernels}\label{sec:p_accuracy_kernels}

While the prescription reproduces exactly, by construction, the behavior of the complete solutions in the IR limit, it does introduce inaccuracies for general configurations of the momenta. Nevertheless, it preserves the correct structure of the kernels even far away from the IR, modulo momentum-$\,$dependent corrections that are a priori of $\mathcal{O}(f_\chi^0)$ for the $\chi$ fields and $\mathcal{O}(f_\chi)$ for the $c$ fields. 

\begin{figure}[t]
\centering
\includegraphics[width=\linewidth]{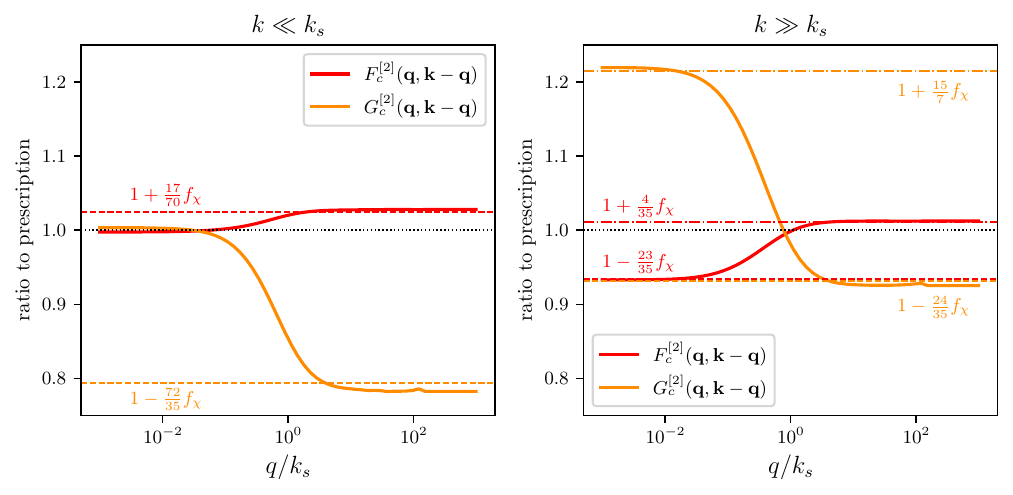}
\vspace{-0.5cm}\caption{Comparison of numerical solutions and analytical limits for the second order kernels of the cold species, $F_c^{[2]}(\vec{q}, \vec{k}-\vec{q}; \tilde{\eta})$ and $G_c^{[2]}(\vec{q}, \vec{k}-\vec{q}; \tilde{\eta})$, normalized to the respective values $F^{[2]}_{c,\,{\rm p}}$ and $G^{[2]}_{c,\,{\rm p}}$ defined by the prescription in Eqs.~\eqref{eq:prescription}. We set $f_\chi = 0.1$ and choose $\bar{\mu} = \vec{k}\cdot \vec{q}/(kq) = 0$ to remove the $1/q$ IR pole, which is exactly reproduced by the prescription. Solid curves correspond to the numerical solution for $p = 0$, while dashed lines show analytical limits. {\it Left:} $k \ll k_s(a)$. The $q\ll k_s(a)$ limits are trivial, recovering $\Lambda$CDM. The $q \gg k_s(a)$ analytical limits are derived from Eqs.~\eqref{eq:c_2_limits_k_ks_q} in Appendix~\ref{app:app_2}. {\it Right:} $k \gg k_s(a)$. The $q\ll k_s(a)$ [$q\gg k_s(a)$] analytical limits are obtained from Eqs.~\eqref{eq:limit_c} [Eqs.~\eqref{eq:c_2nd_hot}].}\label{fig:ana_vs_num}
\end{figure}

\begin{figure}[t]
\centering
\includegraphics[width=\linewidth]{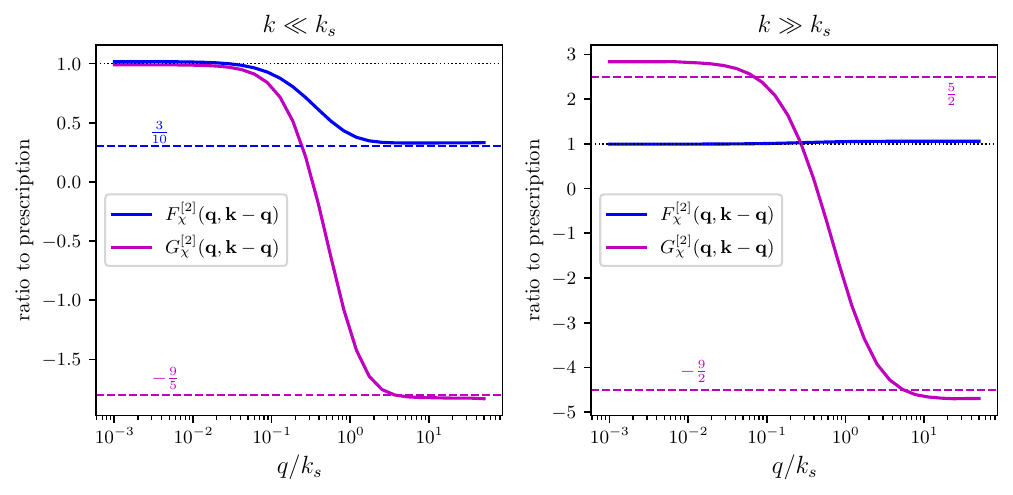}
\vspace{-0.5cm}\caption{Comparison of numerical solutions and analytical limits for the second order kernels of the warm species, $F_\chi^{[2]}(\vec{q}, \vec{k}-\vec{q}; \tilde{\eta})$ and $G_\chi^{[2]}(\vec{q}, \vec{k}-\vec{q}; \tilde{\eta})$, normalized to the respective values $F^{[2]}_{\chi,\,{\rm p}}$ and $G^{[2]}_{\chi,\,{\rm p}}$ defined by the prescription in Eqs.~\eqref{eq:prescription}. We set $f_\chi = 0.1$ and choose $\bar{\mu} = \vec{k}\cdot \vec{q}/(kq) = 0$ to remove the $1/q$ IR pole, which is exactly reproduced by the prescription. Solid curves correspond to the numerical solution for $p = 0$, while dashed lines show analytical limits.  {\it Left:} $k \ll k_s(a)$. The $q\ll k_s(a)$ limits are trivial, recovering $\Lambda$CDM. The $q \gg k_s(a)$ analytical limits are derived from Eqs.~\eqref{eq:chi_UV_2nd} in Appendix~\ref{app:app_2}. {\it Right:} $k \gg k_s(a)$. The $q \ll k_s(a)$ [$q \gg k_s(a)$] analytical limits are obtained from Eqs.~\eqref{eq:limit_chi} [Eqs.~\eqref{eq:chi_UV_2nd} in Appendix~\ref{app:app_2}].}\label{fig:ana_vs_num_Fchi}
\end{figure}

In Fig.~\ref{fig:ana_vs_num} we compare numerical and analytical solutions for the second order kernels of the cold fluid in the $(\vec{q}, \vec{k} - \vec{q})$ configuration of momenta relevant for the $P^{22}$ contribution to the $1$-loop power spectrum. Both numerical and analytical solutions are normalized to the value prescribed by Eqs.~\eqref{eq:Fcn_p} or~\eqref{eq:Gcn_p}. We set $f_\chi = 0.1$ and evaluate the numerical results for $p=0$ (as well as $\gamma = 2$, as we always assumed from Sec.~\ref{sec:analytical} onwards). We choose $\vec{k}$ and $\vec{q}$ to be orthogonal in order to remove the IR pole at small $q$ (by construction, the IR pole is exactly reproduced by our prescription). The figure shows two conceptually distinct aspects.~(i)~First, we observe that the numerical solutions (solid curves) are in good agreement with the analytical approximations (dashed lines) when all momenta are separated from $k_s(a)$. The small residual differences are due to corrections of $\mathcal{O}(f_\chi^2)$ that we have not calculated. This validates the analytical results presented in Sec.~\ref{sec:analytical}. (ii) Second, the deviations of the solid curves from unity show that our prescription does indeed introduce errors of $\mathcal{O}(f_\chi)$ with respect to the exact numerical result, as expected. This is the price we accept to pay, in exchange for the fast evaluation of the $1$-loop corrections illustrated by Eq.~\eqref{eq:P_c_Thchi_1l_prescr}.

In Fig.~\ref{fig:ana_vs_num_Fchi} we perform an analogous comparison, but for the second order kernels of the warm species $\chi$. Here we observe that: (i) numerical solutions (solid curves) and analytical limits (dashed lines) agree, modulo $\mathcal{O}(f_\chi)$ corrections which we have not calculated, thus confirming the results of Sec.~\ref{sec:analytical}; and (ii) the prescription introduces $\mathcal{O}(f_\chi^0)$ errors with respect to the exact numerical result in some regions of momenta, as the deviations of the solid curves from unity show. The deviations are especially large for $G_\chi^{[2]}$ (solid purple), whereas the prescription provides a better approximation to $F_\chi^{[2]}$ (solid blue).

\subsubsection{Ultraviolet behavior and counterterms}\label{sec:CT_galaxy}
Next, we focus on the behavior of the prescription in the UV limit. For the $1$-loop power spectra of matter density perturbations we find
\begin{align} 
\big\{P^{13}_{\delta_c \delta_\chi}(k) ,\, &\,P^{13}_{\delta_\chi \delta_\chi}(k),\, P^{13}_{\delta_c \delta_c}(k) \big\}_{\mathrm{p}} \nonumber\\ 
&\xrightarrow{\mathrm{UV}} - \,\frac{61k^2}{630\pi^2}  \big\{  P_{\delta_c \delta_\chi}^{L}(k),\, P_{\delta_\chi \delta_\chi}^{L}(k),\, P_{\delta_c \delta_c}^{L}(k) \big\} \int^{\Lambda_{\rm UV}}  \mathrm{d}q P_{\delta_c \delta_c}^{L}(q)\,. \label{eq:UV_prescription}
\end{align}
By comparing these to the exact perturbative results for the UV limits in Eqs.~\eqref{P13_c_chi}, \eqref{P13_chi_chi} and~\eqref{eq:Pcc-UV}, respectively, we see that the prescription does not capture the additional scale-dependent effects seen in those equations. Nevertheless, we expect these additional scale-dependent effects to be small for two separate reasons. For power spectra involving the $\chi$ fields, where the additional scale dependence formally arises at $\mathcal{O}(f_\chi^0)$, because the $\chi$ transfer functions are suppressed for $k\gtrsim k_s(a)$. For power spectra involving only the CDM+baryon perturbations, because the additional scale dependence appears only at $\mathcal{O}(f_\chi)$. 

Owing to the presence of both $\delta_c$ and $\delta_\chi$ in the two-fluid bias expansion of Eq.~\eqref{eq:delta_g} and in accordance with the UV behavior obtained with the prescription, Eq.~\eqref{eq:UV_prescription}, our prediction for the galaxy power spectrum multipoles will need to include the following counterterms
\begin{equation}\label{eq:all_counterterms}
    P^{\,\mathrm{ctr}}_{\ell}(k) = -\, 2 c_{cc, \ell}\,k^2 P_{\delta_c \delta_c}^L(k) - 2 c_{c \chi, \ell}\,k^2 P^L_{\delta_c \delta_\chi}(k) - 2c_{\chi\chi, \ell}\,k^2 P^L_{\delta_\chi \delta_\chi}(k)\,.
\end{equation}
When $k_s \gg k_\mathrm{NL}$ these reduce to the usual counterterms of $\Lambda$CDM analyses~\cite{Ivanov:2019pdj} expressed in terms of the total matter power spectrum, namely $-\,2 c_{mm,\ell}\,k^2 P_{\delta_m \delta_m}^L(k)$. Conversely, when $k_s \ll k_\mathrm{NL}$ the power spectra involving $\chi$ are strongly suppressed and one is left with $-\,2 c_{cc,\ell}\,k^2 P_{\delta_c \delta_c}^L(k)$. This is the relevant situation for the standard massive neutrinos~\cite{Aviles:2021que}. 

Qualitatively similar results are obtained for power spectra of the velocity fields, for instance
\begin{equation} \label{eq:UV_prescription_vel}
P^{13}_{\delta_c \Theta_\chi}(k)_{\mathrm{p}} \;\xrightarrow{\mathrm{UV}}\; - \,\frac{25k^2}{126\pi^2} P_{\delta_c \Theta_\chi}^{L}(k) \int^{\Lambda_{\rm UV}}  \mathrm{d}q P_{\delta_c \delta_c}^{L}(q)\,,
\end{equation}
which should be compared with the exact perturbative expression in Eq.~\eqref{P13_deltac_Thchi}. In principle, the presence of the velocity divergences in the bias expansion of Eq.~\eqref{eq:delta_g} would then require the introduction of new counterterms. In practice, since $f(k) = f_{\Lambda\mathrm{CDM}}$ and $h(\tilde{\eta}) \sim g(\tilde{\eta})$ at zeroth order in $f_\chi$, we expect the set of counterterms in Eq.~\eqref{eq:all_counterterms} to adequately capture all the UV sensitivity.

\subsubsection{Accuracy for power spectra}\label{sec:p_accuracy_P}
\begin{figure*}[t]
    \includegraphics[width=\textwidth]{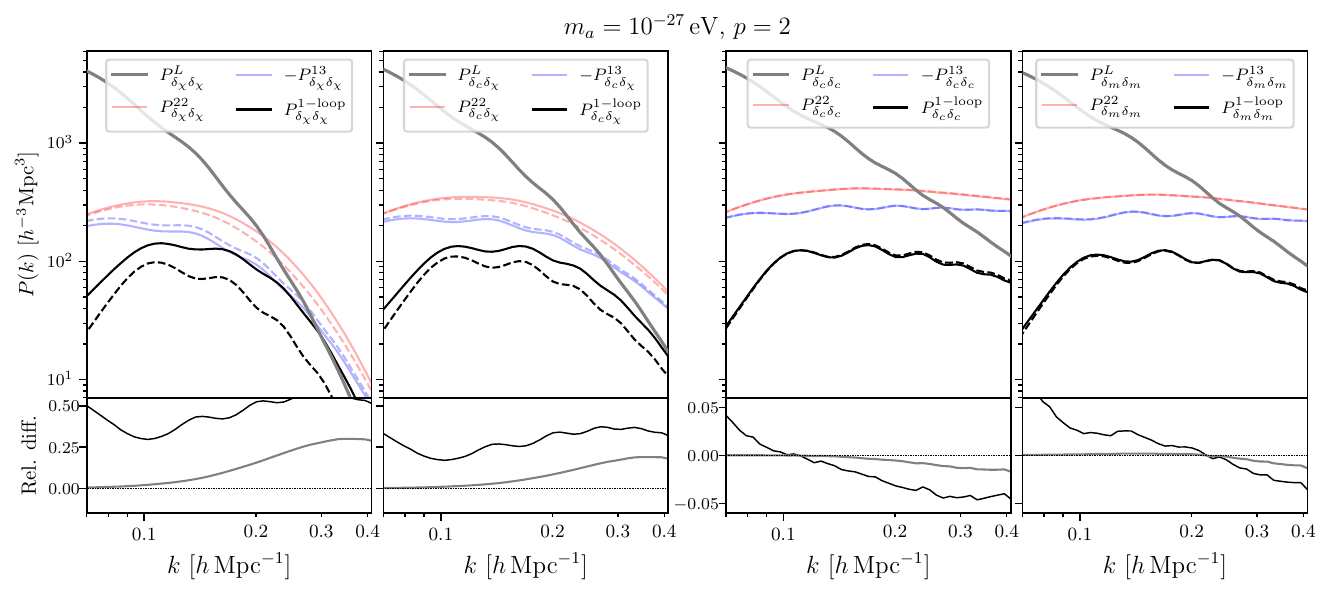}
    \includegraphics[width=\textwidth]{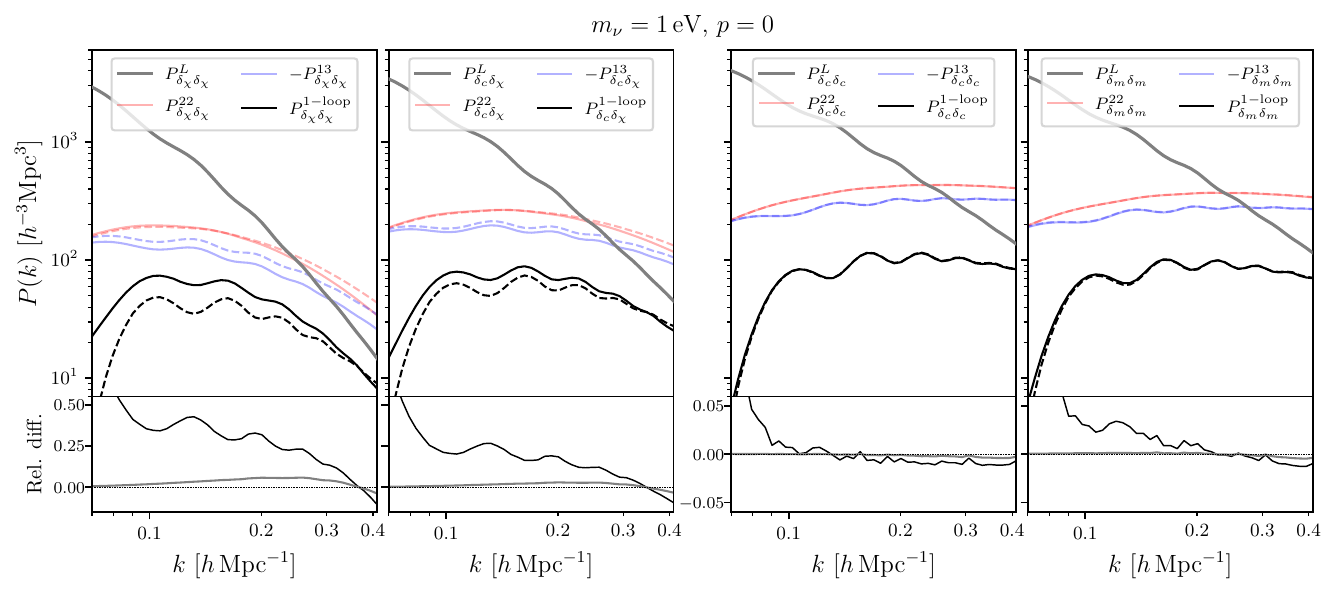}
    \caption{{\it Top row:} Comparison of the computations of the $1$-loop terms of the matter power spectra using our prescription for the nonlinear kernels, Eqs.~\eqref{eq:prescription} (dashed lines) to the full numerical results (solid lines). From left to right: $\chi \chi, c \chi, c c$ and $m m$ density power spectra at $z_{\rm obs} = 0.5$, representative of the redshift observed by galaxy surveys. Red, blue and black correspond to the $22$, (minus) $13$ and total loop contributions. The solid gray line indicates the linear power spectrum. The bottom inset shows $ (P^{1\text{-}\mathrm{loop}} - P^{1\text{-}\mathrm{loop}}_{\rm p} ) / P^{1\text{-}\mathrm{loop}}$ in black and $ (P^{1\text{-}\mathrm{loop}} - P^{1\text{-}\mathrm{loop}}_{\rm p} ) / ( P^L + P^{1\text{-}\mathrm{loop}} )$ in gray. We have used as inputs the linear power spectra for an ULA cosmology with $m_a=10^{-27}\;\mathrm{eV}$ and $f_\chi = 0.1$, corresponding to $p = 2$ and $k_s(z_{\rm obs}) \approx 0.29\, h\,\mathrm{Mpc}^{-1}$. {\it Bottom row:} Same as in the top row, but using as inputs the linear power spectra for a massive neutrino cosmology. We assume a single neutrino species with $m_\nu = 1$~eV, corresponding to $f_\chi \approx 0.08$, $p = 0$ and $k_s(z_{\rm obs}) \approx 0.30\, h\,\mathrm{Mpc}^{-1}$.}
    \label{fig:P_comp_ULA_nu}
\end{figure*}

In Fig.~\ref{fig:P_comp_ULA_nu} we compare the $1$-loop matter power spectra computed with our prescription to those obtained from exact numerical solutions of the fluid equations. We do so for the two benchmark cosmologies we have already considered in Sec.~\ref{sec:linear}, namely an ULA with $m_a = 10^{-27}\;\mathrm{eV}$ (top row) and a neutrino with $m_\nu = 1\;\mathrm{eV}$ (bottom row). The low-redshift characteristic scale is $k_s(z_{\rm obs}) \sim 0.3\,h\,\mathrm{Mpc}^{-1} \sim k_{\rm NL}$ for both models, corresponding to the physical regime where accurately modeling the nonlinear corrections is most challenging.

We discuss first the ULA cosmology. The quantitative comparison confirms the expectations based on the previous discussion: the relative deviations of the 1-loop terms $ (P^{1\text{-}\mathrm{loop}} - P^{1\text{-}\mathrm{loop}}_{\rm p} ) / P^{1\text{-}\mathrm{loop}}$ observed in the top row of Fig.~\ref{fig:P_comp_ULA_nu} are somewhat smaller than $1$ for the $\chi\chi$ and $c\chi$ power spectra and somewhat smaller than $f_\chi$ for the $cc$ and $mm$ power spectra. This happens because, although the prescription introduces inaccuracies of $\mathcal{O}(f_\chi^0)$ in the $\chi$ kernels and $\mathcal{O}(f_\chi)$ in the $c$ kernels, by construction these inaccuracies do not affect the IR limits of the kernels, which give the largest contributions to the loop integrals. 

\begin{figure}[t]
\centering
\includegraphics[width=\linewidth]{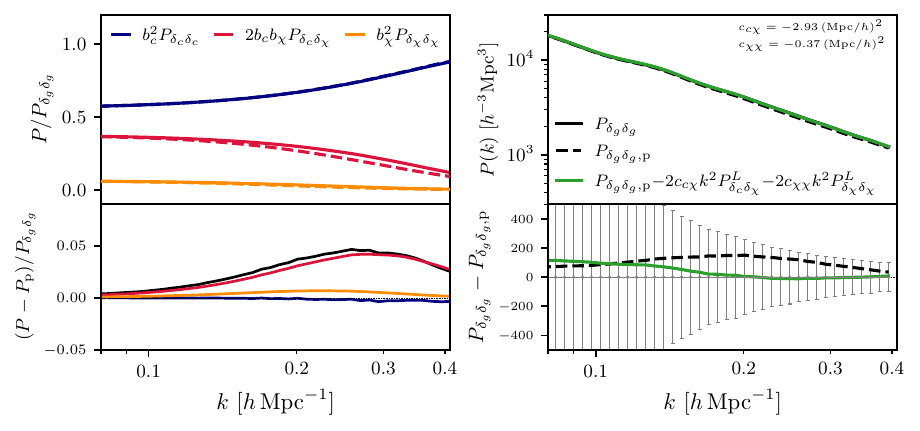}
\vspace{-0.2cm}\caption{{\it Left:} Fractional contributions of the $cc$, $c\chi$ and $\chi\chi$ components to the real space power spectrum of $\delta_g = b_c \delta_c +b_\chi \delta_\chi$ up to $1$-loop order. Solid lines show the full numerical results, e.g.~$b_c^2 P_{\delta_c \delta_c}/P_{\delta_g \delta_g}$, whereas dashed lines correspond to evaluating the $1$-loop terms in the numerator with the prescription, e.g.~$b_c^2 P_{\delta_c \delta_c,\mathrm{p}}/P_{\delta_g \delta_g}$. The bottom inset shows the difference between solid and dashed curves, e.g.~$b_c^2 (P_{\delta_c \delta_c}^{1\text{-}\mathrm{loop}} - P_{\delta_c \delta_c,\mathrm{p}}^{1\text{-}\mathrm{loop}})/P_{\delta_g \delta_g}$. In black we show the difference for the full galaxy power spectrum, $(P_{\delta_g \delta_g}^{1\text{-}\mathrm{loop}} - P_{\delta_g \delta_g,\mathrm{p}}^{1\text{-}\mathrm{loop}})/P_{\delta_g \delta_g}$, which is dominated by the $c\chi$ component and would remain constant even for $f_\chi\to 0$.  {\it Right:} Comparison between the full numerical result for $P_{\delta_g \delta_g}$, its value obtained with the prescription, and the additional effect of summing $c\chi$ and $\chi\chi$ counterterms with fitted coefficients. The bottom inset shows the differences between the predictions, compared to the BOSS $Q_0$ errorbars~\cite{Ivanov:2021fbu} for reference. In both panels we have set $b_c = 1.5$, $b_\chi = 0.5$ and used as inputs the linear power spectra for an ULA cosmology with $m_a = 10^{-27}\,\mathrm{eV}$ and $f_\chi = 0.02$.}\label{fig:P_gg_breakdown}
\end{figure}

As a result, our prescription ensures a relative accuracy {\it better} than $f_\chi P^{1\text{-}\mathrm{loop}}/P^L$ in the predictions of the $cc$ and $mm$ power spectra. For the benchmark ULA cosmology with $f_\chi = 0.1$ shown in the top row of Fig.~\ref{fig:P_comp_ULA_nu}, we obtain a $\lesssim 2\%$ accuracy up to $k = 0.4\, h\,\mathrm{Mpc}^{-1}$. This implies, in particular, that our prescription can be used to accurately evaluate lensing observables, which are only sensitive to $P_{\delta_m\delta_m}$. On the other hand, for the $\chi\chi$ and $c\chi$ power spectra the prescription gives a relative accuracy that is better than $P^{1\text{-}\mathrm{loop}}/P^L$, but is {\it not} proportional to $f_\chi$. In the top row of Fig.~\ref{fig:P_comp_ULA_nu} we find $\lesssim 20\%$ for $c\chi$ and $\lesssim 30\%$ for $\chi\chi$ (recall that $f_\chi = 0.1$ here). These results are confirmed by Fig.~\ref{fig:P_gg_breakdown}, where the same ULA mass as in the top row of Fig.~\ref{fig:P_comp_ULA_nu} is assumed but the fraction is smaller, $f_\chi = 0.02$. The real-space galaxy power spectrum is evaluated up to $1$-loop order, starting from the simplified two-fluid linear bias expansion $\delta_g = b_c \delta_c + b_\chi \delta_\chi$. In the left panel of Fig.~\ref{fig:P_gg_breakdown} we observe that the prescription reproduces the $cc$ component of $P_{\delta_g \delta_g}$ with negligible error. However, since based on the discussion in Sec.~\ref{sec:nonlinear_zspace} we conservatively assume $b_\chi \sim \mathcal{O}(1)$, the prescription introduces relative errors in the $c\chi$ and $\chi\chi$ components that are not proportional to $f_\chi$, and would remain constant even for vanishingly small $f_\chi$.

Nonetheless, this limitation of the prescription does not hinder our ability to extract reliable cosmological constraints. The reason is that the spurious terms introduced by the prescription in the $c\chi$ and $\chi\chi$ components of the galaxy power spectrum are largely reabsorbed into the free coefficients of the $k^2 P_{\delta_c \delta_\chi}^L(k)$ and $k^2 P_{\delta_\chi \delta_\chi}^L(k)$ counterterms in Eq.~\eqref{eq:all_counterterms}. This is demonstrated by the right panel of Fig.~\ref{fig:P_gg_breakdown}: after summing $c\chi$ and $\chi\chi$ counterterms with fitted coefficients to $P_{\delta_g \delta_g, \mathrm{p}}(k)$, the residual difference with the full numerical calculation of $P_{\delta_g\delta_g}(k)$ is entirely negligible relative to the BOSS error bars. The $c\chi$ counterterm gives the most important effect. We view this as strong evidence that our prescription does not introduce unwanted spurious effects in the derivation of constraints on $f_\chi$ from the galaxy power spectrum, when $b_\chi$ is treated as an $\mathcal{O}(1)$ coefficient in the two-fluid bias expansion.\footnote{If $b_\chi \sim \mathcal{O}(f_\chi)$ is assumed, the prescription automatically yields an accuracy on $P_{\delta_g\delta_g}$ comparable to the one for the total matter power spectrum.} The prescription remains valid even when intrinsic nonlinearities associated to the wave-like nature of ULAs are included, see Appendix~\ref{app:app_4_ULAs} for a discussion.

In the bottom row of Fig.~\ref{fig:P_comp_ULA_nu} we check the accuracy of the prescription for our benchmark massive neutrino cosmology. Relative to the ULA scenario, the massive neutrino yields a smaller $P^{1\text{-}\mathrm{loop}}/P^L$ ratio for the $c\chi$ and $\chi\chi$ power spectra, hence our prescription performs better. Nevertheless, the genuine $1$-loop inaccuracy $ (P^{1\text{-}\mathrm{loop}} - P^{1\text{-}\mathrm{loop}}_{\rm p} ) / P^{1\text{-}\mathrm{loop}}$ has comparable size to the ULA scenario and all the lessons we learned there still apply. Furthermore, the prescription is robust against the inclusion of intrinsic nonlinearities originating from the Vlasov equation, as shown in Appendix~\ref{app:app_4_Vlasov}.

\subsubsection{Redshift space treatment}\label{sec:prescription_zspace}
The $1$-loop contribution to the redshift-space galaxy power spectrum, which was given in Eq.~\eqref{eq:Pggs_1loop}, simplifies considerably after the prescription~\eqref{eq:prescription} is applied to the redshift-space galaxy kernels~\eqref{eq:RSD-kernels}. The resulting form can be worked out explicitly by means of the formulas for the real-space galaxy kernels, which we report in Eqs.~\eqref{eqs:Fgn_Ggn} of Appendix~\ref{app:app_2} (retaining only linear bias operators for brevity). Even after the application of the prescription, the evaluation of momentum integrals over the scale-dependent linear functions that appear in $F_g^{[1]}$ and $G_g^{[1]}$ is still required. For instance, the third line of Eq.~\eqref{eq:Fgs3} yields the following contribution to the 13 piece,
\begin{align}
P_{\delta_g \delta_g, s}^{13}(k,\mu)_{\rm p} &\,\supset 4\, F_{g,s}^{[1]}(\vec{k};\tilde{\eta}) P_{\delta_c \delta_c}^L (k)  f_{\Lambda\mathrm{CDM}}^2  k^2 \mu^2 \nonumber \\
&\,\times \int \frac{d^3 q}{(2\pi)^3} \frac{\mu_{\vec{q}}}{q}\frac{\mu_{\vec{k}-\vec{q}}}{|\vec{k}-\vec{q}|}  G_g^{[1]}(\vec{q};\tilde{\eta}_q) G_g^{[1]}(\vec{k}-\vec{q};\tilde{\eta}_{|\vec{k}-\vec{q}|}) G_2(\vec{k}, - \vec{q}) P_{\delta_c \delta_c}^L (q)\,,\label{eq:zspace_integral_ex1}
\end{align}
where the expression of $G_g^{[1]}$ was given in Eq.~\eqref{eq:Gg1}. Analogous terms appear in the 22 piece. Additionally, there are terms built out of linear fields only, which are insensitive to the application of the prescription. In the 13 piece, these arise from the fourth and fifth lines of Eq.~\eqref{eq:Fgs3} and yield
\begin{align}
P_{\delta_g \delta_g, s}^{13}(k,\mu) \supset  -\,F_{g,s}^{[1]}(\vec{k}; \tilde{\eta}) P_{\delta_c \delta_c}^L (k) f_{\Lambda\mathrm{CDM}}^2 k^2 \mu^2 &\Big[    F_g^{[1]}(\vec{k};\tilde{\eta}) + f_{\Lambda\mathrm{CDM}} \mu^2 G_g^{[1]}(\vec{k};\tilde{\eta}) \Big]  \nonumber\\
\times&\; \int\hspace{-1mm} \frac{d^3 q}{(2\pi)^3} \frac{\mu_{\vec{q}}^2}{q^2} \big[  G_g^{[1]}(\vec{q};\tilde{\eta}_q) \big]^2 P_{\delta_c \delta_c}^L (q)\,.\label{eq:zspace_integral_ex2}
\end{align}
In principle, the fast evaluation of momentum convolution integrals such as those in Eqs.~\eqref{eq:zspace_integral_ex1} and~\eqref{eq:zspace_integral_ex2} could be achieved through a new application of FFTLog methods~\cite{McEwen:2016fjn,Schmittfull:2016jsw,Simonovic:2017mhp}. A similar strategy was followed in Ref.~\cite{Aviles:2021que} for massive neutrino cosmologies. In the analysis of Sec.~\ref{sec:constraints}, however, we adopt a simplified approach where the scale dependence of the linear functions is fully retained outside the $q$ integrals, but the scale-independent expressions for standard CDM are used inside the integrals. Namely, in the integrands we set $g, h \to 1$ and $f\to f_{\Lambda \mathrm{CDM}}$. This implies in particular $G_g^{[1]}\to 1$, hence the integrals in Eqs.~\eqref{eq:zspace_integral_ex1} and~\eqref{eq:zspace_integral_ex2} reduce to standard expressions for which the FFTLog procedure is well known.\footnote{In fact, Eq.~\eqref{eq:zspace_integral_ex2} is fully degenerate with a linear combination of counterterms and it is therefore dropped from the theory prediction.}

\section{Planck plus BOSS Constraints on Ultra-Light Axions}\label{sec:constraints}
The in-depth discussion of the previous sections on the extension of the EFTofLSS in the presence of multiple fluids with finite sound speed provides the theoretical framework to investigate the parameter spaces of such models with cosmological data. The two benchmark scenarios considered so far in this work are ULAs and light thermal relics. In the latter case the relic abundance and the suppression of the matter and galaxy power spectra depend on the mass of the new particle, the number of its bosonic or fermionic degrees of freedom and its temperature. For illustrative purposes, in Sec.~\ref{sec:lin_sol} the latter two were identified with those of the standard neutrinos. A complete analysis, which includes varying the freeze-out temperature and matching to relevant particle physics models, goes beyond the scope of this work and will be presented in a forthcoming publication~\cite{modelspaper}. For this reason, in this first proof-of-principle analysis we only consider ULAs. Our goal is to constrain their energy density $\Omega_a h^2$ as a function of their mass $m_a$. 

\subsection{Data specification and modeling choices}
The data considered in this work are the CMB temperature, polarization and lensing power spectra as measured by Planck~\cite{Planck:2018vyg,Planck:2018lbu}, combined with the measurement of the full shape of the galaxy power spectrum performed by spectroscopic surveys. To enable a direct comparison with the previous cosmological constraints on ULAs presented in Ref.~\cite{Rogers:2023ezo}, we restrict ourselves to SDSS/BOSS data~\cite{BOSS:2015zan,Philcox:2021kcw} and do not use the recent BAO measurements from DESI~\cite{DESI:2025zgx}.\footnote{At the time of writing, the covariance matrices necessary for a likelihood analysis of DESI full-shape data are not yet publicly available.} The BOSS sample consists of approximately one million galaxies divided in two redshift slices, centered at $z_{\rm eff} = 0.38$ and $z_{\rm eff} = 0.61$, and in two regions above and below the galactic plane, for a total of four sets of measured galaxy power spectra.

Similarly to previous full-shape analyses of the $\Lambda$CDM model \cite{DAmico:2019fhj,Ivanov:2019pdj,Philcox:2021kcw,Chen:2021wdi}, when inferring cosmological parameters we marginalize over four bias parameters, one EFT counterterm for each multipole of the power spectrum, and three free parameters to account for the amplitude and the scale dependence of the galaxy shot noise. These nuisance parameters completely characterize the dependence of the galaxy fluctuations on the CDM+baryon density and velocity fields (labeled by $c$ in this work).

In addition, and drawing from the discussion in Sec.~\ref{sec:nonlinear_real}, we include the following new ingredients. The galaxy bias model contains one extra free parameter quantifying the response of the galaxy number density to the ULA density field,
\begin{equation}
    \delta_{g}\, \supset\, b_c \delta_c + b_\chi \delta_\chi\label{eq:our-delta-g-r}\,,
\end{equation}
where the ULA is labeled by $\chi$ consistently with previous sections. Importantly, in Eq.~\eqref{eq:our-delta-g-r} the axion density field $\delta_\chi$ is computed up to $n$-th order in perturbation theory using the prescription in Eqs.~\eqref{eq:prescription}, i.e.
\begin{equation}
    \label{eq:prescription_ula}
    \delta_\chi^{[n]}(\vec{k})=\left(\frac{\delta_\chi^{[1]}(\vec{k})}{\delta_c^{[1]}(\vec{k})}\right)_{\hspace{-1mm}\texttt{CLASS}} \delta_c^{[n]}(\vec{k})
\end{equation}
in terms of the linear theory transfer functions.\footnote{The linear power spectra of the CDM+baryon component, $\delta_c$, and of the axion field, $\delta_{\chi}$, are obtained directly from the Einstein-Boltzmann solver \texttt{AxiCLASS} \cite{Poulin:2018dzj,Smith:2019ihp}.} As noted in Sec.~\ref{sec:analytical}, for a proper parameter exploration that involves $\delta_\chi$ it is imperative to include nonlinearities of this field, at least in an approximate way. Neglecting those, e.g.~keeping $\delta_\chi$ only at linear level, would lead to spurious effects in the regime of small deviations from pure standard CDM, which is where precision cosmological data are the most powerful. As discussed in Sec.~\ref{sec:nonlinear_real}, we expect Eq.~\eqref{eq:our-delta-g-r} evaluated according to Eq.~\eqref{eq:prescription_ula} to capture the main novelty of the two-fluid description.

The new dependence of the galaxy field on $\delta_\chi$ also requires the inclusion of two new classes of counterterms proportional to the linear power spectra $P^L_{\delta_c \delta_\chi}(k)$ and $P_{\delta_\chi \delta_\chi}^L(k)$, see Eq.~\eqref{eq:all_counterterms}. Then all other effects arising at the nonlinear level are very degenerate with $b_\chi$, or too small given BOSS errorbars. A similar argument applies to any difference between our prescription in Eq.~\eqref{eq:prescription_ula} and the numerical solution, see Sec.~\ref{sec:p_accuracy_P}.
In total we vary, at fixed $m_a$, the six $\Lambda$CDM cosmological parameters plus the energy density in the axion, $\Omega_a h^2$, and 16 nuisance parameters for each of the four sets of BOSS galaxy power spectra. Finally, the new scale dependences introduced by the $\chi$ perturbations and by the linear growth rate of the cold species are neglected inside redshift space loops, as discussed in Sec.~\ref{sec:prescription_zspace}. A more detailed description of the setup, code and priors on cosmological and nuisance parameters used in the analysis is reported in Appendix~\ref{app:app_3}.

\subsection{Discussion}
\begin{figure*}[t]
\centering
\includegraphics[width=0.85\textwidth]{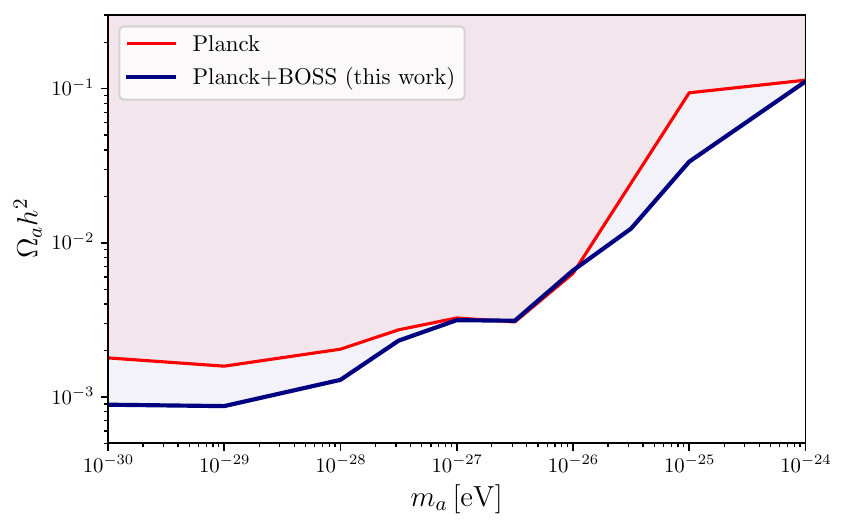} 
\caption{Our final constraints on ULAs from Planck+BOSS (blue), compared to those from Planck only (red). For each value of the ULA mass, $m_a$, we report the $95\%$~CL upper bound on its energy density today, $\Omega_a h^2$.}
\label{fig:ULAconstraints-final}
\end{figure*}

The constraints from Planck CMB data in the $(m_a, \Omega_a h^2)$ plane are shown by the red curve in Fig.~\ref{fig:ULAconstraints-final} and reported in Table~\ref{tab:constraints}. The shape of these bounds is well known~\cite{Hlozek:2014lca,Hlozek:2016lzm,Hlozek:2017zzf,Rogers:2023ezo} and represents the baseline which the addition of LSS data can further improve upon. 

For $m_a \gg H(a_{\rm eq}) \sim 10^{-28}\;\mathrm{eV}$ the ULAs transition to a DM behavior before matter-radiation equality. At the high-mass end the transition happens early during radiation domination, thus leaving the pre- and post-recombination physics unchanged compared to a $\Lambda$CDM Universe. Consequently, the constraining power of the primary CMB evaporates quickly, as we observe in Fig.~\ref{fig:ULAconstraints-final} for $m_a \gtrsim 10^{-26}$ eV. For lighter ULAs, such that the transition from $w_a \simeq -1$ to $w_a \simeq 0$ happens close to equality and hence recombination, the diffusion damping scale is modified and the evolution of the potential wells in which the baryon-photon plasma propagates is affected, leading to an early Integrated Sachs-Wolfe (ISW) contribution. These effects impact the relative heights of the acoustic peaks. The CMB lensing power spectrum is not measured with sufficient precision by Planck to alter significantly the above conclusions, though this is expected to change~\cite{Hlozek:2016lzm} with the recent ACT results~\cite{ACT:2023dou} and future data from the Simons Observatory~\cite{SimonsObservatory:2025wwn} and CMB-S4~\cite{CMB-S4:2016ple}.

\renewcommand{\tabcolsep}{4.5pt}
\begin{table}[b]
\centering
\begin{tabular}{cccl}
$m_a/\mathrm{eV}$ & $\Omega_a h^2 <$ (Planck) & $\Omega_a h^2 <$ (Planck+BOSS)   \\\hline
$10^{-24}$   & $0.11$ & $0.11$   \\ 
$10^{-25}$   &  $0.099$  & $0.034$   \\ 
 $10^{-26}$   &  $0.0063$  & $0.0066$   \\ 
 $10^{-27}$   &   $0.0033$ & $0.0031$  \\ 
$10^{-28}$    &  $0.0020$ &  $0.0013$  \\
$10^{-29}$    &  $0.00158$ &  $0.00087$  \\
$10^{-30}$    &  $0.00179$ &  $0.00089$  \\\hline
\end{tabular}
\caption{$95\%$ CL upper bounds on the ULA energy density obtained in this work, as a function of the ULA mass. We report bounds from Planck CMB data alone and from the combination of Planck with BOSS galaxy power spectrum measurements.}
\label{tab:constraints}
\end{table}

For $m_a < H(a_{\rm eq})$ the ULAs retain $w_a \simeq -1$ until some time during matter domination (or even $\Lambda$ domination, for the lightest masses). If the redshift of matter-radiation equality is kept fixed, the presence of the axion modifies the expansion history post-recombination and in particular the distance to the last scattering surface, thereby shifting the location of the CMB acoustic peaks. This shift can be compensated by a different value of the Hubble constant $H_0$, and consequently (assuming flatness) of the energy density in $\Lambda$, leaving a residual modification of the late ISW amplitude. These effects lead to the strongest Planck constraints on ULAs, $f_\chi \lesssim 0.01$ for $10^{-30} < m_a/\mathrm{eV} < 10^{-28}$. For even smaller masses (not shown in our plots), the modification of the distance to last scattering is progressively reduced and the constraints arise primarily from the ISW effect, resulting in weaker sensitivity~\cite{Lague:2021frh, Rogers:2023ezo}. Finally, as $m_a$ approaches $H_0\sim 10^{-33}\;\mathrm{eV}$ the constraint on $\Omega_a h^2$ disappears as the ULAs cannot be distinguished from a Cosmological Constant.

\begin{figure*}[t]
    \includegraphics[width=0.46\textwidth]{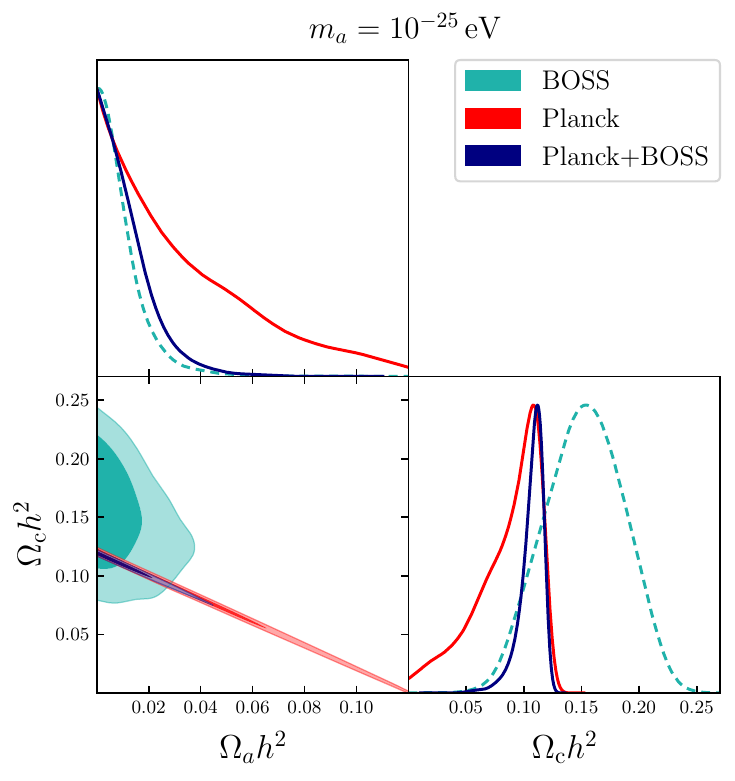} 
        \quad
    \includegraphics[width=0.466\textwidth]{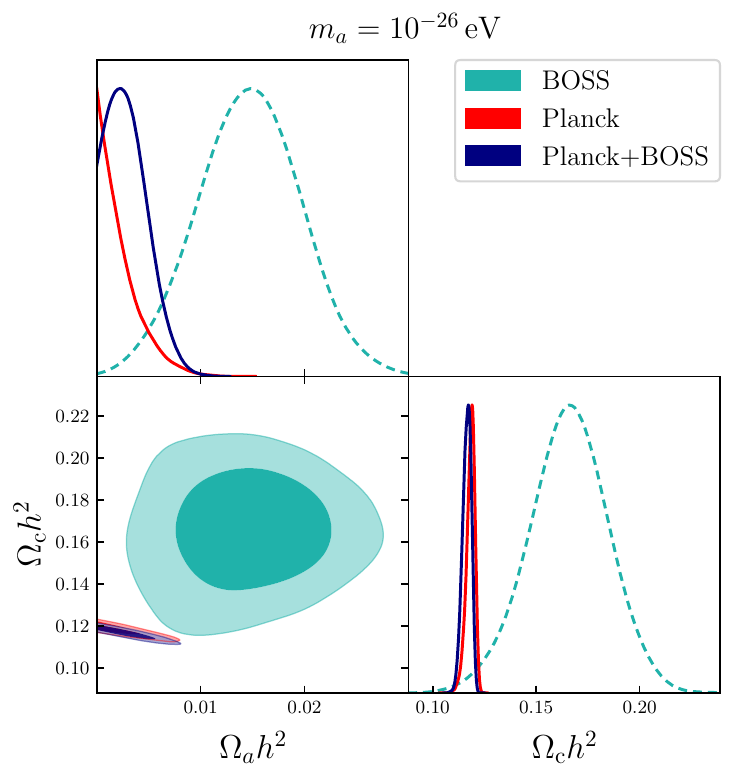} 
    \caption{Constraints on the parameter space of ULAs, at $68\%$ and $95\%$~CL, for $m_a=10^{-25}\, \mathrm{eV}$ (\textit{left}) and $m_a=10^{-26}\, \mathrm{eV}$ (\textit{right}). Red contours correspond to Planck alone, cyan contours to BOSS only, and blue contours to our final Planck+BOSS combination.}
    \label{fig:ULA-PB-m25-m26}
\end{figure*}

The LSS data are sensitive to a similar range of masses, bracketed by the ULA becoming indistinguishable from a Cosmological Constant at the lower end and from standard CDM at the upper end. The latter happens when the scale $k_{\rm drop}$ (and therefore also the late-time Jeans scale $k_s(z_{\rm obs})$) becomes larger than the nonlinear scale $k_{\rm NL}$, hence the ULA behaves as CDM in the whole perturbative range where the EFTofLSS applies. Our Planck+BOSS constraints are shown by the blue line in Fig.~\ref{fig:ULAconstraints-final} and listed in Table~\ref{tab:constraints}. We find that the addition of LSS data improves the bounds by roughly a factor of $2$ for lower ULA masses, mostly driven by a better determination of the matter density, which helps reduce the geometric degeneracies of CMB data. At the higher end, by setting $k_{\rm drop} \sim k_{\rm NL} \sim 0.4\,h\;\mathrm{Mpc}^{-1}$ we estimate that $m_a \sim 10^{-25}\;\mathrm{eV}$ is the largest mass that can be meaningfully constrained, as confirmed by Fig.~\ref{fig:ULAconstraints-final}. A subset of the contour plots for $m_a = 10^{-25}$~eV is shown in the left panel of Fig.~\ref{fig:ULA-PB-m25-m26}: the constraints are dominated by BOSS (cyan), while Planck data cannot distinguish between ULAs and standard CDM, as shown by the very elongated red contour in the direction of constant $\Omega_a h^2 + \Omega_c h^2$. The BOSS sensitivity arises from shape information, as it probes the suppression of the galaxy power spectrum at the largest perturbative $k$ included in our analysis (see the left panel of Fig.~\ref{fig:Q0} in Appendix~\ref{app:app_3}). The combination of both datasets is slightly less constraining than BOSS alone, which we attribute to projection effects. Similarly to Ref.~\cite{Rogers:2023ezo}, we find a preference for non-zero $\Omega_a h^2$ at $m_a = 10^{-26}$~eV, whose posterior is shown in the right panel of Fig.~\ref{fig:ULA-PB-m25-m26}. This is likely due again to projection effects after marginalizing over the bias and EFT nuisance parameters~\cite{Rogers:2023ezo}, since not only is the value of $\Omega_a h^2$ preferred by BOSS incompatible with Planck, but also the other $\Lambda$CDM parameters, especially $A_s$ and $n_s$, present significant shifts compared to the best fit CMB values.

\begin{figure*}[t]
\centering
\includegraphics[width=0.85\textwidth]{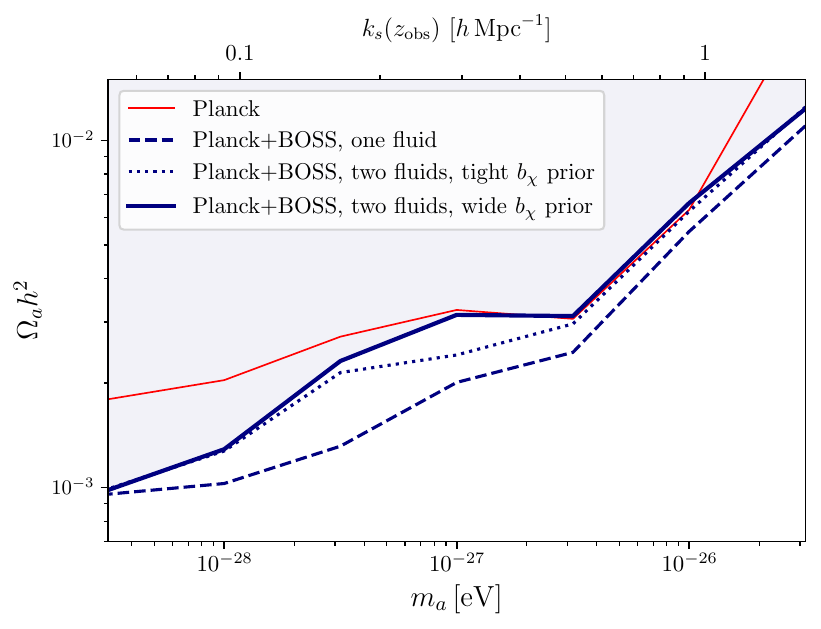} 
\caption{Impact of our refined theoretical modeling of the galaxy power spectrum on the constraints on the ULA energy density. All curves show $95\%$~CL upper bounds. Results from the Planck+BOSS combination are reported in blue, while the red curve corresponds to Planck alone. Relative to Fig.~\ref{fig:ULAconstraints-final}, we zoom in on the mass range where the effect of our two-fluid EFTofLSS extension is important. The solid blue curve shows our baseline results, whereas the dashed blue curve corresponds to a single-fluid analysis following Ref.~\cite{Rogers:2023ezo}. The dotted blue curve reports the constraints when restricting the prior on $b_\chi$ from $\mathcal{O}(1)$ to a tighter interval of $\mathcal{O}(f_\chi)$, see the main text for details. The Jeans scale $k_s$ is evaluated at $z_{\rm obs} =0.5$.}
    \label{fig:ULAconstraints-theory}
\end{figure*}

In Fig.~\ref{fig:ULAconstraints-theory} we quantify the effect on the bounds of our refined model of the galaxy power spectrum. In the displayed ULA mass range, the addition of the new bias parameter $b_\chi$ and of the accompanying new counterterms (solid blue curve, matching Fig.~\ref{fig:ULAconstraints-final}) markedly reduces the improvement over Planck brought by a single-fluid analysis of LSS data (dashed blue curve) we performed following closely those already present in the literature~\cite{Rogers:2023ezo,Lague:2021frh}.\footnote{We checked that our single fluid analysis reproduces to high accuracy the results presented in Ref.~\cite{Rogers:2023ezo}.} As expected, the effect of marginalizing over a larger set of EFT nuisance parameters starts being important once the Jeans scale is well inside the region of wavenumbers probed by galaxy surveys, $k_s(z_{\rm obs})\gtrsim 0.05\,h\,\mathrm{Mpc}^{-1}$. The effect is most pronounced for $k_s(z_{\rm obs}) \sim k_{\rm NL}$, corresponding to $m_a \sim 10^{-27}\;\mathrm{eV}$, and eventually becomes negligible again when the mass is so large that $k_{\rm drop}$ is beyond the smallest scales included in our analysis, hence BOSS cannot distinguish the ULA fluid from standard CDM ($k_{\rm drop} > 0.4\,h\,\mathrm{Mpc}^{-1}$, corresponding to $k_s (z_{\rm obs}) \gtrsim 3\, h\,\mathrm{Mpc}^{-1}$).

Finally, some of the information gain over Planck is recovered if one is able to impose a less conservative prior on the linear bias parameter $b_
\chi$, which one could expect to be of $\mathcal{O}(f_\chi)$ based on the discussion in Sec.~\ref{sec:nonlinear_zspace}. This is shown by the dotted curve in Fig.~\ref{fig:ULAconstraints-theory}, obtained by restricting the prior on $b_\chi$ from $\mathcal{U}(-3, 3)$, as assumed in our baseline analysis, to $\mathcal{U}(-0.2, 0.2)$. Complete details on the analysis presented in this section, as well as additional contour plots, are provided in Appendix~\ref{app:app_3}.

Since we focus on the region of ULA masses where our novel two-fluid modeling has an impact, we do not show results for $m_a < 10^{-30}\;\mathrm{eV}$. In this range $k_s(z_{\rm obs})$ is so small that a single-fluid analysis is fully appropriate. Moreover, the accidental suppression of the coefficients multiplying new momentum structures in, e.g.,~Eqs.~\eqref{eq:c_2nd_hot} implies that using the standard EdS kernels supplemented by the $1 - 3 f_\chi/5$ suppression of the linear growth rate suffices for all practical purposes.

\section{Outlook}\label{sec:outlook}
Having already tested the endurance of the reader to a significant degree, we will keep this outlook to a minimum, by only mentioning some of the many directions in which this work can be expanded. 

Our two-fluid extension of the EFTofLSS lays the foundation to test mixed DM scenarios with future LSS data from DESI, Euclid and Rubin. This includes the ultra-light axions we have discussed here, but also massive thermal relics, which we have only considered briefly and will be the focus of a separate publication~\cite{modelspaper}, as well as models where a fraction of DM develops a high temperature due to interactions with other species. 

More generally, the framework we have introduced opens up the possibility to derive ``model-independent'' constraints in the space of mixed DM models. For each choice of the parameters $\gamma$ and $p$ that determine the time- and scale-dependence of the sound speed (see Eq.~\eqref{eq:cs_eff2}), one may express, with minimal assumptions, the cosmological constraints in terms of the fraction of the energy density of the non-cold species, $f_\chi$, versus its characteristic scale, $k_s$. Such constraints could then be projected onto any microscopic models of interest that feature a sound speed with the assumed scaling dependences. This strategy is part of a broader program aimed at identifying a representative set of scenarios for BSM physics in the dark sector, which can be tested with LSS data. Carrying it out will likely require one to solve the nonlinear equations considered in this work for other time dependences of the sound speed term, going beyond the $\gamma = 2$ case we have focused on. 

Furthermore, the exact analytical expressions we have derived for nonlinear kernels in certain kinematic configurations provide the starting point to investigate the signatures of mixed DM scenarios in higher-point correlation functions. Finally, this work can serve as a stepping stone to investigate the effects of BSM dynamics beyond the perfect fluid approximation. We intend to return to several of the above questions in the near future. 

\section*{Acknowledgments}
We thank Diego Redigolo for collaboration in the early stages of this work and for many helpful discussions. We acknowledge insightful conversations with Emilio Bellini, Mikhail Ivanov, Marco Marinucci and Oliver Philcox. We also thank the participants of the workshops ``New Physics from Galaxy Clustering III'' in Parma, ``Cosmological Probes of New Physics'' at the University of Notre Dame and ``New Physics from Galaxy Clustering at GGI'' in Florence for stimulating interactions. F.~V. and E.~Se.~acknowledge support from the Theory Grant 2023 ``NeuMass'' of the Italian National Institute for Astrophysics (INAF). E.~Sa.~was supported in part by the Science and Technology Facilities Council under the Ernest Rutherford Fellowship ST/X003612/1. E.~Sa.~was also partly supported by the grant RYC2023-042775-I, funded by the Spanish Ministry of Science, Innovation and Universities (MCIU) through the Spanish State Research Agency (AEI, 10.13039/501100011033) and by the FSE+.

\appendix

\section{Universality of Perturbation Dynamics in Scale-free Beyond-$\Lambda$CDM\\ Scenarios}\label{sec:app_1}
The ``hot'' limit of the two-fluid setup we consider, namely $k$ modes for which the clustering of $\chi$ is entirely negligible, is a regime where the dynamics of the CDM+baryon fluid $c$ is affected non-trivially by new physics, but without the appearance of a new scale. Here we show that, remarkably, the dynamics in this regime is tightly related to the one found in Refs.~\cite{Archidiacono:2022iuu,Bottaro:2023wkd} for another example of scale-free new dynamics affecting the CDM+baryon fluid: DM self-interactions mediated by a massless (scalar) field. Such tight relation is present even though the background dynamics is modified away from $\Lambda$CDM in the self-interacting DM scenario, while it is not modified in the mixed warm/cold DM scenario we study here. This result suggests a universal nature of the dynamics of cosmological perturbations in the scale-free regime, even beyond $\Lambda$CDM.

For $k$ modes where $\chi$ does not cluster, $\delta_\chi \to 0$, the Eqs.~\eqref{eq:eta_3} and~\eqref{eq:eta_4} describing the dynamics of the cold species read, after setting $\Omega_m/ f_{\Lambda \rm CDM}^2 \to 1$,
\begin{subequations}
\begin{align}
\partial_\eta \delta_c  - \Theta_c =&\, \nabla_i (\delta_c V_c^i)\,,  \\
\partial_\eta \Theta_c + \frac{1}{2} \Theta_c - \frac{3}{2}\delta_c (1 - f_\chi) =&\, \nabla_i (V_c^j \nabla_j V_c^i )\,.
\end{align}
\end{subequations}
We now define $V_c^i \equiv (1 - 3f_\chi / 5) \bar{V}_c^i$, leading to the following form for the equations up to $\mathcal{O}(f_\chi)$,
\begin{subequations}
\begin{align}
\frac{\partial \delta_c}{\partial [ \big( 1 - \tfrac{3}{5}f_\chi \big) \eta ]}  - \bar{\Theta}_c =&\, \nabla_i (\delta_c \bar{V}_c^i)\,, \label{eq:hot_c_delta}  \\
\frac{\partial \bar{\Theta}_c}{\partial [ \big( 1 - \tfrac{3}{5}f_\chi \big) \eta ]} + \frac{1}{2} \bar{\Theta}_c \Big( 1 + \frac{3}{5}f_\chi \Big) - \frac{3}{2}\delta_c \Big( 1 + \frac{1}{5}f_\chi \Big) =&\, \nabla_i (\bar{V}_c^j \nabla_j \bar{V}_c^i ) \,. \label{eq:hot_c_theta}
\end{align}
\end{subequations}
Next, we consider the evolution of the matter perturbations in a scenario where $100\%$ of DM $\chi$ self-interacts via an effectively massless scalar mediator $s$. We start from the first two of Eqs.~(A5) in Ref.~\cite{Bottaro:2023wkd} for  $\delta_m = \digamma_{\hspace{-0.75mm}\chi} \delta_\chi + (1 - \digamma_{\hspace{-0.75mm}\chi})\delta_b$ and the analogously-defined $v_m^i$, where to avoid confusion we have introduced one modification to the notation used in that reference: here we denote $\digamma_{\hspace{-0.75mm}\chi} = \bar{\rho}_\chi / (\bar{\rho}_\chi + \bar{\rho}_b)$. We define $V_m^i \equiv  - v_m^i / (f_m \mathcal{H})$ and take $\eta_m \equiv \log D_{1m}$ as time variable, obtaining
\begin{subequations}
\begin{align}
\frac{\partial \delta_m}{\partial \eta_m}  - \Theta_m =&\, \nabla_i (\delta_m V_m^i)\,,  \\
\frac{\partial \Theta_m}{\partial \eta_m} + \Theta_m \Big[\frac{3}{2}\frac{\Omega_m}{f_m^2}(1 + \digamma_{\hspace{-0.75mm}\chi} \varepsilon) - 1 \Big] - \frac{3}{2} \frac{\Omega_m}{f_m^2}\delta_m (1 + \digamma_{\hspace{-0.75mm}\chi} \varepsilon) =&\, \nabla_i (V_m^j \nabla_j V_m^i ) \,,
\end{align}
\end{subequations}
where $\varepsilon = \beta \widetilde{m}_s^2 \digamma_{\hspace{-0.75mm}\chi}$. Here $\beta$ is the self-interaction strength normalized to gravity and $\widetilde{m}_s = d \log m_\chi(s) / d s$, where $m_\chi(s)$ is the field-dependent DM mass. Finally, in matter domination we have $\Omega_m = (\bar{\rho}_\chi + \bar{\rho}_b)/\bar{\rho}_{\rm tot} =  1 - \digamma_{\hspace{-0.75mm}\chi} \varepsilon/3$ and $f_m = 1 +  \digamma_{\hspace{-0.75mm}\chi} \varepsilon$~\cite{Bottaro:2023wkd},\footnote{We thank the authors of Ref.~\cite{Costa:2025kwt} for pointing out to us that in the analytical treatment of sub-horizon perturbations in Refs.~\cite{Archidiacono:2022iuu,Bottaro:2023wkd}, $\Omega_m$ was incorrectly set to $1$. Rectifying this leads to small changes in the coefficients of some formulas. These corrections have already been made in arXiv v4 of Ref.~\cite{Archidiacono:2022iuu}, and will also be implemented in an upcoming v3 of Ref.~\cite{Bottaro:2023wkd}.} leading us to the expressions
\begin{subequations}
\begin{align}
\frac{\partial \delta_m}{\partial \eta_m}  - \Theta_m =&\, \nabla_i (\delta_m V_m^i)\,, \label{eqs_massless_5F_delta}  \\
\frac{\partial \Theta_m}{\partial \eta_m} + \frac{1}{2} \Theta_m \Big( 1 - 4\, \digamma_{\hspace{-0.75mm}\chi} \varepsilon \Big) - \frac{3}{2} \delta_m \Big(1 - \frac{4}{3}\, \digamma_{\hspace{-0.75mm}\chi} \varepsilon \Big) =&\, \nabla_i (V_m^j \nabla_j V_m^i ) . \label{eqs_massless_5F_theta}
\end{align}
\end{subequations}
We observe that making the following replacements in Eqs.~\eqref{eqs_massless_5F_delta} and~\eqref{eqs_massless_5F_theta} gives exactly Eqs.~\eqref{eq:hot_c_delta} and~\eqref{eq:hot_c_theta}:
\begin{equation}
\delta_m \to \delta_c\,, \quad \Theta_m \to \bar{\Theta}_c\,, \quad D_{1m}\to e^{(1 - \frac{3}{5}f_\chi)\eta}\,, \quad \digamma_{\hspace{-0.75mm}\chi} \varepsilon \to - \frac{3}{20}f_\chi\,.
\end{equation}
Thus, at all orders in perturbation theory, the solutions for the hot limit of the scenario we study in this paper are immediately obtained from the results for the $m$ perturbations given in Ref.~\cite{Bottaro:2023wkd}, by substituting
\begin{equation} \label{eq:replacements_App}
\delta_m \to \delta_c\,, \quad - \frac{\theta_m}{f_m \mathcal{H}} \to \Big(1 + \frac{3}{5}f_\chi \Big) \Theta_c\,, \quad D_{1m}\to e^{(1 - \frac{3}{5}f_\chi)\eta}\,, \quad \digamma_{\hspace{-0.75mm}\chi} \varepsilon \to - \frac{3}{20}f_\chi\,,
\end{equation}
where we recall that $\digamma_{\hspace{-0.75mm}\chi} \varepsilon$ was called $f_\chi \varepsilon$ in Ref.~\cite{Bottaro:2023wkd}.

\section{More Analytical Expressions for Nonlinear Kernels}\label{app:app_2}
This appendix contains additional analytical results for the nonlinear kernels in the two-fluid EFTofLSS, which did not find space in the main text. We start by recalling the expressions of the standard EdS second-order kernels in terms of the mode-coupling functions $\alpha_s$ and $\beta$,
\begin{equation}
F_2 (\vec{k}_1, \vec{k}_2) = \frac{5}{7}\alpha_s (\vec{k}_1, \vec{k}_2) + \frac{2}{7} \beta (\vec{k}_1, \vec{k}_2)\,, \quad G_2 (\vec{k}_1, \vec{k}_2) = \frac{3}{7}\alpha_s (\vec{k}_1, \vec{k}_2) + \frac{4}{7} \beta (\vec{k}_1, \vec{k}_2)\,.
\end{equation}
At second perturbative order, the configurations of the $\chi$ kernels that enter the $1$-loop power spectra have the following UV behavior at $\mathcal{O}(f_\chi^0)$,
\begin{subequations}\label{eq:chi_UV_2nd}
\begin{align}
   F_\chi^{(0)[2]}(\vec{q},\vec{k} -\vec{q}; \tilde{\eta}) & \stackrel{k_s(a)\, \ll \, q}{=} \frac{t^{(0)} (\tilde{\eta})F_2(\vec{q}, \vec{k} - \vec{q})}{g^{(0)}(\tilde{\eta}_q)\, g^{(0)}(\tilde{\eta}_{|\vec{k} - \vec{q}|})} \,,\\
   G_\chi^{(0)[2]}(\vec{q},\vec{k} -\vec{q};\tilde{\eta}) & \stackrel{k_s(a)\, \ll \, q}{=} \frac{(\partial_{\tilde{\eta}}t^{(0)}+2 t^{(0)})(\tilde{\eta}) F_2 (\vec{q}, \vec{k} - \vec{q})}{g^{(0)}(\tilde{\eta}_q)\, g^{(0)}(\tilde{\eta}_{|\vec{k} - \vec{q}|})}\,,
\end{align}
\end{subequations}
where 
\begin{equation}
t^{(0)}(\tilde{\eta}) = T^{(0)}(\sqrt{6} e^{- \tilde{\eta}/2}), \quad \mathrm{with}\quad T^{(0)}(x) = \frac{3}{10} -\frac{x^2}{20}G^{(0)}(x).
\end{equation}
The ratio $t^{(0)}/g^{(0)}\, [(\partial_{\tilde{\eta}}t^{(0)} + 2 t^{(0)})/h^{(0)}]$ smoothly transitions from $3/10~[3/5]$ at $k \ll k_s(a)$ to $1~[3/2]$ at $k \gg k_s(a)$. 

For the cold species at $\mathcal{O}(f_\chi)$ we find, in the $k \ll k_s(a) \ll q$ configuration,
\begin{subequations}\label{eq:c_2_limits_k_ks_q}
\begin{align}
   F_c^{(1)[2]}(\vec{q},\vec{k} -\vec{q}; \tilde{\eta}) &\, = \frac{3}{490} \big[19 \alpha_s (\vec{q},\vec{k} -\vec{q}) + 2\beta(\vec{q},\vec{k} -\vec{q})\big] \,, \\
   G_c^{(1)[2]}(\vec{q},\vec{k} -\vec{q};\tilde{\eta}) &\, = - \frac{6}{245} \big[\alpha_s (\vec{q},\vec{k} -\vec{q}) + 13\beta(\vec{q},\vec{k} -\vec{q})\big] \,,
\end{align}
\end{subequations}
whereas in the $k_s(a) \ll k, q$ configuration one finds the results already reported in Eq.~\eqref{eq:c_2nd_hot}, with $\vec{k}_1 = \vec{q}$ and $\vec{k}_2 = \vec{k} - \vec{q}$. Equations~\eqref{eq:c_2_limits_k_ks_q} and~\eqref{eq:chi_UV_2nd} correspond to some of the analytical limits shown in Figs.~\ref{fig:ana_vs_num} and~\ref{fig:ana_vs_num_Fchi} in the main text, respectively.

Considering for brevity only linear bias parameters, the real space galaxy kernels at second and third order are found to be
\begin{subequations}\label{eqs:Fgn_Ggn}
\begin{align}
F_g^{[2]}(\vec{k}_1, \vec{k}_2;\tilde{\eta}) =&\; b_c F_c^{[2]} (\vec{k}_1, \vec{k}_2;\tilde{\eta}) + b_\chi F_\chi^{[2]} (\vec{k}_1, \vec{k}_2;\tilde{\eta})g(\tilde{\eta}_{k_1})g(\tilde{\eta}_{k_2}) \nonumber \\ 
&\qquad + b_\Theta \Big[ G_\chi^{[2]} (\vec{k}_1, \vec{k}_2;\tilde{\eta})g(\tilde{\eta}_{k_1})g(\tilde{\eta}_{k_2}) - G_c^{[2]} (\vec{k}_1, \vec{k}_2;\tilde{\eta}) \Big]\,,  \\
G_g^{[2]}(\vec{k}_1, \vec{k}_2;\tilde{\eta}) =&\; (1 - b_v) G_{c}^{[2]}(\vk_1,\vk_2; \tilde{\eta}) + b_v G_{\chi}^{[2]}(\vk_1,\vk_2; \tilde{\eta}) g(\tilde{\eta}_{k_1})g(\tilde{\eta}_{k_2})\,, \\
F_g^{[3]}(\vec{k}_1, \vec{k}_2, \vec{k}_3;\tilde{\eta}) =&\; b_c F_c^{[3]} (\vec{k}_1, \vec{k}_2, \vec{k}_3;\tilde{\eta}) + b_\chi F_\chi^{[3]} (\vec{k}_1, \vec{k}_2,\vec{k}_3;\tilde{\eta})g(\tilde{\eta}_{k_1})g(\tilde{\eta}_{k_2}) g(\tilde{\eta}_{k_3}) \nonumber \\
+&\, b_\Theta \Big[ G_\chi^{[3]} (\vec{k}_1, \vec{k}_2, \vec{k}_3;\tilde{\eta})g(\tilde{\eta}_{k_1})g(\tilde{\eta}_{k_2})g(\tilde{\eta}_{k_3}) - G_c^{[3]} (\vec{k}_1, \vec{k}_2, \vec{k}_3;\tilde{\eta}) \Big]\,,  \\
G_g^{[3]}(\vec{k}_1, \vec{k}_2, \vec{k}_3 ;\tilde{\eta}) =&\; (1 - b_v) G_{c}^{[3]}(\vk_1,\vk_2,\vk_3; \tilde{\eta}) + b_v G_{\chi}^{[3]}(\vk_1,\vk_2, \vk_3; \tilde{\eta}) g(\tilde{\eta}_{k_1})g(\tilde{\eta}_{k_2})g(\tilde{\eta}_{k_3})\,.
\end{align}
\end{subequations}
We recall that our definition of $\tilde{\eta}$ always refers to the external momentum $\vec{k} = \vec{k}_1 + \ldots + \vec{k}_n$. Applying our prescription~\eqref{eq:prescription} to Eqs.~\eqref{eqs:Fgn_Ggn} is the first step needed to derive the exemplary redshift-space expression shown in Eq.~\eqref{eq:zspace_integral_ex1} in the main text.

\section{Details of the Ultra-Light Axion Analysis}\label{app:app_3}
In this appendix we discuss the technical background to the constraints on ULAs presented in Sec.~\ref{sec:constraints}. We combine CMB data from Planck with full-shape galaxy power spectrum data from BOSS. 
For the former we use Planck 2018 temperature, polarization and lensing angular power spectra~\cite{Planck:2018vyg,Planck:2018lbu} in the same setup as Ref.~\cite{Rogers:2023ezo}. In particular, Ref.~\cite{Rogers:2023ezo} has shown that even for ULAs whose masses are large enough that they cluster on observationally relevant scales, the impact of non-linearities~\cite{Vogt:2022bwy} on the lensing potential power spectrum can be safely ignored in the considered multipole range $L\leq 400$ and for Planck errorbars. For galaxy clustering data we follow Refs.~\cite{Rogers:2023ezo, Moretti:2023drg} in using the windowless measurements of the BOSS DR12 samples~\cite{BOSS:2015zan} at $z_{\rm eff} = 0.38$ and $z_{\rm eff} = 0.61$ provided in Ref.~\cite{Philcox:2021kcw}. As in Ref.~\cite{Rogers:2023ezo} we fit the galaxy power spectrum multipoles $P_{\ell}$ up to $k_\mathrm{max}=0.2\,h\,\mathrm{Mpc}^{-1}$ and the RSD-free combination $Q_0$ \cite{Ivanov:2021fbu} for $k\in [0.2,0.4]\,h\,\mathrm{Mpc}^{-1}$. Differently from Ref.~\cite{Rogers:2023ezo}, we do not use post-reconstruction BAO information from BOSS or BAO data from other surveys. Theoretical predictions for the galaxy power spectrum are obtained with the \texttt{PBJ} code~\cite{Moretti:2023drg, Oddo:2021iwq}, which has been integrated in the \texttt{Cobaya} framework~\cite{Torrado:2020dgo} for the purpose.

Relative to the previous analysis~\cite{Rogers:2023ezo},
our theoretical model has one new bias parameter, $b_\chi$, and additional counterterms. In practice, we fix $m_a$ and vary the parameters
\begin{align}
    \big(\Omega_a h^2,\, \Omega_c h^2,\, \Omega_b h^2,\,&  h,\, \log\,(10^{10}A_s),\, n_s,\, \tau_{\rm reio}\big)\nonumber\\
    &\times \prod_{i\, =\, 1}^4 \left(b_c,\,b_\chi,\, b_2,\, b_{\mathcal{G}_2},\, b_{\Gamma_3},\, c_{cc,\ell},\,c_{c\chi,\ell},\,c_{\chi\chi,\ell^\prime}  ,\,P_\mathrm{shot},\,a_0,\,a_2\right)_i
    \label{eq:varied-parameters}
\end{align}
where $i$ indexes the redshift/sky cut bins, while $\ell = 0,2,4$ and $\ell^\prime = 0,2$ run over the power spectrum multipoles. We assume one massive neutrino with fixed $m_\nu = 0.06\;\mathrm{eV}$. The first line of Eq.~\eqref{eq:varied-parameters} contains the cosmological parameters that are varied in the fit, whereas the second line contains the EFTofLSS parameters. The linear bias coefficients $b_c$ and $b_\chi$ are defined in Eq.~\eqref{eq:our-delta-g-r}. We set to zero $b_\Theta$ and $b_v$ (see Eqs.~\eqref{eq:delta_g} and~\eqref{eq:v_g}, respectively) because, based on the discussion in Sec.~\ref{sec:nonlinear_real}, we expect that their effects can be largely absorbed into redefinitions of $b_c$ and $b_\chi$. The nonlinear bias coefficients $b_2, b_{\mathcal{G}_2}$ and $b_{\Gamma_3}$ are given in the basis of Ref.~\cite{Chudaykin:2020aoj}; we recall that in our analysis only the perturbations of the cold species $c$ (but not $\chi$) are used to build the associated nonlinear operators. The counterterms are introduced in Eq.~\eqref{eq:all_counterterms}, those proportional to the $c_{c\chi, \ell}$ and $c_{\chi\chi, \ell^\prime}$ coefficients being new additions associated to our two-fluid treatment. The $c_{\chi\chi, 4}$ counterterm is not included in our analysis, because the associated operator $P_{\delta_g\delta_g, s}(k,\mu) \supset \mu^4 k^2 P_{\delta_\chi \delta_\chi}^L(k) $ is not generated under our assumption that $b_\Theta = b_v = 0$ (see the discussion in Sec.~\ref{sec:prescription_zspace}). We have checked that adding also $c_{\chi\chi, 4}$ or, vice versa, removing $c_{\chi\chi, 2}$, affects the bounds on $\Omega_a h^2$ by less than $5\%$.

Stochastic effects are included through~\cite{Philcox:2021kcw}
\begin{equation}\label{eq:stoch}
P_{\delta_g \delta_g}^{\,\rm stoch}(k,\mu) = \bar{n}^{-1}\bigg[ 1 + P_{\rm shot} + a_0 \Big( \frac{k}{k_{\rm NL}}\Big)^2 + a_2 \mu^2 \Big( \frac{k}{k_{\rm NL}} \Big)^2 \bigg]\,,
\end{equation}
where $k_{\rm NL} = 0.45 \,h\,\mathrm{Mpc}^{-1}$ and the background galaxy number densities are $\bar{n}^{-1} = 3500$ $[\mathrm{Mpc}/h]^3$ for the low $z_{\rm eff}$ samples and $\bar{n}^{-1} = 5000 \;[\mathrm{Mpc}/h]^3$ for the high $z_{\rm eff}$ samples~\cite{Rogers:2023ezo}.

We adopt wide uninformative priors for the cosmological parameters, as they are all well constrained by our combination of observables. By contrast, the priors we choose for the EFTofLSS parameters require a detailed discussion.

\begin{figure*}[t]
\centering
\includegraphics[width=0.49\textwidth]{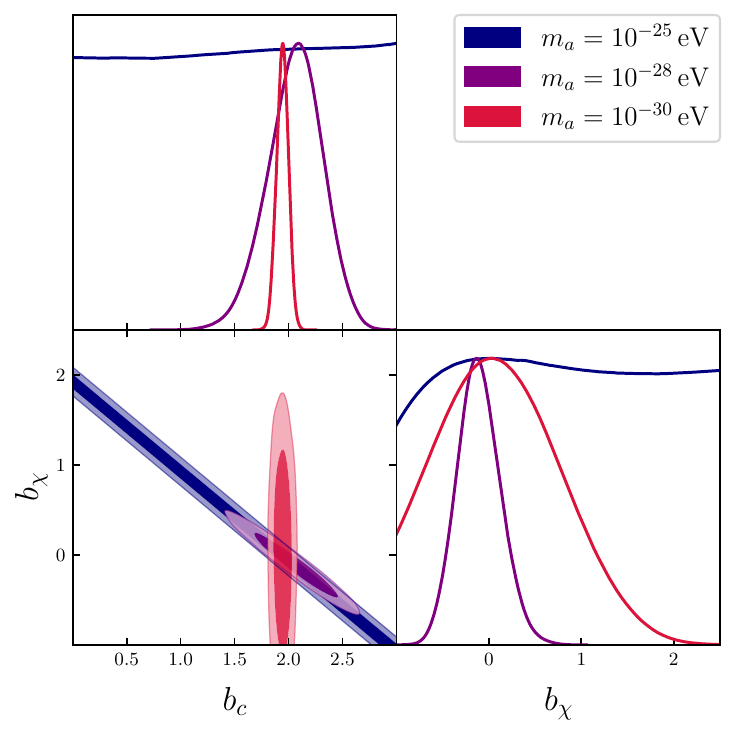}
\caption{Posteriors for the linear bias coefficients $b_c$ and $b_\chi$ of one redshift/sky cut bin ($z_{\rm eff} = 0.38$, NGC), for different ULA masses, in our Planck+BOSS analysis. The different degeneracy directions are manifest, depending on whether $k_s \gg k_{\rm NL}$~(blue), $k_s \sim k_{\rm NL}$~(purple) or $k_s \ll k_{\rm NL}$~(red).}
\label{fig:b1ba}
\end{figure*}

For the linear bias coefficients, by default we assume the wide uniform priors $b_c \sim \mathcal{U}(-1.5, 3)$ and $b_\chi \sim \mathcal{U}(-3, 3)$. This allows us to capture, in particular, the perfect degeneracy between the two parameters for large values of $m_a$: since all observationally relevant scales satisfy $k\ll k_s$, both underlying fluids are effectively cold and only the sum $b_c + b_\chi$ is constrained. This is shown in Fig.~\ref{fig:b1ba} for $m_a = 10^{-25}\;\mathrm{eV}$, corresponding to $k_s(z_{\rm obs}) \approx 3\; h\,\mathrm{Mpc}^{-1} \gg k_{\rm NL}$. For comparison we also show the degeneracy for a much smaller mass, $m_a = 10^{-30}\;\mathrm{eV}$: in this case $k \gg k_s(z_{\rm obs}) \approx 0.01\; h\,\mathrm{Mpc}^{-1}$ for all relevant scales, hence the ULA clustering is nearly negligible and only $b_c$ is significantly constrained. Finally, $m_a = 10^{-28}\;\mathrm{eV}$ shows an intermediate situation, with $k_s(z_{\rm obs}) \approx 0.1\; h\,\mathrm{Mpc}^{-1}$ falling in the middle of the observational window: in this case both linear bias parameters are significantly constrained. The impact of modifying the prior on $b_\chi$ to a narrow informative one, $b_\chi \sim \mathcal{U}(-0.2, 0.2)$, is discussed in Sec.~\ref{sec:constraints}. 

For the nonlinear bias coefficients we adopt the same Gaussian priors as in $\Lambda$CDM analyses~\cite{Philcox:2021kcw}, as we also do for the counterterms of the cold species $c_{cc,\ell}$. For the new counterterms, on the other hand, we perform tests to select priors that are wide enough to be uninformative. In summary, for the counterterms we assume
\vspace{0.5mm}
\begin{equation}
\frac{c_{cc,0},c_{cc,4}}{[\mathrm{Mpc}/h]^2} \sim \mathcal{N}(0,30^2) \,,\quad\;\;\frac{c_{cc,2}}{[\mathrm{Mpc}/h]^2} \sim \mathcal{N}(30,30^2) \,, \quad
\frac{c_{c\chi,\ell}, c_{\chi\chi,\ell^\prime}}{[\mathrm{Mpc}/h]^2} \sim \mathcal{N}(0,60^2)\,.
\end{equation}

\begin{figure*}[t]
\centering
\includegraphics[width=0.49\textwidth]
{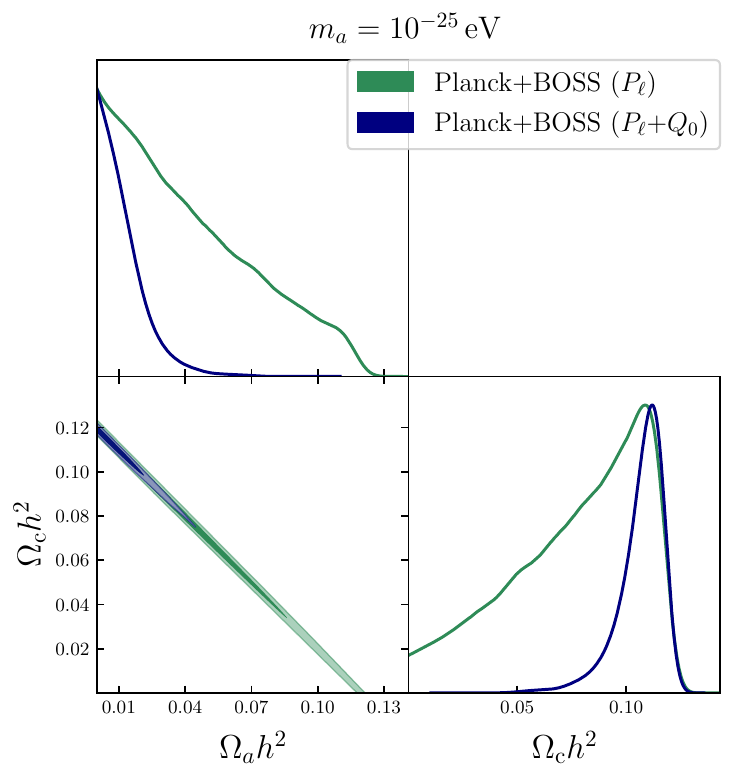}
\includegraphics[width=0.49\textwidth]
{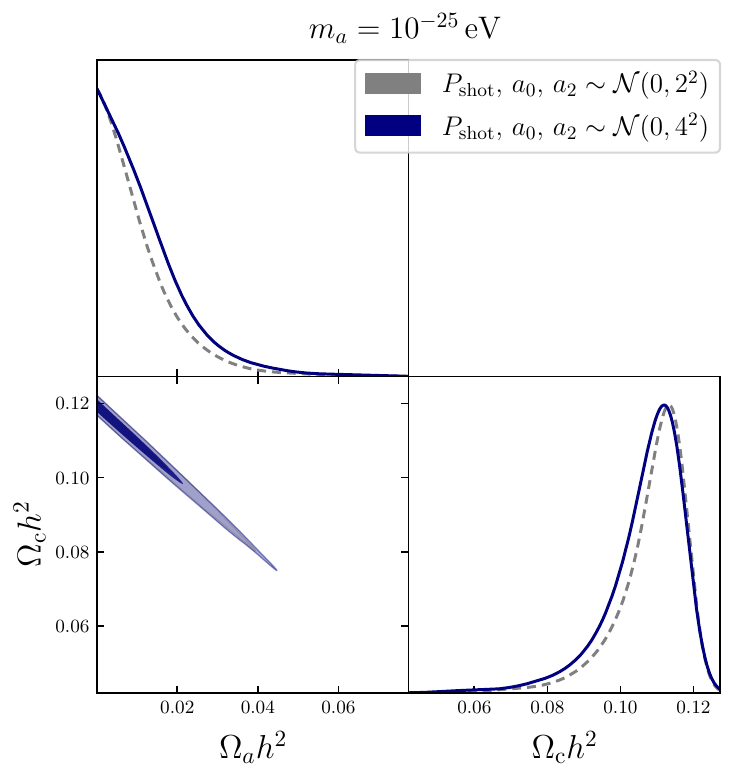}
\caption{Planck+BOSS constraints on ULAs with $m_a = 10^{-25}\;\mathrm{eV}$. \textit{Left:} Comparison of the combinations Planck+$P_\ell$ and Planck+$P_\ell$+$Q_0$, showing that the constraining power comes dominantly from the RSD-free observable $Q_0$ due to the larger wavenumbers it probes. \textit{Right:} Impact of enlarging the priors for the stochastic parameters, for the combination Planck+$P_\ell$+$Q_0$. Due to a partial degeneracy with the imprint of the ULAs, doubling the priors used for $\Lambda$CDM analyses~\cite{Philcox:2021kcw} gives a weaker constraint on $\Omega_a h^2$.}
\label{fig:Q0}
\end{figure*}

\vspace{0.5mm}
\noindent Finally, we discuss our priors on the stochastic parameters in Eq.~\eqref{eq:stoch}. For the largest ULA masses we consider, which yield a suppression of power at the smallest scales probed by our analysis, the effect of scale-dependent stochasticity is found to be partly degenerate with the one of $\Omega_a h^2$. This is illustrated in Fig.~\ref{fig:Q0} for $m_a = 10^{-25}\;\mathrm{eV}$. For this mass we estimate $k_{\rm drop} \approx 0.44 \,h\,\mathrm{Mpc}^{-1}$, hence the galaxy power spectrum is (mildly) suppressed relative to $\Lambda$CDM only in the $k$ window probed by the $Q_0$ observable, which provides all the constraining power (left panel).\footnote{Although $k_{\rm drop}$ for this ULA mass is {\it larger} than $k = 0.4\,h\,\mathrm{Mpc}^{-1}$, the largest wavenumber for which we fit $Q_0$, with our definition of $k_{\rm drop}$ the linear matter power spectrum is somewhat suppressed already at $k\lesssim k_{\rm drop}$ (this can be observed in the left panel of Fig.~\ref{fig:comparison_CLASS} for a different $m_a$), hence meaningful constraints are still obtained.} In the right panel we show how enlarging the priors on the stochastic parameters results in a weaker constraint on $\Omega_a h^2$. We checked that doubling the standard deviation of these Gaussian priors with respect to the $\Lambda$CDM analysis of Ref.~\cite{Philcox:2021kcw} saturates this effect, therefore we adopt $P_\mathrm{shot},\, a_0,\, a_2 \sim \mathcal{N}(0,4^2)$. This highlights how standard assumptions may need to be revised when considering beyond-$\Lambda$CDM dynamics.

Notice that we perform an analytical marginalization over the parameters that enter linearly in the model. This is the case of $b_{\Gamma_3}$, all the counterterm coefficients $c_{cc,\ell}$, $c_{c\chi,\ell}$ and $c_{\chi\chi,\ell^\prime}$ and the stochastic parameters $P_\mathrm{shot},\, a_0$ and $a_2$. 

As a validation of our setup, we reproduced with high accuracy the Planck constraints presented in Ref.~\cite{Rogers:2023ezo}. In addition, we reproduced with high accuracy their Planck+BOSS constraints by restricting the theoretical model presented above to a single-fluid description and setting the priors to the standard $\Lambda$CDM values~\cite{Philcox:2021kcw,Rogers:2023ezo}.

\section{Impact of Intrinsic Nonlinearities}\label{app:app_4}
In this work we described the warm species as a perfect fluid characterized by a sound speed $c_s^2$ at linear level, and with nonlinearities of pure gravitational origin. This is, however, an approximation, as in general each microscopic model can predict intrinsic nonlinearities, too. In this Appendix we discuss such intrinsic nonlinearities for the two benchmark models studied in this paper: ULAs, where they arise from the wave nature of the scalar field, and light but massive thermal relics, where they originate from the Vlasov (or collisionless Boltzmann) equation. In both cases, we demonstrate quantitatively that neglecting the intrinsic nonlinearities has a minor impact on our results.

\subsection{Wave Nonlinearities for Ultra-Light Axions}\label{app:app_4_ULAs}

In this work we approximated ULAs as a perfect fluid characterized by a sound speed $c_s^2 = k^2/ (4m_a^2 a^2)$ at linear level, and with nonlinearities given only by the gravitational mode coupling. However, it is well known that, after taking the nonrelativistic approximation for the axion field $\phi$, the resulting Schr\"odinger-Poisson system can be recast into a fluid dynamics formulation that also includes nonlinearities intrinsic to the wave-like nature of the ULAs (see Refs.~\cite{Li:2018kyk,Kousha:2023jzt} for previous discussions in the scenario where heavier scalars with $m_a \sim 10^{-22}\;\mathrm{eV}$ form the entirety of DM). In our framework, this formulation is reproduced by making in the Euler equation for $\chi$, Eq.~\eqref{eq:Euler_chi}, the substitution
\begin{equation} \label{eq:substitution_QP}
+\frac{1}{\bar{\rho}_\chi} \nabla^2 \delta \mathcal{P}_\chi \quad \to \quad -\, \frac{1}{2 m_a^2 a^2}\, \nabla^2 \frac{\nabla^2 \sqrt{\rho_\chi}}{\sqrt{\rho_\chi}} = -\, \frac{1}{4 m_a^2 a^2} \nabla^2 \bigg[ \frac{\nabla^2 \delta_\chi}{1+\delta_\chi} - \frac{1}{2} \frac{\nabla_i \delta_\chi \nabla^i \delta_\chi}{(1 + \delta_\chi)^2} \bigg]\,.
\end{equation}
The continuity equation~\eqref{eq:continuity_chi} is unchanged. The quantity $- \nabla^2 \sqrt{\rho}_\chi\, / \sqrt{\rho_\chi}$ is known as quantum pressure and arises from a stress-energy tensor containing off-diagonal terms~\cite{Li:2018kyk}. In this paper we retained only the linear piece of the quantum pressure in Eq.~\eqref{eq:substitution_QP}, which led us to the perfect fluid formulation. Here, we quantitatively show that neglecting the intrinsic wave nonlinearities has a minor effect on our results.

After changing the time variable to $\eta = \log D_{\Lambda \mathrm{CDM}}$, setting $\Omega_m / f_{\Lambda \rm CDM}^2 \to 1$ and going to Fourier space, we find that Eq.~\eqref{eq:eta_kspace_nonlinear_initial_chi_euler} is extended to 
\begin{align}
&\partial_{\eta} \Theta_\chi(\vec{k},\eta) +  \frac{1}{2}\Theta_\chi(\vec{k},\eta) -\frac{3}{2} [f_\chi \delta_\chi(\vec{k},\eta) + (1 - f_\chi) \delta_c(\vec{k},\eta)] \nonumber \\
&=\int_{\bf{k}} \mathrm{d} k_{12} \beta(\vec{k}_1, \vec{k}_2)\Theta_\chi(\vec{k}_1, \eta)\Theta_\chi(\vec{k}_2, \eta) \nonumber \\ 
&\quad+ \frac{c_s^2 k^2}{\mathcal{H}^2 \Omega_m} \sum_{r\, =\, 1}^{\infty} (-1)^r \int_{\vec{k}} \mathrm{d} k_{1\ldots r}\,  P_r (\vec{k}_1, \ldots, \vec{k}_r) \delta_\chi (\vec{k}_1, \eta) \ldots \delta_\chi (\vec{k}_r, \eta)\,, \label{eq:eta_kspace_nonlinear_initial_chi_euler_QP}
\end{align}
where we defined
\begin{equation}
P_r (\vec{k}_1, \ldots, \vec{k}_r) \equiv \frac{1}{2r} \,\bigg[ 1 + \frac{\sum_{i \,=\, 1}^r k_i^2}{ (\sum_{i \,=\, 1}^r \vec{k}_i)^2} \bigg]\,.
\end{equation}
The $r = 1$ term in the third line of Eq.~\eqref{eq:eta_kspace_nonlinear_initial_chi_euler_QP} reproduces the linear sound speed term in Eq.~\eqref{eq:eta_kspace_nonlinear_initial_chi_euler}, whereas the $r\geq 2$ terms represent the intrinsic ULA nonlinearities. 

Finally, switching to the $k$-dependent time variable $\tilde{\eta}$ we arrive at the extended version of Eq.~\eqref{eq:eta_kspace_nonlinear_chi_euler}, 
\begin{align}
&\partial_{\tilde{\eta}} \Theta_\chi(\vec{k},\tilde{\eta}) +  \frac{1}{2}\Theta_\chi(\vec{k},\tilde{\eta}) + \frac{3}{2} \, e^{-\tilde{\eta}} \delta_\chi(\vec{k},\tilde{\eta}) -\frac{3}{2} [f_\chi \delta_\chi(\vec{k},\tilde{\eta}) + (1 - f_\chi) \delta_c(\vec{k},\tilde{\eta})] \nonumber \\
&=\int_{\bf{k}} \mathrm{d} k_{12} \beta(\vec{k}_1, \vec{k}_2)\Theta_\chi(\vec{k}_1, \tilde{\eta})\Theta_\chi(\vec{k}_2, \tilde{\eta}) + \frac{3}{2}e^{-\tilde{\eta}}  \int_{\vec{k}} \mathrm{d}k_{12} P_2 (\vec{k}_1, \vec{k}_2) \delta_\chi (\vec{k}_1, \tilde{\eta}) \delta_\chi (\vec{k}_2, \tilde{\eta}) \nonumber\\
&\quad - \frac{3}{2}e^{-\tilde{\eta}}  \int_{\vec{k}} \mathrm{d}k_{123} P_3(\vec{k}_1, \vec{k}_2, \vec{k}_3) \delta_\chi (\vec{k}_1, \tilde{\eta}) \delta_\chi (\vec{k}_2, \tilde{\eta}) \delta_\chi (\vec{k}_3, \tilde{\eta}) + \ldots\label{eq:eta_kspace_nonlinear_chi_euler_QP} 
\end{align}
where we used the fact that $\gamma =2$ for ULAs. The dots in Eq.~\eqref{eq:eta_kspace_nonlinear_chi_euler_QP} indicate terms with at least four fields, which we omitted since they are not needed for the calculation of the power spectra up to $1$-loop order. In summary, to include intrinsic wave nonlinearities in a mixed DM scenario containing a fraction $f_\chi$ of ULAs, one needs to solve the four coupled equations~\eqref{eq:eta_kspace_nonlinear_chi_cont},~\eqref{eq:eta_kspace_nonlinear_chi_euler_QP},~\eqref{eq:eta_kspace_nonlinear_c_cont} and~\eqref{eq:eta_kspace_nonlinear_c_euler}.

To gain some insight on the impact of the intrinsic ULA nonlinearities, we derive analytical solutions for the kernels at second perturbative order. For $n=2$, Eq.~\eqref{eq:eta_kspace_nonlinear_chi_euler_QP} reads
\begin{align}
&\Big(\partial_{\tilde{\eta}} + \frac{1}{2} + \frac{h}{g}(\tilde{\eta}_{k_1})+\frac{h}{g}(\tilde{\eta}_{k_2}) \Big) G_\chi^{[2]} + \frac{3}{2}\Big(e^{-(\gamma - 1)\tilde{\eta}} - f_\chi\Big) F_\chi^{[2]} -\frac{3}{2}(1 - f_\chi) \frac{F_c^{[2]}}{g(\tilde{\eta}_{k_1}) g(\tilde{\eta}_{k_2})} \nonumber \\ \label{eq:chi_2nd_B_QP}
&\hspace{6.0cm} = \beta(\vec{k}_1, \vec{k}_2) \frac{h}{g}(\tilde{\eta}_{k_1})\frac{h}{g} (\tilde{\eta}_{k_2}) + \frac{3}{2}e^{-\tilde{\eta}} P_2 (\vec{k}_1, \vec{k}_2)\,,
\end{align}
which extends Eq.~\eqref{eq:chi_2nd_B}. At $\mathcal{O}(f_\chi^0)$ and considering the $k_1 \ll k_s \ll k_2$ kinematic configuration, the solutions for the $\chi$ kernels in Eqs.~\eqref{eq:limit_chi} are extended to
\begin{subequations} \label{eq:limit_chi_QP}
\begin{align}\label{eq:limit_chi_A_QP}
F_\chi^{(0)[2]}(\vec{k}_1, \vec{k}_2; \tilde{\eta}) =\;& F_2(\vec{k}_1, \vec{k}_2) + P_2 (\vec{k}_1, \vec{k}_2) \,, \\ G_\chi^{(0)[2]}(\vec{k}_1, \vec{k}_2; \tilde{\eta}) =\;& 2\hspace{0.3mm} G_2(\vec{k}_1, \vec{k}_2) - \frac{3}{14}\alpha_s(\vec{k}_1, \vec{k}_2) + \frac{1}{2}\alpha_a (\vec{k}_1, \vec{k}_2) - \frac{2}{7}\beta(\vec{k}_1, \vec{k}_2) + 3 P_2 (\vec{k}_1, \vec{k}_2)  \,,\label{eq:limit_chi_B_QP}
\end{align}
\end{subequations}
whereas the solutions for the cold species in Eqs.~\eqref{eq:limit_c} are unaffected, because they are sourced only by gravity in this configuration. The solutions for $c$ in the hot limit, Eqs.~\eqref{eq:c_2nd_hot}, are also unaffected. Importantly, since the $P_2(\vec{q}, \vec{k} - \vec{q})$ kernel does not have a $1/q$ pole, the IR structure discussed in Sec.~\ref{sec:IR} is not altered. The IR structure is unchanged even at higher perturbative orders, hence one still arrives at the prescription we presented in Sec.~\ref{sec:prescription}. 

Finally, we revisit the UV structure of the second-order $\chi$ kernels. At $\mathcal{O}(f_\chi^0)$, Eqs.~\eqref{eq:chi_UV_2nd} are extended to
\begin{subequations}\label{eq:chi_UV_2nd_QP}
\begin{align}
   F_\chi^{(0)[2]}(\vec{q},\vec{k} -\vec{q}; \tilde{\eta}) & \stackrel{k_s\, \ll \, q}{=} \frac{t^{(0)} (\tilde{\eta})F_2(\vec{q}, \vec{k} - \vec{q})}{g^{(0)}(\tilde{\eta}_q)\, g^{(0)}(\tilde{\eta}_{|\vec{k} - \vec{q}|})} + c^{(0)}(\tilde{\eta}) P_2 (\vec{q}, \vec{k} - \vec{q})  \,,\\
   G_\chi^{(0)[2]}(\vec{q},\vec{k} -\vec{q};\tilde{\eta}) & \stackrel{k_s\, \ll \, q}{=} \frac{(\partial_{\tilde{\eta}}t^{(0)}+2 t^{(0)})(\tilde{\eta}) F_2 (\vec{q}, \vec{k} - \vec{q})}{g^{(0)}(\tilde{\eta}_q)\, g^{(0)}(\tilde{\eta}_{|\vec{k} - \vec{q}|})} + (\partial_{\tilde{\eta}} c^{(0)} + 4 c^{(0)} )(\tilde{\eta})   P_2 (\vec{q}, \vec{k} - \vec{q})\,,
\end{align}
\end{subequations}
where
\begin{equation} \label{eq:c0_def}
c^{(0)}(\tilde{\eta}) = C^{(0)}(\sqrt{6} e^{- \tilde{\eta}/2}), \quad \mathrm{with}\quad C^{(0)}(x) = \frac{x^2}{42} - \frac{x^4}{840} + \frac{x^6}{5040} G^{(0)}(x)\,. 
\end{equation}
We find that $c^{(0)}\, [\partial_{\tilde{\eta}} c^{(0)} + 4 c^{(0)}] \to e^{-\tilde{\eta}}/7\,\, [3 e^{-\tilde{\eta}}/7]$ for $k \ll k_s$ and $\to 1\,[4]$ for $k \gg k_s$. We observe that, even in the $k\gg k_s$ regime where the wave nonlinearities are relevant, the new pieces on the right-hand sides of Eqs.~\eqref{eq:chi_UV_2nd_QP} are strongly suppressed by the linear transfer functions, relative to the perfect-fluid solutions proportional to $F_2$. We thus conclude that the wave nonlinearities give negligible contributions to the kernels in the UV limit.

\begin{figure*}[b]
    \includegraphics[width=\textwidth]{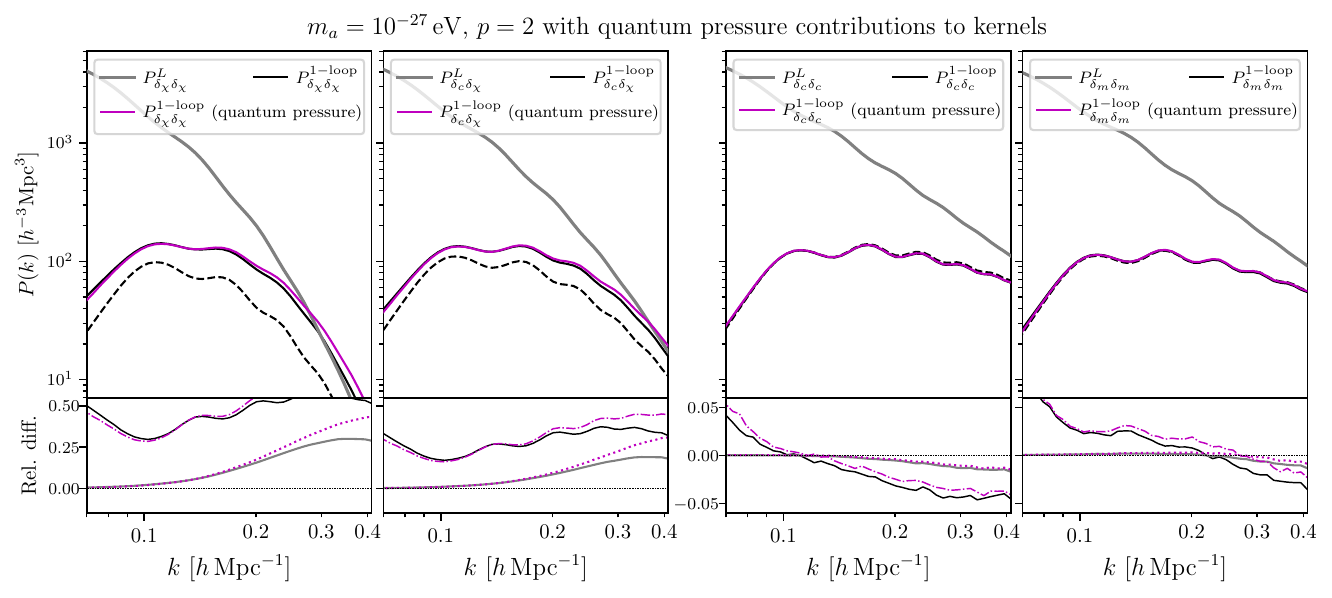}
    \caption{Impact of including intrinsic wave nonlinearities due to quantum pressure in the computation of the $1$-loop terms of the matter power spectra, for a mixed DM scenario containing ULAs. Black and gray lines are identical to the top row of Fig.~\ref{fig:P_comp_ULA_nu}, while the purple lines correspond to the full numerical solutions including wave nonlinearities. From left to right: $\chi \chi, c \chi, c c$ and $m m$ density power spectra at $z_{\rm obs} = 0.5$, representative of the redshift observed by galaxy surveys. We have used as inputs the linear power spectra for an ULA cosmology with $m_a=10^{-27}\;\mathrm{eV}$ and $f_\chi = 0.1$, corresponding to $p = 2$ and $k_s(z_{\rm obs}) \approx 0.29\, h\,\mathrm{Mpc}^{-1}$.}
    \label{fig:P_comp_ULA_QP}
\end{figure*}

In summary, by solving the second-order equations we have shown that the intrinsic wave nonlinearities do not modify the IR and UV limits of the kernels with respect to the perfect fluid description assumed in this paper. As illustrated by Eqs.~\eqref{eq:limit_chi_QP}, away from the IR and UV limits modifications of $\mathcal{O}(1)$ are expected for the $\chi$ kernels, as long as the external momentum is comparable to or larger than the characteristic scale. We expect these results to hold even at higher orders. This is confirmed by Fig.~\ref{fig:P_comp_ULA_QP}, where we compare the $1$-loop matter power spectra obtained by solving numerically the equations with (purple) and without (black and gray) the wave nonlinearities associated with the quantum pressure. As expected, we observe mild differences in the $\chi\chi$ and $c\chi$ power spectra at $k\gtrsim k_s$, which will be anyway reabsorbed by counterterms (recall Fig.~\ref{fig:P_gg_breakdown}). There are also small differences at $k\lesssim 0.1\,h\,\mathrm{Mpc}^{-1}$, but these are due to the finite numerical precision. The impact on the $cc$ power spectrum, instead, is tiny. These results demonstrate that our prescription for the evaluation of the $1$-loop galaxy power spectrum remains valid even when wave nonlinearities are considered. Nonetheless, it would be interesting to study the impact of the latter on other observables that can be calculated exactly, such as the tree-level bispectrum. We leave this for future work.

\subsection{Nonlinearities from the Vlasov Equation}\label{app:app_4_Vlasov}
To examine the role of intrinsic nonlinearities for light thermal relics, we start from the Vlasov equation for the particle phase space density $f_\chi (\vec{x}, t, \vec{p})$,
\begin{equation}
\frac{\mathrm{d} f_\chi}{\mathrm{d}t} = \frac{\partial f_\chi}{\partial t} + \frac{\mathrm{d} x^i}{\mathrm{d}t} \frac{\partial f_\chi}{\partial x^i} + \frac{\mathrm{d} p^i}{\mathrm{d}t} \frac{\partial f_\chi}{\partial p^i} = 0\,,
\end{equation}
where $\vec{x}$ denotes comoving coordinates while $\vec{p}$ is the physical momentum. In the nonrelativistic limit and in the sub-horizon regime, taking the first two moments of the Vlasov equation yields
\begin{align}
\rho_\chi^\prime + 3 \mathcal{H} \rho_\chi + \nabla_i (\rho_\chi v_\chi^i) =&\; 0\,, \\
(\rho_\chi v_\chi^i)^\prime + \nabla_j \Sigma^{ij}_\chi + 4 \mathcal{H} \rho_\chi v_\chi^i + \rho_\chi \nabla^i \Psi =&\; 0\,,
\end{align}
where primes denote conformal time derivatives, $\rho_\chi = m_\chi \int \mathrm{d}^3 p f_\chi / (2\pi)^3$ is the energy density, $v_\chi^i = \rho_\chi^{-1} \int \mathrm{d}^3 p p^i f_\chi / (2\pi)^3$ is the peculiar velocity, and $\Sigma_\chi^{ij} = m_\chi^{-1} \int \mathrm{d}^3 p p^i p^j f_\chi/ (2\pi)^3$ is related to the stress tensor $\sigma_\chi^{ij}$ by $\Sigma^{ij}_\chi = \rho_\chi v_\chi^i v_\chi^j + \sigma_\chi^{ij}$. From these one obtains the standard continuity equation~\eqref{eq:continuity_chi} and the Euler equation in the form
\begin{equation}
\theta_\chi^\prime + \mathcal{H} \theta_\chi + \nabla_i \bigg( \frac{\nabla_j \sigma_\chi^{ij}}{\rho_\chi} \bigg) + \nabla^2 \Psi = - \nabla_i (v_\chi^j \nabla_j v_\chi^i)\,.
\end{equation}
Extensions of perturbation theory for CDM, where the velocity dispersion tensor $\sigma_\chi^{ij}/\rho_\chi$ is included as a dynamical variable, were developed in Refs.~\cite{McDonald:2009hs,Aviles:2015osc,Cusin:2016zvu,Erschfeld:2018zqg,Garny:2022tlk,Garny:2022kbk}.

We now assume that the stress tensor of the warm species contains a non-negligible pressure perturbation, while dropping viscous terms (and the negligible background pressure): $\sigma_\chi^{ij} = \delta \mathcal{P}_\chi \delta^{ij}$. After application of the Poisson equation, our final form of the Euler equation is Eq.~\eqref{eq:Euler_chi} with the substitution
\begin{equation} \label{eq:substitution_Vlasov}
+\frac{1}{\bar{\rho}_\chi} \nabla^2 \delta \mathcal{P}_\chi \quad \to \quad \, + \frac{1}{\bar{\rho}_\chi} \nabla_i \bigg( \frac{\nabla^i  \delta \mathcal{P}_\chi}{1 + \delta_\chi} \bigg) = + \frac{1}{\bar{\rho}_\chi} \bigg[ \frac{\nabla^2 \delta\mathcal{P}_\chi}{1+\delta_\chi} - \frac{\nabla_i \delta \mathcal{P}_\chi \nabla^i \delta_\chi}{(1 + \delta_\chi)^2} \bigg] \,.
\end{equation}
In this work, only the linear piece of the pressure term on the right-hand side of Eq.~\eqref{eq:substitution_Vlasov} was retained. We now demonstrate that including the additional intrinsic nonlinearities has a minor impact on our results.

After changing the time variable to $\eta$, setting $\Omega_m / f_{\Lambda \rm CDM}^2 \to 1$ and going to Fourier space, we assume that the $\chi$ sound speed is $k$-independent, $\delta \mathcal{P}_\chi (\vec{k},\eta) = c_s^2 (\eta) \delta \rho_\chi (\vec{k},\eta)$. This is appropriate, in particular, when the perturbations are adiabatic, as it is the case for thermal relics (recall that in Sec.~\ref{eq:relics} we took $c_s^2$ proportional to the adiabatic sound speed). We thus arrive at an equation that is formally identical to Eq.~\eqref{eq:eta_kspace_nonlinear_initial_chi_euler_QP} as derived above for ULAs, but with a different (and simpler) form for the intrinsic nonlinear kernels, $P_r (\vec{k}_1, \ldots, \vec{k}_r) = 1/r$. Notice that this result was already obtained in Ref.~\cite{Shoji:2009gg}. Since $\gamma = 2$ for thermal relics, Eq.~\eqref{eq:eta_kspace_nonlinear_chi_euler_QP} also holds. Then, the solutions to the equations including intrinsic nonlinearities for ULAs, which were given in Eqs.~\eqref{eq:limit_chi_QP}$\,$-$\,$\eqref{eq:c0_def}, also apply here provided the appropriate form of the $P_r$ kernels is used. As we did for ULAs, we can thus conclude that the prescription we propose in this paper remains valid even when nonlinearities intrinsic to the Vlasov equation are considered.

\bibliography{warmLSS}
\end{document}